\documentclass{kapedbk} 
\usepackage{edbkps}
\usepackage{epsfig}     
\usepackage{times}     
\usepackage{graphicx,color}       
\usepackage{epsfig,rotating}      
\usepackage{graphicx,dcolumn,xspace}
\usepackage{amsmath}
\usepackage{graphicx}
\usepackage{amsfonts}
\usepackage{amssymb}
\newcolumntype{.}{D{.}{.}{1}}
\newcommand{\outline}[1]{}
\newcommand {\gtsim} {\ {\raise-.5ex\hbox{$\buildrel>\over\sim$}}\ }
\newcommand {\ltsim} {\ {\raise-.5ex\hbox{$\buildrel<\over\sim$}}\ }
\def\Valf{\hbox{$V_{\rm A}$}}
\def\Vsound{\hbox{$V_{\rm S}$}}
\def\chiH{\hbox{$\chi$(H)}}
\def\chiHe{\hbox{$\chi$(He)}}
\def\chiN{\hbox{$\chi$(N)}}
\def\chiRX{\hbox{$N {\rm (X^+)}$/($N{\rm (X^\circ)}$+$N{\rm (X^+)) }$}}
\def\lsrstd{\hbox{LSR$_{Std}$}}
\def\lsrhip{\hbox{LSR$_{Hip}$}}
\def\sigV{$\sigma$$_{\rm v}$}
\def\Bis{\hbox{$B_{\rm IS}$}}

\def\T{\hbox{$T$}}
\def\Pth{\hbox{$P_{\rm Th }$}}
\def\Eth{\hbox{$E_{\rm Th }$}}
\def\Ptu{\hbox{$P_{\xi }$}}
\def\Pb{\hbox{$P_{\rm B }$}}
\def\Eb{\hbox{$E_{\rm B }$}}

\def\ddot{\hbox{...}}

\def\bdop{$b_{\rm D}$}
\def\glong{$\ell$}
\def\glat{$b$}
\def\ff{$f$}
\def\T{$T$}
\def\Vbf{$V_{\rm CLIC}$}
\def\U{$V_{\rm x}$}
\def\V{$V_{\rm y}$}
\def\W{$V_{\rm z}$}
\def\ix{$\hat{x}$}
\def\iy{$\hat{y}$}
\def\iz{$\hat{z}$}

\def\Ts{$T_{\rm s}$}
\def\gl{$l$}
\def\gb{$b$}

\def\cop{\hbox{{\emph{Copernicus}}}}

\def\ebv{\hbox{$E(B-V)$}}

\def\Bpar{\hbox{$B_{\parallel}$}}

\def\Br{\hbox{$B_{\rm r}$}}
\def\Bu{\hbox{$B_{\rm u}$}}
\def\B{\hbox{$B$}}
\def\Lya{\hbox{L${\alpha}$}}
\def\Halpha{\hbox{H${\alpha}$}}

\def\Tk{\hbox{$T_{\rm k}$}}

\def\kms{\hbox{km s$^{\rm -1}$}}
\def\NDI{\hbox{$N$(D$^{\rm o }$)}}

\def\NeI{\hbox{Ne$^{ \rm o }$}}
\def\NI{\hbox{N$^\mathrm{o}$}}
\def\NHI{\hbox{$N$(H$^{\rm o }$)}}
\def\NMgI{\hbox{$N$(Mg$^{\rm o }$)}}
\def\RMg{\hbox{$N$(Mg$^{\rm + }$)/$N$(Mg$^{\rm o }$)}}
\def\RC{\hbox{$N$(C$^{\rm + }$)/$N$(C$^{\rm + * }$)}}

\def\nHI{\hbox{$n$(H$^{\rm o }$)}}
\def\NHeI{\hbox{$N$(He$^{\rm o }$)}}
\def\nHeI{\hbox{$n$(He$^{\rm o }$)}}
\def\logNHI{\hbox{log$N$(H$^{\rm o }$)}}
\def\NHII{\hbox{$N({\rm H^+})$}}
\def\NH2{\hbox{$N$(H$_{\rm 2}$)}}
\def\NH{\hbox{$N$(H)}}
\def\NHtot{\hbox{$N$(H$_{\rm tot }$)}}

\def\NCII{\hbox{$N({\rm C^+})$}}
\def\NCIIstar{\hbox{$N({\rm C^{+ *}})$}}
\def\H2{\hbox{H$_2$}}
\def\nH{\hbox{$n_{\rm H}$}}

\def\nHII{\hbox{$n({\rm H^+})$}}

\def\ArI{\hbox{${\rm Ar^{ \rm o }}$}}
\def\ArII{\hbox{${\rm Ar^{ \rm + }}$}}

\def\no{\hbox{$n_{\rm o}$}}
\def\nHIavg{\hbox{$\langle  n$(H$^\circ$)$ \rangle$}}
\def\nHIavgLIC{\hbox{$\langle  n$(H$^\circ$)$ \rangle _\mathrm{LIC}$}}
\def\nHIavgtwo{$n_{HI,0.2}$}

\def\HH2{\hbox{H$_{\rm H^{ \rm o } + 2H_2}$}}
\def\NHH2{\hbox{$N$(H$_{\rm H^{ \rm o } + 2H_2}$)}}
\def\NHH{\hbox{ $N$(H$_2$)} }

\def\HH{\hbox{H$_2$}}

\def\N{\hbox{$N$}}
\def\n{\hbox{$n$}}
\def\el{\hbox{\rm e${\rm ^-}$}}

\def\cmtwo{\hbox{cm$^{-2}$}}

\def\cc{\hbox{cm$^{-3}$}}
\newcommand{\deeg}{$^\circ$}
\def\nel{\hbox{$n {\rm (e)}$}}
\def\np{\hbox{$n {\rm (p)}$}}
\def\HI{\hbox{H$^{ \rm o }$}}
\def\DI{\hbox{D$^{ \rm o }$}}

\def\CII{\hbox{${\rm C^+}$}}
\def\CIIstar{\hbox{${\rm C^{+ *}}$}}
\def\NII{\hbox{N$^\mathrm{+}$}}
\def\OI{\hbox{O$^{\rm o}$}}
\def\OII{\hbox{O$^{\rm +}$}}
\def\OVI{\hbox{O$^{+5}$}}
\def\CIV{\hbox{${\rm C^{+3}}$}}
\def\NFeII{\hbox{$N{\rm (Fe^+)}$}}

\def\MgI{\hbox{${\rm Mg^o }$}}
\def\MgII{\hbox{${\rm Mg^+}$}}
\def\SII{\hbox{${\rm S^+}$}}
\def\HII{\hbox{${\rm H^+}$}}
\newcommand{\Htwo}{H\,\textsc{II}}
\def\NCaII{\hbox{$N$(Ca$^+$)}}
\def\CaII{\hbox{Ca$^+$}}
\def\CaIII{\hbox{Ca$^{++}$}}
\def\SiII{\hbox{Si$^+$}}
\def\SiIII{\hbox{Si$^{+2}$}}
\def\NaI{\hbox{Na$^{\rm o }$}}
\def\NaII{\hbox{Na$^{\rm + }$}}
\def\KI{\hbox{K$^{\rm o }$}}
\def\HeI{\hbox{He$^{\rm o }$}}
\def\HeII{\hbox{He$^{\rm + }$}}
\def\FeII{\hbox{Fe$^+$}}
\def\TiII{\hbox{Ti$^+$}}

\definecolor{edit}{cmyk}{0,1.0,1.0,0}  
\let\footnote\savefootnote
\let\footnotetext\savefootnotetext
\let\footnoterule\savefootnoterule

\sectionequationnumber 
\upperandlowercase
\chapsectrunningheads
\chaptersection

\begin{document}
\pagenumbering{arabic}

\newcommand\adsr{{Adv.~Space~Res.}}%
\newcommand\jatp{{J.~Atmos.~Terres.~Phys.}}%
\newcommand\aj{{Astron.~J.}}%
\newcommand\actaa{{Acta Astron.}}%
\newcommand\areps{{Ann.~Rev.~Earth \& Plan.~Sci.}}%
\newcommand\araa{{Ann.~Rev.~Astron.~\& Astrophys.}}%
\newcommand\apj{{Astrophys.~J.}}%
\newcommand\apjl{{Astrophys.~J.~Let.}}%
\newcommand\apjs{{Astrophys.~J.~Supl.}}%
\newcommand\ao{{Appl.~Opt.}}%
\newcommand\apss{{Astrophys.~\& Space Sci.}}%
\newcommand\aap{{Astron.~\& Astrophys.}}%
\newcommand\aapr{{Astron.~ \& Astrophys.~Rev.}}%
\newcommand\aaps{{Astron.~ \& Astrophys. Supl.}}%
\newcommand\azh{{Astron.~Zh.}}%
\newcommand\baas{{Bull.~Amer.~Astron.~Soc.}}%
\newcommand\caa{{Chinese Astron. Astrophys.}}%
\newcommand\cjaa{{Chinese J. Astron. Astrophys.}}%
\newcommand\icarus{{Icarus}}%
\newcommand\jcap{{J. Cosmology Astropart. Phys.}}%
\newcommand\jrasc{{JRASC}}%
\newcommand\memras{{Mm.~Roy.~Astron.~Soc.}}%
\newcommand\mnras{{Mon.~Not.~ Roy.~Astron.~Soc.}}%
\newcommand\na{{New Astron.}}%
\newcommand\nar{{New Astron.~Rev.}}%
\newcommand\pra{{Phys.~Rev.~A}}%
\newcommand\prb{{Phys.~Rev.~B}}%
\newcommand\prc{{Phys.~Rev.~C}}%
\newcommand\prd{{Phys.~Rev.~D}}%
\newcommand\pre{{Phys.~Rev.~E}}%
\newcommand\prl{{Phys.~Rev.~Lett.}}%
\newcommand\pasa{{PASA}}%
\newcommand\pasp{{Pub.~Astron.~Soc.~Pac.}}%
\newcommand\pasj{{PASJ}}%
\newcommand\qjras{{QJRAS}}%
\newcommand\rmxaa{{Rev. Mexicana Astron. Astrofis.}}%
\newcommand\skytel{{S\&T}}%
\newcommand\solphys{{Sol.~Phys.}}%
\newcommand\sovast{{Soviet~Ast.}}%
\newcommand\ssr{{Space~Sci.~Rev.}}%
\newcommand\zap{{Zeit.~Astrophy.}}%
\newcommand\nat{{Nature}}%
\newcommand\iaucirc{{IAU~Circ.}}%
\newcommand\aplett{{Astrophys.~Lett.}}%
\newcommand\apspr{{Astrophys.~Space~Phys.~Res.}}%
\newcommand\bain{{Bull.~Astron.~Inst.~Netherlands}}%
\newcommand\fcp{{Fund.~Cosmic~Phys.}}%
\newcommand\gca{{Geochim.~Cosmochim.~Acta}}%
\newcommand\grl{{Geophys.~Res. ~Lett.}}%
\newcommand\jcp{{J.~Chem.~Phys.}}%
\newcommand\jgr{{J.~Geophys.~Res.}}%
\newcommand\jqsrt{{J.~Quant.~Spec.~Radiat.~Transf.}}%
\newcommand\memsai{{Mem.~Soc.~Astron.~Italiana}}%
\newcommand\nphysa{{Nucl.~Phys.~A}}%
\newcommand\physrep{{Phys.~Rep.}}%
\newcommand\physscr{{Phys.~Scr}}%
\newcommand\planss{{Planet.~Space~Sci.}}%
\newcommand\procspie{{Proc.~SPIE}}%


\pagenumbering{arabic}  
\setcounter{chapter}{5}
\articletitle[Short-term Variations in the Galactic Environment of the Sun]{Short-term
  Variations in the \\ Galactic Environment of the Sun }
\chaptitlerunninghead{Short-term Variations in Galactic Environment}
\author{Priscilla C.~Frisch}
\affil{University of Chicago}
\email{frisch@oddjob.uchicago.edu}
\and
\author{Jonathan D.~Slavin}
\affil{Harvard-Smithsonian Center for Astrophysics}
\email{jslavin@cfa.harvard.edu}


\vspace{0.19in}
\noindent
{\bf Table of Contents:}
\begin{table}[h!]
\begin{tabular}{l l r}
6.1 & Overview & 2 \\
6.2 & The Solar Journey through Space:  The Past $10^4$ to $10^6$ Years   &  14 \\
6.3 & Neighborhood ISM:  Cluster of Local Interstellar Clouds &  19 \\
6.4 & Radiative Transfer Models of Partially Ionized Gas   &  32 \\
6.5 & Passages through Nearby Clouds   &  38 \\
6.6 & The Solar Environment and Global ISM   &  43 \\
6.7 & Summary   &  50 \\
& References  &  52 \\
\end{tabular}
\end{table}
\begin{abstract}

The galactic environment of the Sun varies over short timescales as
the Sun and interstellar clouds travel through space.  Small
variations in the dynamics, ionization, density, and magnetic field
strength in the interstellar medium (ISM) surrounding the Sun can
lead to pronounced changes in the properties of the heliosphere.  The
ISM within $\sim$30 pc consists of a group of cloudlets that flow
through the local standard of rest with a bulk velocity of
$\sim$17--19 \kms, and an upwind direction suggesting an origin
associated with stellar activity in the Scorpius-Centaurus
association.  The Sun is situated in the leading edge of this flow, in
a partially ionized warm cloud with a density of $\sim$0.3 \cc.
Radiative transfer models of this tenuous ISM show that the fractional
ionization of the ISM, and therefore the boundary conditions of the
heliosphere, will change from radiative transfer effects alone as the
Sun traverses a tenuous interstellar cloud.  Ionization equilibrium is
achieved for a range of ionization levels, depending on the ISM
parameters.  Fractional ionization ranges of \chiH=0.19--0.35
and \chiHe=0.32--0.52 are found for tenuous clouds in equilibrium.
In addition, both temperature and velocity vary between clouds.
Cloud densities derived from these models permit primitive estimates
of the cloud morphology, and the timeline for the Sun's passage
through interstellar clouds for the past and future $\sim$10$^5$
years.  The most predictable transitions happen when the Sun
emerged from the near vacuum of the Local Bubble interior and entered
the cluster of local interstellar clouds flowing past the Sun, which
occurred sometime within the past 140,000 years, and again when the Sun
entered the local interstellar cloud now surrounding and inside of the
solar system, which occurred sometime within the past 44,000 years,
possibly a 1000 years ago.  Prior to $\sim$140,000 years ago, no
interstellar neutrals would have entered the solar system, so the
pickup ion and anomalous cosmic ray populations would have been
absent.  The tenuous ISM within 30 pc is similar to low column density
observed globally.  In this chapter, we review the factors important
to understanding short-term variations in the galactic environment of
the Sun.  Most ISM within 40 pc is partially ionized warm material,
but an intriguing possibility is that tiny cold structures may be
present.

\end{abstract}


\section[Overview]{Overview}\label{sec:intro}

In 1954 Spitzer noted that the ``Study of the stars is one of
[mankind's] oldest intellectual activities.  Study of the matter
between stars is one of the youngest.''  \nocite{Spitzer:1954}
Comparatively, the study of interstellar matter (ISM) at the
heliosphere is an infant.  Still younger is the study of the
interaction between the heliosphere and ISM that forms the galactic
environment of the Sun.  The total pressure of the ISM at the solar
system is counterbalanced by the solar wind ram pressure, but since
both the Sun and clouds move through space, this balance is perturbed
as the Sun passes between clouds with different velocities or physical
properties.  In this chapter, we focus on variations in the galactic
environment of the Sun over timescales of $\pm$3 Myrs, guided by data
and models of local interstellar clouds.  In essence, we strive to
provide a basis for understanding the ``galactic weather'' of the
solar system over geologically short timescales, and in the process
discover recent changes in the Sun's environment that affect particle
populations inside of the heliosphere, and perhaps the terrestrial
climate.  The heliosphere, the bubble containing the solar wind plasma 
and magnetic field, dances in the wind of interstellar gas drifting past
the Sun.  This current of tenuous partially ionized low density ISM
has a velocity relative to the Sun of $\sim$26 \kms.  It would have
taken less than 50,000 years for this gas to drift from the vicinity of the
closest star $\alpha$ Cen, near the upwind direction in the local
standard of rest (LSR, Tables \ref{tab:velocity}), and into the solar
system.

Star formation disrupts the ISM.  The nuclear ages of massive nearby stars in
Orion (at a distance of $\sim 400$ pc) and Scorpius-Centaurus (at $\sim 150$
pc) are 4-15 Myr, so the solar system has been bombarded by high energy photons
and particles many times over the past 10$^7$ years.  Recent nearby supernova
events include the formation of the Geminga pulsar $\sim 250,000$ years ago,
and possibly the release of $\zeta$ Oph as a runaway star $\sim$0.5-1 Myr ago.
The dynamically evolving interstellar medium inhibits the precise description
of the solar Galactic environment on timescales longer than $\sim 1-3$ Myrs.

Variations in the galactic environment of the Sun for time scales of
$\sim$3 Myrs are short compared to the vertical oscillations of the Sun
through galactic plane, and short compared to the disruption of
local interstellar clouds by star formation.  The oscillation of the
Sun in the Galactic gravitational potential carries the Sun through
the galactic plane once every $\sim 34$ Myrs, and the Sun
will reach a maximum height above the plane of $\sim 78$ pc in about 14
Myrs (\nolinebreak \cite{Bash:1986,Vandervoort:1993}, also see Chapter 5 by Shaviv).  The motion of the Sun
compared to the kinematics of some ensemble of nearby stars is known
as the solar apex motion, which defines a hypothetical closed circular
orbit around the galactic center known as the Local Standard of Rest
(LSR, \S \ref{sec:lsr}).  Note that the LSR definition is sensitive to
the selection of the comparison stars, since the mean motions and
dispersion of stellar populations depend on the stellar masses.  The
transformation of the solar trajectory into the LSR for comparison with
spatially defined objects, such as the Local Bubble, introduces
uncertainties related to the LSR.

\nocite{HTII}
\nocite{Frisch:1993a,Frisch:1994,Frisch:1995,Frisch:1997,ZankFrisch:1999}

A striking feature in the solar journey through space is the recent
emergence of the Sun from a near vacuum in space, the Local Bubble,
with density $\rho < 10^{-26}$ g~\cc\
(\cite{Frisch:1981,FrischYork:1986}, \S \ref{sec:journey}).  We will
show that the Sun has spent most of the past $\sim$3 Myrs in this
extremely low density region of the ISM, known as the Local
Bubble.  The Local Bubble extends to distances of over 200 pc in parts of the
third galactic quadrant ($l = 180^\circ \rightarrow 270^\circ$,
\cite{FrischYork:1983}), and is part of an extended interarm region
between the Sagittarius/Carina and the Perseus spiral arms.  The
proximity of the Sun to the Local Bubble has a profound effect on
the solar environment, and affects the local interstellar radiation
field, and dominates the historical solar galactic environment of the
Sun (\S \ref{sec:lbradiation}).  Neutral ISM was absent from the bubble, and
byproducts of the ISM-solar wind interaction, such as dust, pickup
ions and anomalous cosmic rays, would have been absent from the
heliosphere interior.

The Local Bubble is defined by its geometry today, so the solar space motion is
a variable in estimating the departure of the Sun from this bubble.  
For all reasonable solar apex motions, sometime between 1000 and 140,000 
years ago the Sun encountered material denser than the nearly empty 
bubble.  During the late Quaternary
geological period, the Sun found itself in a flow of warm low density ISM, $n
\sim 0.3$ \cc, originating from the direction of the Scorpius-Centaurus
Association (\S \S \ref{sec:clic}, \ref{sec:origin}).  The Sun is presently
surrounded by this warm, $T \sim 7,000$ K, tenuous gas, which is similar to the
dominant form of ISM in the solar neighborhood.  The physical characteristics
of this cluster of local interstellar cloudlets (CLIC) are discussed in \S
\ref{sec:clic}.

\begin{figure}[!th]
\caption[Distribution of ISM Creating the Local Bubble.]{\label{fig:lb} 
SEE 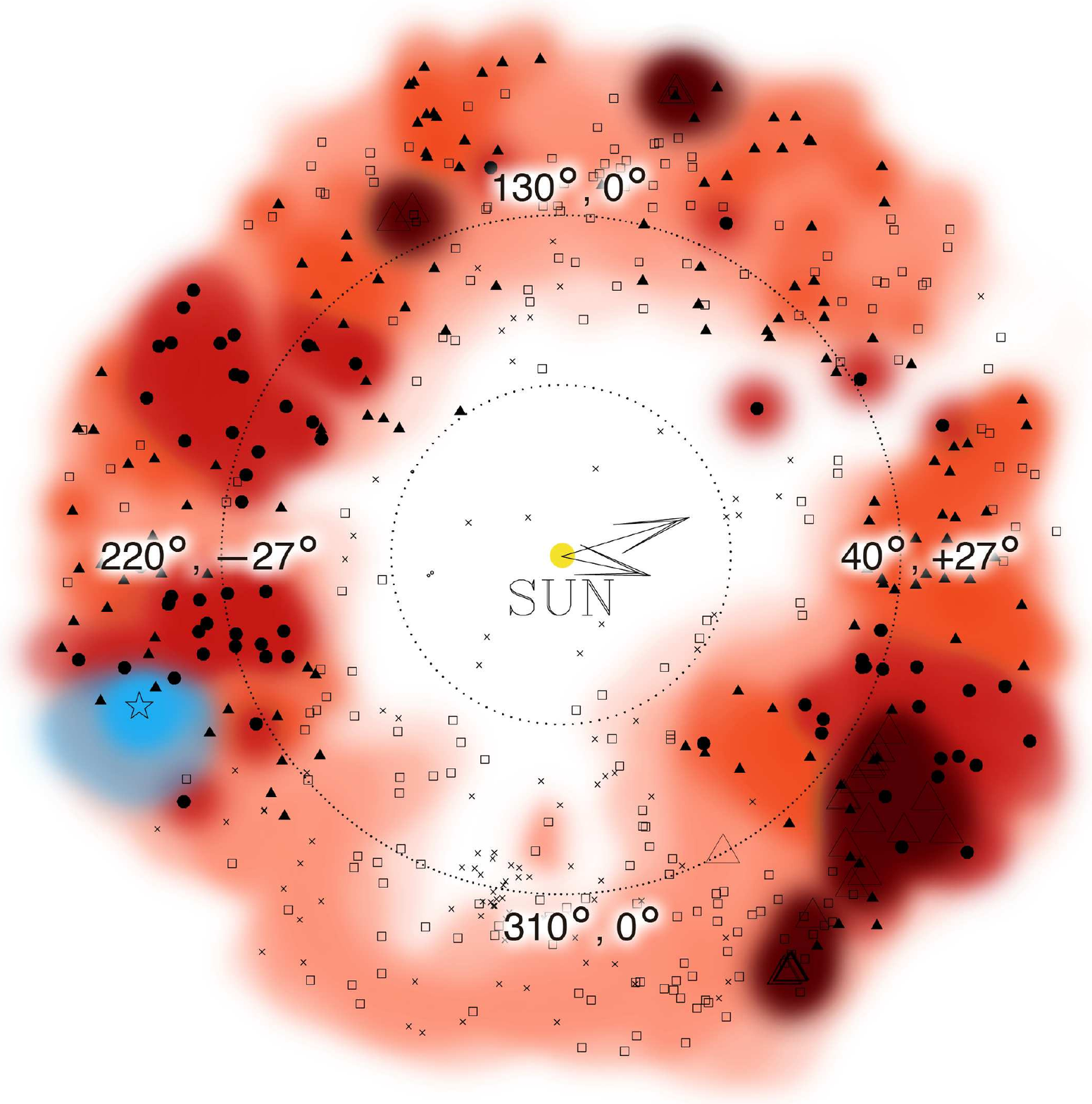 and 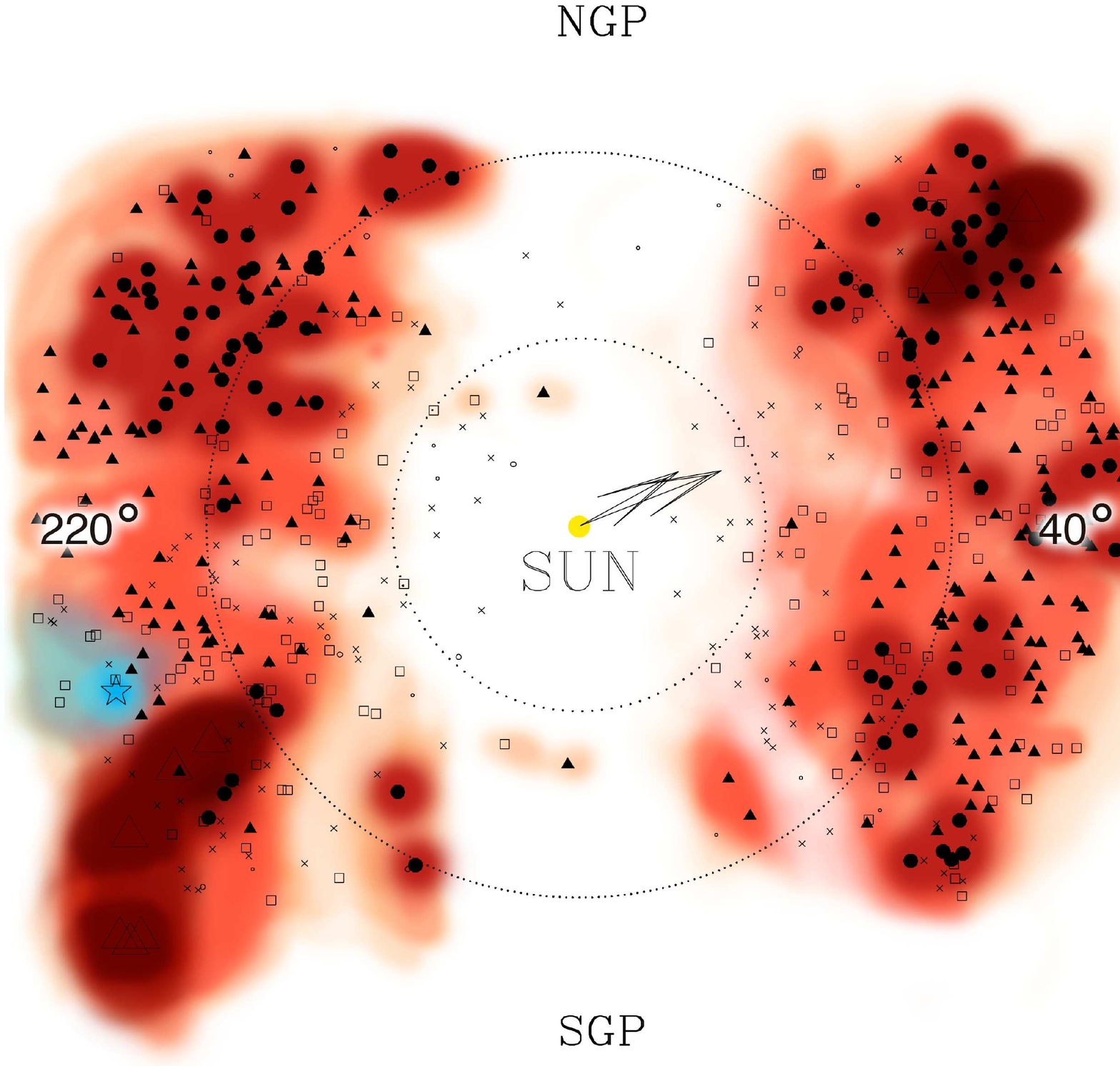.  The solar position and motion are compared to the Local Bubble for two
planes aligned with the solar apex motion (Table \ref{tab:velocity}).  The left
figure gives the Local Bubble ISM distribution in a plane that is tilted by
27\deeg\ to the Galactic plane with the northern surface normal pointing
towards (220\deeg, $+$63\deeg). Stars with latitudes within about $\pm$15\deeg
of that plane are plotted.  The right figure shows a meridian slice
perpendicular to the galactic plane, and aligned with the \glong=40\deeg ---
220\deeg\ axis.  Stars with longitudes within $\pm$25\deeg\ of the meridian
slice are plotted.  These ISM distributions were constructed from cleaned and
averaged photometric and astrometric data for O, B, and A stars in the
Hipparcos catalog.  For a solar LSR velocity of $13-20$ pc/Myrs, the Sun has
been within the Local Bubble for over 3 Myrs.  The symbols show \ebv\ values
0.017--0.051 mag (tiny x's), 0.051--0.088 mag (boxes), 0.088--0.126 mag (dots), and
$>0.126$ mag (triangles) so that shading levels 
give \NH$\sim 10^{20.4} \rightarrow 10^{20.7},~ \sim 10^{20.7} \rightarrow 10^{20.9},~\mathrm{and}
> 10^{20.9}$,
after assuming N(\HI+2\HH)/\ebv=21.76 \cmtwo
mag$^{-1}$.  The solar motion is shown for both Standard and Hipparcos values.
The large intensely shaded triangles show dust clouds from Dutra and Bica
(2002).\nocite{DutraBica:2002} The CLIC column densities are below the minimum
value of \NHI$\sim 10^{20.40}$ \cmtwo\ displayed here.}
\end{figure}

The solar entry into the CLIC would correspond to the appearance of
interstellar dust, pickup ions and anomalous cosmic rays in the heliosphere.
The dust filtration by the solar wind and in the heliosheath is sensitive both
to dust mass and the phase of the solar cycle (see Landgraf chapter), so that
these two effects are separable.  In principle the solar journey through clouds
could be traced by comparative studies of interstellar dust settling onto the
surfaces of inner versus outer moons.

In \S \ref{sec:enterlic} we discuss the Sun's entry into, and exit from, the
interstellar cloud now surrounding the solar system, and show that it can
be determined from a combination of data and theoretical models.  The velocity
of this cloud, known as the local interstellar cloud (LIC), has been
determined by observations of interstellar \HeI\ inside of the solar system (\cite{Witte:2004}).  We test
several possibilities for the three-dimensional (3D) shape of the LIC, and
conclude that the Sun entered the surrounding cloud sometime within the past
40,000 years, and possibly very recently, in the past $\sim$1000 years.  

The velocity difference between interstellar \HeI\ inside of the solar
system, and interstellar \DI, \FeII\ and other ions towards stars in
the upwind direction, including the nearest star $\alpha$ Cen,
together indicate the Sun will emerge from the cloud now surrounding
the Sun within $\sim$4,000 years.  This raises the question of ``What
is an interstellar cloud?'', which is difficult to answer for low
column density ISM such as the CLIC, where cloudlet velocities may
blend at the nominal UV resolution of 3 \kms.

Determining the times of the solar encounter with the LIC, and other
clouds, requires a value for \nHI\ in the cloud because distance
scales are determined by \NHI/\nHI.  Theoretical radiative transfer
models of cloud ionization provide \nHI\ and \nHII\ for tenuous ISM
such as the LIC, and supply important boundary constraints for
heliosphere models.  CLIC column densities are low, \N$\ltsim 10
^{18}$, and radiative transfer effects are significant (\S
\ref{sec:rt}).  Models predict \nHI$\sim$0.2 \cc\ and \nHII$\sim$0.1
\cc\ at the Sun, with ionization levels rising slowly to the cloud
surface so that \nHIavgLIC$\sim$0.17 \cc.

In \S \ref{sec:wim} we discuss the relative significance of diffuse warm
partially ionized gas (WPIM) and traditional Str\"omgren spheres for the
heliosphere.  Distinct regions of dense fully ionized ISM are infrequent
near the Sun, although regions of ionized gas surround distant massive
stars bordering the Local Bubble, and surround $\beta$ Cen and $\lambda$
Sco, which helps energize the Loop I superbubble interior.  The Sun's path
is unlikely to traverse, or have traversed, an \Htwo\ region around an
O--B1 star for $ \pm $4 Myrs because no such stars or \Htwo\ regions are
within 80 pc of the Sun.  Hot white dwarf stars are more frequent
nearby, and may in some cases be surrounded by small Str\"omgren
spheres.  In contrast, diffuse ionized gas is widespread and has been
observed extensively using weak optical recombination lines from \HII\
and other species, synchrotron emission, and by the interaction of
pulsar wave packets with ionized interstellar gas.  The CLIC and diffuse
ionized gas are part of a continuum of ionization levels predicted and
observed for low density interstellar gas.

The heliosphere should vary over geologically short timescales, for
reasons unrelated to the Sun or solar activity, when encountering the
small scale structure of the nearby ISM.  These variations in the cosmic
environment of the Sun may be traced by terrestrial radioisotope
records, or by variable fluxes of interstellar dust grains deposited on
planetary moons (see other chapters in this volume).  Over 3 Myrs
timescales the Sun travels 40-60 pc through the LSR, and the interval
between cloud traversals should be less than $\le 80,000$ years.
Differing physical properties of cloudlets in the flow will affect the
heliosphere and interplanetary medium.  If cloud densities are
relatively uniform at \nHI$\sim$0.2 \cc, then data on ISM towards nearby
stars indicate these clouds fill $\sim$33\% of space within 10 pc, with
mean cloud lengths of $\sim 1.0$ pc, and with a downwind hemisphere that
is emptier than the upwind hemisphere (\S \ref{sec:clicmorphology}).

The ISM within $\sim$2 pc is inhomogeneous.  The ratio \FeII/\DI\
increases as the viewpoint sweeps from the downwind to upwind
direction, signaling either grain destruction in an upwind interstellar
shock, or increased cloud ionization.  The \FeII/\DI\ gradient is consistent
with the destruction of CLIC dust grains in a fragment of the
expanding superbubble around the Scorpius-Centaurus Association (\S
\ref{sec:origin}).  There are several possibilities for the next cloud
to be encountered, and the most likely candidate is the cloud towards
the nearest star $\alpha$ Cen (\S \ref{sec:next}), where only modest
consequences for the heliosphere are expected, since the temperature
and velocity of the cloud differ only by $\sim$800 K and $\sim$3 \kms,
respectively.

The properties of the CLIC are similar to low column density ISM
observed in the solar vicinity, and observed in the Arecibo Millennium
survey of very low \NHI\ clouds (\cite{HTIII}).  The warm neutral
medium (WNM) in the Arecibo survey appears to be similar to the CLIC, but
so far, counterparts of the exotic cold, low column density, neutral
clouds, $T \sim 50$ K and \NHI$\sim 10 ^{18}$ \cmtwo, are not found
locally (\S \ref{sec:global}).  If they are hidden in the upwind ISM,
they would be far from ionization equilibrium (having much lower ionization than the surrounding warm, ionized gas), in contrast to the LIC
gas (\S \ref{sec:rt}).  Cold ISM towards 23 Ori indicate densities of
$>$10 \cc\ for the cold neutral medium gas (CNM).

We do not discuss high-velocity, $v \sim 100$ \kms, low column density
radiative shocks.  Such disturbances would cross the heliosphere on
timescales of $\sim5-10$ years, seriously perturbing it
(\cite{SonettJokipii:1987,Frisch:1999,Muelleretal:2005}), and would be
evolving as they cool.   Low column density high velocity shocked gas,
\NH$\sim 10^{16} - 10 ^{17}$ \cmtwo, \no$\sim$0.5--5 \cc, could be
common in ISM voids.  However, with high velocities of $\sim$100 \kms,
such gas will traverse voids such as the Local Bubble in less than 1 Myr.
Similar high-velocity gas is associated with the Orion's Cloak
superbubble shell around Orion and Eridanus, with $v \sim 100$ \kms\ and
\NH$\sim 10^{16} - 10 ^{17}$ \cmtwo\ (\cite{Welty23:1999}).  Rapidly
moving low column density shocks similar to these would pass over the
heliosphere quickly, in 50 -- 1,500 years, with strong implications for
heliosphere structure (Zank et al., this volume).  The fairly coherent
velocity and density structure and relatively weak turbulence in the CLIC
argues against any such disturbance having occurred recently (see
\S\ref{sec:pressure}).  We also do not discuss dense molecular clouds,
which are discussed in the Shaviv chapter.

As a way of summarizing short term solar environment, we show the
solar position inside of the Local Bubble in Figure \ref{fig:lb} (see
\ref{sec:localbubble}).  We will now briefly review relevant fundamental
material.

\subsection[Fundamental Concepts]{Fundamental Concepts}\label{sec:basic}

Basic concepts needed for understanding the following discussions are
summarized here. 
We use the local standard of rest, LSR, based on the
Standard solar motion, unless otherwise indicated.  Velocities
quoted as heliocentric (HC) are
with respect to the Sun, while LSR velocities are
obtained by correcting for the motion of the Sun with
respect to the nearest stars, or the solar apex motion.
The basic principles underlying obtaining the properties of interstellar
clouds are summarized in \S \ref{sec:absline}.  

The ultraviolet and X-ray parts of the spectrum play an essential role
in the equilibrium of interstellar clouds.  The spectral regions relevant to
the following discussions are the ultraviolet (UV), with wavelengths
$1200\,\mathrm{\AA} < \lambda <  3000\,\mathrm{\AA}$; the far ultraviolet
(FUV), $912\,\mathrm{\AA} < \lambda < 1200\,\mathrm{\AA}$; the extreme
ultraviolet (EUV), $100\,\mathrm{\AA} < \lambda < 912\,\mathrm{\AA}$; and
the soft X-ray, $ \lambda< 100\,$\AA.  The spectrum of the soft X-ray
background (SXRB), photon energies $<0.5$ keV, has been
an important diagnostic tools for the distribution of ISM with column density
\NHI$> 10 ^{19.5}$ \cmtwo.  The first ionization potential (FIP) of an
atom is the energy required to ionize the neutral atom.  The FIP of H is
13.6 eV.  Elements with FIP $< 13.6$ eV are predominantly ionized in all
clouds, since the ISM opacity to photons with energies $E<13.6$ eV is
very low.  The ionization potential, IP, of several ions are also
important.  For instance IP(\CaII)=11.9 eV, and $N$(\CaIII)/\NHI\ may be
large in warm clouds.  For $T>4,000$ K and \nel$<$0.13 \cc,
\CaIII/\CaII$>$1.  Variations in the spectrum of the ionizing radiation
field, as it propagates through a cloud caused by the wavelength
dependence of the opacity, $\tau(\lambda)$, are referred to as
radiative transfer (RT) effects.

Color excess is given by \ebv, and represents the observed star color, in
magnitudes, measured in the UBV system, compared to intrinsic stellar
colors (e.g.\ \cite{Cox:2000}).  Color excess represents the intrinsic star color
combined with reddening by foreground dust.  The column density, $N$(X), is the
number of atoms X contained in a column of material with a cross
section of 1 \cmtwo\ and bounded at the outer limit by an arbitrary
point such as a star. The volume density, $n$(X), is the number of
atoms \cc.  The notation \nHIavgtwo\ indicates the 
\HI\ space density in units of 0.2 \cc.  The fractional
ionization of an element X is \chiRX.  The filling factor, \ff, gives
the percentage of the space that is filled by ISM
denser than the hot gas in the Local Bubble (i.e.\ warm or cold gas).

Among the various acronyms we use are the following:  The interstellar
cloud feeding neutral ISM and interstellar dust grains (ISDG) into the
solar system is the local interstellar cloud (LIC).  A cluster of local
interstellar clouds (CLIC) streams past the Sun, forming absorption
lines in the spectra of nearby stars.  Alternative names for the CLIC are
the ``Local Fluff'' and the very local ISM, and the LIC is a member of the CLIC.
The upwind portion of the CLIC has been called the ``squall line''  
(\cite{Frisch:1995}).  In the global ISM we find cold neutral
medium (CNM), warm neutral medium (WNM), and warm ionized medium
(WIM).  Both observations and our radiative transfer models of tenuous
ISM, \NH$\ltsim 10 ^{18}$ \cmtwo, indicate that warm partially ionized
medium (WPIM) is prevalent nearby.

The Scorpius-Centaurus Association (SCA) is a region of star formation
that dominates the distribution of the ISM in the upwind direction of
the CLIC flow, including the Loop I supernova remnant.  The Local
Bubble (LB) is a cavity in the interstellar dust and gas around the Sun that
may merge in some regions with the bubble formed by stellar evolution in
the SCA, Loop I, near the solar location (\cite{Frisch:1995}).  The
Local Bubble is used here to apply to those portions of nearby
space, within $\sim$70 pc, with densities $\rho <10^{-26}$ g \cc.

The densities of neutral interstellar gas at the solar location are
provided by data on pickup ions (PUI) and anomalous cosmic rays (ACR).
PUIs are formed from interstellar neutrals inside the heliosphere,
which have become ionized and captured by the solar wind (Moebius et al.,
this volume).  ACRs are PUIs that are accelerated in the solar
wind and termination shock to energies of $\le$1 GeV/nucleon (Fahr et al.,
Florinski et al., this volume).

Element abundances in the CLIC gas are similar to warm gas in the 
Galactic disk, and the missing atoms are believed to be incorporated in
dust grains (\cite{Frischetal:1999,Welty23:1999}).  We will assume
that the solar abundance
pattern provides the benchmark abundance standard for the local ISM,
although this issue has been debated (\cite{Snow:2000}), and
although the solar abundance of O and other elements are
controversial (e.g. \cite{Lodders:2003}).  Using solar
abundances as a baseline, the depletion $\delta$ of an element
represents the amount of the element that is missing from the gas. 
Depletion is defined as $\delta_{\rm X} = {\rm log}~ \frac {[ {\rm X} ]_{\rm
ISM}} {[ {\rm X} ]_{\rm Solar}} $ for an element {\rm X} with
abundance $[ {\rm X} ]$ (\cite{SavageSembach:1996}).  
The depletion and condensation temperature of
an element correlate fairly well.  This is shown by the three observed 
interstellar depletion groups 
$\delta_{\mathrm Ti,Ca} \sim -3.3$, 
$\delta_{\mathrm Fe,Cr,Co,Ni,V} \sim -2.3$, and 
$\delta_{\mathrm Mg,Si} \sim -1.3$, which condense at temperatures 
of $\sim 1,500$ K, $\sim 1,330$ K, and $\sim 1,300$ K respectively
during mineral formation in
an atmosphere of solar composition and pressure (\nolinebreak \cite{Ebel:2000}).
The interstellar gas-to-dust mass ratio and grain composition can be
reconstructed from the atoms missing from the gas phase
(\nolinebreak \cite{Frischetal:1999}).  Depletions vary between warm tenuous and
cold dense ISM, an effect originally parameterized by velocity and known
as the Routly-Spitzer effect (\nolinebreak \cite{RoutlySpitzer:1952}),
and now attributed to partially ionized warm ISM that has been shocked (\S \ref{sec:results}).  
The largest variations are found for refractory elements, and result from 
ISDG destruction by interstellar shocks
(\cite{Jones:1994,Slavinetal:2004}).  Depletion estimates are
sensitive to uncertainties in \NH, which in cold clouds must include
\NHH \nolinebreak, and in warm tenuous clouds must include \HII, both of which can
be hard to directly measure.

Satellites that have revolutionized our understanding of the galactic
environment of the Sun include \cop, the International Ultraviolet
Explorer, IUE, the Extreme Ultraviolet Explorer, EUVE, the shuttle
launched Interstellar Medium Absorption Profile Spectrograph, IMAPS,
the Far Ultraviolet Explorer, FUSE, and the Hubble Space Telescope,
HST (with the Goddard High Resolution Spectrometer, GHRS, and the
Space Telescope Imaging Spectrograph, STIS).


Several papers in the literature have provided important insights to
the physical properties of the local ISM, and are referred to
throughout.  Among them are: Frisch (1995, hereafter F95), which reviews the
physical characteristics of the Local Bubble and the CLIC; Gry and
Jenkins (2001, hereafter GJ) and Hebrard et al. (1999), which present the
physical characteristics of the two nearby interstellar clouds
observed in the downwind direction, the LIC and blue-shifted cloud;
Slavin and Frisch (2002, hereafter SF02) and Frisch and Slavin (2003, 2004, 2005,
hereafter FS), which present detailed radiative transfer model calculations of
the surrounding ISM; Frisch et al. (2002, hereafter FGW), which probes the
kinematics of nearby ISM; and a series of papers by Redfield and
Linsky (2000, 2002, 2003, 2004, hereafter RL,), and Wood et al. (2005,
hereafter W05), which assemble a wide range of UV data on the CLIC.  Recent
papers presenting results from FUSE also contribute significantly to
our understanding of the tenuous ISM close to the Sun.
\nocite{Frisch:1995,GryJenkins:2001,Hebrardetal:1999,SlavinFrisch:2002,FrischSlavin:2003}
\nocite{FrischSlavin:2005cospar,RedfieldLinsky:2000,RLI,RLII,RLIII}

\subsection[Solar Apex Motion and Local Standard of Rest]{The Solar Apex Motion
and Local Standard of Rest} \label{sec:lsr}

Comparisons between the solar position and spatially defined objects,
such as the LB, require adoption of a velocity frame of reference.
The ``Local Standard of Rest'' (LSR) represents an instantaneous
inertial frame for a corotating group of nearby stars ($ <$500 pc) on
a closed circular orbit around the galactic center, and is commonly
used for this purpose.  The solar motion with respect to the LSR,
known as the solar ``apex motion'', is dynamically defined with
respect to a sample of nearby stars.  All reasonable values for the
LSR indicate that the Sun is traveling away from the ISM void in the
third quadrant of the Galaxy, \glong=180\deeg $\rightarrow$ 270\deeg,
at a velocity of $\sim$13--20 pc/Myrs.

The Sun is $\sim$8 kpc from the galactic center.  The mean orbital
motion around the galactic center of nearby stars is $\sim$220 \kms,
and the solar velocity is 225$\pm$20 \kms\ (e.g., see proper motion
studies of extragalactic radio sources beyond the galactic center,
\cite{Reidetal:1999}).  Accurate astrometric data for stars is
provided by the $Hipparcos$ catalog of proper motions and positions of
$\sim$120,000 nearby stars (\cite{Perrymanetal:1997}).  For the
coordinate system \ix, \iy, \iz, which represent unit vectors towards
the galactic center, direction of galactic rotation, and north
galactic pole respectively, the speeds \U, \V, and \W\ then represent
the mean solar velocity in the \ix, \iy, and \iz, directions compared
to some arbitrarily selected set of nearby stars.  A kinematically
unbiased set of stars from the Hipparcos Catalog shows that the mean
vertical and radial components of nearby star velocities have no
systematic dependence on stellar mass (\cite{DehnenBinney:1998}).
However \V\ determined for stars hotter than A0 ($B-V <$0.0 mag) 
exceeds the mean LSR by $\sim$6 \kms, and the velocity dispersion for
these relatively young stars is smaller than for evolved low mass
stars.  Since molecular clouds may share the motions of massive stars,
the result is a lack of clarity concerning the best LSR transformation
for studying the solar motion with respect to spatially defined
objects.

The general practice is to transform radio data into the LSR velocity
frame using the ``Standard'' solar motion, which is based on the
weighted mean velocity for different populations of bright nearby stars
irrespective of spectral class.  The velocity of the Sun with respect to
the Standard LSR (\lsrstd) is (\U, \V, \W)=(10.4, 14.8, 7.3) \kms\
(\cite{Mihalas:1981}), giving a \emph{solar apex motion}, of 19.5 \kms\
towards \gl=56\deeg, \gb=+23\deeg\ (or \lsrstd, Table
\ref{tab:velocity}).  The solar motion with respect to a kinematically
unbiased subsample of the \emph{Hipparcos} catalog (omitting stars with $B-V <
0.0$ mag) is (\U,\V,\W)=(10.0,5.3, 7.2) \kms, corresponding to a solar
speed of $V$=13.4 \kms\ towards the apex direction \glong=27.7\deeg\ and
\glat=32.4\deeg\ \kms\ (or \lsrhip).  
\begin{table}[h]
\caption[LSR Velocities of the Sun and Nearby Interstellar Clouds.]{The LSR Velocities of the Sun and Nearby Interstellar Clouds}
\begin{tabular}{l c c c }
& \multicolumn{1}{c}{HC} & \multicolumn{1}{c}{\lsrstd} &
\multicolumn{1}{c}{\lsrhip} \\ 
&$V$, \glong\ \glat & $V$, \glong\ \glat & $V$, \glong\ \glat \\
&(\kms, \deeg\ \deeg) & (\kms, \deeg\ \deeg) & (\kms, \deeg\ \deeg) \\
\hline
Sun & \ddot   & 19.5, 56\deeg\ 23\deeg & 13.4, 27.7\deeg\ 32.4\deeg \\
CLIC  & --28.1$\pm$4.6, 12.4\deeg\ 11.6\deeg& --19.4, 331.0\deeg\
--5.1\deeg & --17.0, 357.8\deeg\ --5.1\deeg \\
LIC  &--26.3$\pm$0.4, 3.3\deeg\ 15.9\deeg &--20.7, 317.8\deeg\
--0.5\deeg  & --15.7, 346.0\deeg\ 0.1\deeg\\
GC& --29.1, 5.3\deeg\ 19.6\deeg & --21.7, 323.6\deeg\ 6.3\deeg & --17.7,
351.2\deeg\  8.5\deeg \\
Apex&--35.1, 12.7\deeg\ 14.6\deeg & --24.5, 341.3\deeg\ 3.4\deeg &
--23.3, 5.5\deeg\ 4.1\deeg \\
\hline
\end{tabular}
\newline
Notes:  
The first row lists the Standard and Hipparcos values for the solar apex
motion (\S \ref{sec:lsr}).  The LIC, CLIC, GC (G-cloud), and Apex
velocity vectors refer to the upwind direction.
Columns 2, 3, and 4 give the heliocentric velocity vector, and the LSR
velocities using corrections based on the Standard and Hipparcos-derived
solar apex motions, respectively.  
\label{tab:velocity}
\end{table}

\subsection[Finding Cloud Physics from Absorption Lines]{Finding Cloud 
Physics from Interstellar Absorption Lines} \label{sec:absline}

Sharp optical and UV absorption lines, formed in interstellar clouds
between the Sun and nearby stars, are the primary means for determining
the physical properties of the CLIC.  UV data provide the best look at
cloud physics because resonant absorption transitions for the primary
ionization state of many abundant elements fall in the UV portion of
the spectrum.  However, results based on UV data are limited by a
spectral resolution, which typically is $\gtsim$3.0 \kms, and are unable to resolve the
detailed cloud velocity structure seen in high resolution 
(typically $\sim$0.3--0.5 \kms) optical data.  Ultimately high resolution UV
data, $R \sim$300,000, are required to probe the CLIC velocity
structure and to separate the thermal and turbulent contributions to
absorption line broadening.  The details of interpreting absorption
line data can be found in Spitzer (1978); see alternatively the
classic application of the ``curve of growth'' technique to
the ISM towards $\zeta$ Oph (\nolinebreak \cite{Morton:1975}).  \nocite{Spitzer:1978}

The classic target objects for ISM studies are hot rapidly rotating O,
B and A stars, where sharp interstellar absorption features stand out
against broad stellar lines.  However O and B stars are relatively
infrequent nearby, and interstellar line data can be contaminated by
sharp lines formed in circumstellar shells and disks around some A and
B stars.  Hence care is required to separate ISM and circumstellar 
features (e.g. \cite{Ferlet:1993}).  White dwarf stars of DA and DO
spectral types provide a hot far UV continuum that can, in many
cases, be observed in the interval 912--1200~\AA\
(e.g. \cite{Lehneretal:2003}).  Cool stars are very frequent, and
there are $\sim 30$ times more G and K stars than A stars.  The cool stars 
have a weak UV flux and active chromospheres.  Strong chromospheric emission lines 
can be used as continuum sources for interstellar absorption lines,
although great care is required in the analysis
(e.g. \cite{Woodetal:2005}).  The
disadvantage of cool stars is that uncertainties arise from blending
of the interstellar absorption and stellar chromosphere emission features.  
For example, only the Sun has an unattenuated \HI\ Lyman-$\alpha$ emission
feature, and its strength and shape vary with the phase of solar magnetic 
activity cycle.
Heeding these limitations, cool stars provide an important source of
data on local ISM.

\subsubsection[Absorption Line Data]{Absorption Line Data}

Observations of interstellar absorption lines in the interval
912--3000 \AA\ are required to diagnosis the physical properties of
the ISM, such as composition, ionization, temperature, density, and
depletions (e.g.,
\cite{Jenkins:1987,York:1976,Snow:2000,Savage:1995}).  Most elements
in diffuse interstellar clouds are in the lowest electronic energy
states, and have resonant absorption lines in the UV and FUV.
The type of ISM traced by \HI, \DI, \HH, \CII, \CIIstar, \CIV, \NI, 
\NII, \OI, \OII, \MgI, \MgII, \SiII, \SiIII, \ArI, \SII, \FeII\ depends on 
the element FIP and interaction cross-sections.  Neutral gas is 
traced by \DI, \HI, \NI, \OI, and \MgI, although \MgI\ is also 
formed by dielectronic recombination in WPIM (\S \ref{sec:clicwim}).  
Charge exchange couples the ionization
levels of N, O, and H.  The CLOUDY code documentation gives sources
for charge exchange and other reaction rate constants (\S
\ref{sec:rt}).  Ions with FIP$<$13.6 eV, such as \CII, \MgII, \SiII,
\SII, and \FeII, are formed in both neutral and ionized gas.  Highly
ionized gas is traced by \SiIII\ and \CIV.  The ratios \RMg\ and \RC\
are valuable ionization diagnostics (\S \ref{sec:clicwim}).

Local ISM has been studied at resolutions of $2.5-20$ \kms\ for $\sim 30$
years by \emph{Copernicus}, IUE, HST, HUT, EUVE, IMAPS, FUSE, and
various rocket and balloon experiments.  The best source of UV data in
the 1190--3000 \AA\ region are data from the GHRS and STIS on the
Hubble Space Telescope, while FUV data are available from FUSE, IMAPS,
and \emph{Copernicus}.

Ground-based optical observations have an advantage over space data in
two respects --- instruments provide very high resolution,
$\sim$0.3--1.0 \kms, and there are fewer constraints imposed on
telescope time.  Lines from \CaII, \NaI\ and \TiII\ are useful
diagnostics of the nearby ISM. 
The Ca\,\textsc{II} K line ($\lambda$3933
\AA) is usually the strongest interstellar feature for nearby stars
because Ca depletion is lower (abundances are higher) in warm gas (e.g.
\cite{SavageSembach:1996}), and the lines are strong.  Calcium and
titanium are refractory elements with highly variable abundances.  Both
\CaII\ and \NaI\ are trace
ionization states in the local ISM, while \TiII\ (FIP=13.6 eV) and
\HI\ have similar ionizations.  High resolution optical data provide
the best diagnostic of the cloud kinematics of ISM near the Sun and in
the upwind direction.  Optical lines, including Ca\,\textsc{II}, are generally
too weak for observations in the downwind direction.

\subsubsection[Interpreting Absorption Data]{Interpreting Absorption Data}

Interstellar absorption features in a stellar spectrum provide a
fundamental diagnostic of ISM physics.  Cloud temperatures are
determined from these data, but the $\sim$3 \kms\ resolution of the
best UV data is inadequate to separate out closely spaced velocity
components, or fully distinguish turbulent and thermal line
broadening.

 The wavelength-dependent opacity of an absorption line traces the
atomic velocity distribution and the underlying strength of the
transition between the energy levels, the Einstein transition
probability.  The atomic velocity distribution has contributions from
the bulk ISM motion, turbulence, and the thermal dispersion of atomic
velocity.  In principle, thermal and turbulent contributions to the
line broadening can be separated, given data for atoms with a range of
atomic mass.  In practice, data on either \DI\ or \HI, as well as
heavier elements, are needed.  The temperature of the cloud is found
from line widths, so understanding the limitations of this method are
important.
 
Absorption line data are analyzed with the assumption that particles,
of velocity $v$ and mass $m _{\rm A}$, obey a Maxwellian distribution
at a kinetic temperature \Tk, $f(v) \sim T_{\rm K}^{-3/2} \exp(- m
_{\rm A} v^2/2kT_{\rm K})$.  The kinetic temperature also
parameterizes the energy of ion-neutral and neutral-neutral elastic
scattering.  The velocity dispersion is characterized by the Doppler
line broadening parameter, which is \bdop$=(\frac{2kT_{\rm k}}{m_{\rm
A}})^{1/2}$ for a purely thermal distribution of velocities.  The line
full-width-at-half-maximum is FWHM$\sim$1.7 \bdop.  Few interstellar
absorption lines are Maxwellian shaped, however, because the line
opacity is a non-linear function of column density, and because
multiple unresolved clouds may contribute to the absorption.  In this
case, lines are interpreted using a best-fitting set of several
velocity components, each representing a separate cloud or ``velocity
component'', at velocity $v$, that are found to best reproduce the
line shape given the instrumental resolution.

Each velocity component, however, is broadened by both thermal motions and
turbulence.  The non-thermal line broadening is incorporated into a
general term ``turbulence'', $\xi$, so that $b_{\rm D}^2 = (2 k T_{\rm
K} / m_{\rm a}) + \xi^2 $.  The turbulence term $\xi$ is supposed to
trace the mass-independent component of line broadening.  The main
technique for separating turbulence from thermal broadening uses
\bdop\ as a function of atomic mass, although this technique does not
distinguish unresolved velocity structure.

For low opacity lines, column density is directly proportional to the
equivalent width $W$, and $N\sim W / \lambda ^2$.  For high line
center opacities, $\tau_{\rm o} \propto N \lambda /$\bdop, the line
strength increases in a non-linear way, $W \sim \lambda b_\mathrm{D} (\ln
\tau_\mathrm{o} )^{1/2}$. Column densities derived from observations of
partially saturated lines are highly insensitive to line strength.
These important limitations in determining the component properties
from absorption line data are well known (\cite{Spitzer:1978}).

\subsubsection[Unresolved Clouds ]{Resolution Limitations:  Unresolved Clouds}

\begin{figure}[ht]
\begin{center}
\includegraphics [height=3.2in,width=3.2in]{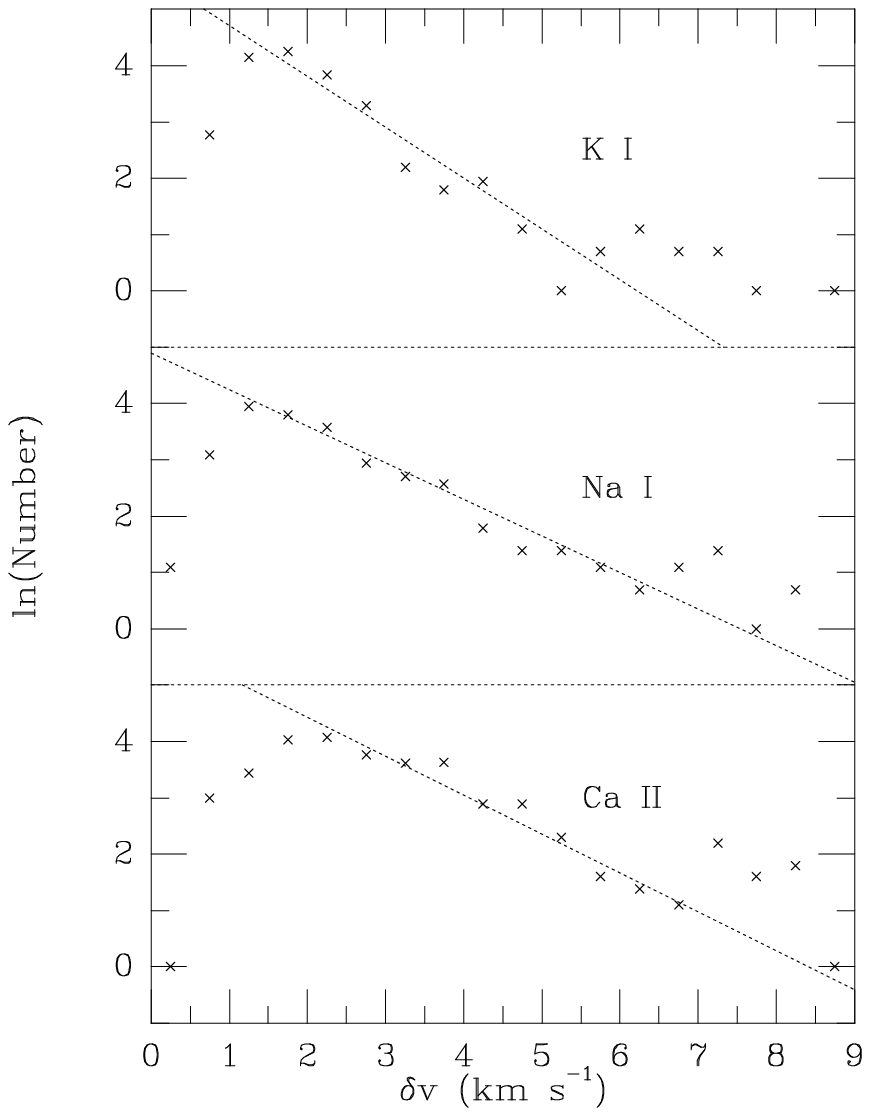}
\end{center}
\caption[Unresolved Interstellar Absorption Lines.]{ \label{fig:welty} The
distribution of velocity separations ($\delta$v) between adjacent
absorption components for \NaI, \KI, and \CaII.  The dotted line shows
the best fit over the range 2 \kms$\leq \delta v \leq 6.5 $ \kms, with
slopes --0.9, --0.6, and --0.7, for \KI, \NaI, and \CaII\
respectively.  The turnover in the numbers of components for small
separations is the result of unresolved velocity structure, while the
increase of component separations for $\delta v > 6-7$ \kms\ shows
that the distribution of cloud velocities is not purely Gaussian
(\cite{WeltyK:2001}).  Figure courtesy of Dan Welty.}
\end{figure}

The resolutions of UV instruments, $R < 10 ^5$, are well below those
of the best optical spectrometers, $R > 3 \times 10^5$.  The result is
the loss of information about the velocity structure of the
clouds forming UV absorption lines.  High resolution, 0.3--0.5
\kms, optical data on \CaII, \NaI, and \KI\ show that the number of
adjacent components separated by velocity $\delta v$ increases
exponentially as $\delta v \rightarrow 0$ \kms\
(\nolinebreak \cite{WeltyNa:1994,WeltyCa:1996,WeltyK:2001}).  The result is that
$\sim$60\% of cold clouds may be missed at the $\sim$3 \kms\
resolution of STIS and GHRS.  Cold clouds have a median \bdop(\NaI)$
\sim$0.73 \kms, and typical temperature $\sim$80 K.
Fig. \ref{fig:welty} shows the distribution of component separations,
$\delta v$, for \NaI, \KI, and \CaII.  The reduction of the numbers of
components with small separations, $\delta v \ltsim 1.5$ \kms,
compared to the best fit exponential distribution, indicate unresolved
velocity structure, while the increase of component separations for
$\delta v > 6-7$ \kms\ indicates that the distribution of cloud velocities
is not purely Gaussian (\cite{WeltyK:2001})

\section[The Solar Journey in Space]{Solar Journey through Space:  The Past 10$^4$ to 10$^6$ Years}
\label{sec:journey} 

The Sun has spent most of Quaternary era, the past 2--3 Myrs, in the
Local Bubble, a region of space with very low ISM density, which extends
to distances of over 200 pc in parts of the third galactic quadrant
(\glong=180\deeg$\rightarrow$270\deeg, \cite{FrischYork:1986}).
Sometime within the past $\sim$140,000 years the Sun encountered a
substantially denser region, and the galactic environment of the Sun
changed dramatically.  The Local Bubble appears to be part of an
extended interarm region between the Sagittarius/Carina and the Perseus
spiral arms (\nolinebreak \cite{Beck:2001}).  For all viable solar apex motions, the
Sun is leaving the emptiest part of this bubble and now entering low
density ISM approaching us from the direction of the Scorpius-Centaurus
Association (SCA, \cite{Frisch:1981}, 1995).  The solar motion with
respect to the LB \nocite{Frisch:1995} and to the CLIC are shown in
Figs. \ref{fig:lb} and \ref{fig:clic}.  The LB dominates the historical
solar galactic environment of the Sun and affects the local interstellar
radiation field.  The absence of interstellar neutrals in the near void
of the LB interior would lead to an absence of pickup ions and anomalous
cosmic rays in the heliosphere for periods when the Sun was in the
bubble.  Reconstructing the solar galactic environment for the past
several million years requires knowledge of both solar motion and the LB
properties.

\subsection[Inside the Local Bubble]{Inside the Local Bubble}
\label{sec:localbubble}

The LB surrounding the Sun was discovered through the absence of
interstellar dust grains, which redden starlight, in the nearest $\sim$70 pc
(\cite{Eggen:1963,Fitzgerald:1968,Lucke:1978,Vergelyetal:1997}).  
Reddening data is typically sensitive to color excesses \ebv$>$0.01 mag, 
corresponding to \NH = $10^{19.76}$ \cmtwo, depending on photometric quality.
The LB overlaps the interior of Gould's Belt
(\cite{FrogelStothers:1977,Grenier:2004}).  About 1\% of the ISM mass is in
dust grains, and interstellar gas and dust are well mixed in space.  

Data showing starlight reddening by interstellar dust are used here to
determine the Local Bubble configuration (LB, Figs. \ref{fig:lb}),
Fig. \ref{fig:lb} displays this configuration for two planes aligned
with the solar apex motion (Table \ref{tab:velocity}).  The right 
figure shows a slice perpendicular to the galactic plane,  where only
stars with longitudes within 25\deeg\ of the plane are plotted.  The left figure shows
the distribution of observed warm gas in a plane that is tilted by
27\deeg\ to the Galactic plane with the northern surface normal pointing
towards (\glong,\glat$)=(220$\deeg, $+63$\deeg) (thus intersecting the
galactic plane along \glong=130\deeg\ and \glong=310\deeg).  Only stars
with latitudes within about $\pm$15\deeg of that plane are plotted.  The
\ebv\ values are smoothed for stars with overlapping distances and
within 13\deeg\ of each other.  Fig. \ref{fig:lb} shows that in no
direction is a column density of \NH$> 10 ^{19.7}$ \cmtwo\ identified
close to the Sun in the anti-apex direction (for the gas-to-dust ratio
$\NH/\ebv = 5.8 \times 10^{21}$ atoms \cmtwo\ mag$^{-1}$,
\cite{BohlinSavageDrake:1978}).  The reddening data are from color
excess values, \ebv, calculated for O, B, and A0--A3 stars in the
Hipparcos catalog, after cleaning the data to omit variable or otherwise
peculiar stars (Hipparcos flag H6=0, \cite{Perrymanetal:1997}).  Both
projections show that an ISM column density of \NH$>10^{19.7}$ \cmtwo\
is not reached in the nearest $\sim$60--70 pc of the Sun along the solar
trajectory, and this conclusion is unaltered by the selection of the
solar apex motion.

The Local Bubble cavity is found in optical, UV, and EUV data, with
different sensitivities to the ISM.  The LB is seen in UV data
sensitive to \NH$> 10^{17}$ \cmtwo\ (\cite{FrischYork:1983}),
in the distribution of white dwarf and cool stars observed in the EUV
and sensitive to \NH$> 10^{18}$ \cmtwo\ (\cite{Warwicketal:1993}), in
\NaII\ absorption line data sensitive to \NH$\gtsim 10^{19}$ \cmtwo\
(\cite{Sfeiretal:1999,Vergelyetal:2001}), and in dust polarization
data sensitive to \NH$\gtsim 10^{19}$ \cmtwo\ (\cite{Leroy:1999}).
The sensitivities and sample densities differ between surveys, but the
results are consistent in showing that there is no slowly moving dense
ISM in the path of the Sun for the previous 3 Myrs, or $\sim$60 pc.
UV and EUV data are more sensitive to low column densities, and
detect upwind CLIC gas.  Figure 1 in Chapter 2 displays the Local
Bubble compared to nearby stellar associations, the Gum Nebula,
which is a giant \Htwo\ region, and the distribution of molecular
clouds in the solar neighborhood.

Are there nearby interstellar clouds that are inside of the Local
Bubble but not shown in Fig. \ref{fig:lb}?  CLIC column densities are
too low for detection in \ebv\ data, but the CLIC is seen in EUV, UV, and
optical data as local ISM with \NH$\ltsim$10$^{19}$ \cmtwo\ (\S
\ref{sec:clic}).  Looking towards the third galactic quadrant, the
only ISM close to the anti-apex direction, and in the distance
interval 5$< d <$80 pc, are \CaII\ components in $\gamma$ Ori (\glong=197\deeg,
\glat=--16\deeg, d=75 pc) showing \NHI$\sim 10^{19.50}$ \cmtwo\ if
\CaII/\HI$\sim$10$^{-8}$.  The LSR motions of these \CaII\ components
correspond to 0$\rightarrow$13 \kms (Standard LSR), so that the Sun
has moved 20--33 pc with respect to this ISM over $\sim$3 Myrs.
Only if one of the components towards $\gamma$ Ori is within 35 pc of
the Sun will they have surrounded the Sun
recently.  This argument is consistent with \CaII\ limits towards the
star $\beta$ Eri (27 pc), near the anti-apex direction, indicating 
\NHI$< 10^{19.30}$ \cmtwo\ (\nolinebreak \cite{FrischSembach:1990}).  There is not enough
data to rule out encounters with very low column density clouds near the
anti-apex direction, \NHI$\ltsim 10^{18}$ \cmtwo, particularly since
the Arecibo survey shows that such ISM may have high LSR velocities
(\S \ref{sec:global}).

Data on CO molecular clouds indicate the Sun is unlikely to have
crossed paths with a large dusty cloud over the past 3 Myrs.  Nearby
CO clouds, $d<$130 pc, typically have low LSR velocities.  Several
dust clouds are found at $\sim$120 pc (large triangles,
Fig. \ref{fig:lb}) towards the anti-apex direction,
\glong$\sim$200\deeg, \glat$\sim$--36\deeg.  They are 3C105.0 ($V$=8.0
\kms), MBM20 ($V$=0.3 \kms), and LDN1642 ($V$=1.3 \kms, Standard LSR
motions).  The clouds would have traveled less than 24 pc over 3 Myrs,
compared to the distance traveled by the Sun of $40 - 60$ pc, so Sun-cloud separations
remained over 40 pc.

\subsection[Radiation and Plasma of the Local Bubble]{Radiation and Plasma of
the Local Bubble}\label{sec:lbradiation}

The location of the Sun inside of a cavity in the ISM has a profound
affect on the immediate solar environment, including the ambient
radiation field and the properties of the interstellar plasma at the
solar location.  Prior to the entry of the Sun into the CLIC, the very
low density material of the Local Bubble surrounded the Sun.  We now
go into some detail about the physical characteristics of that
material, and the resultant radiation field, because the Local Bubble
radiation field is the dominant factor in the ionization state of the
ISM close to the Sun, and because the physical properties of the
heliosphere are extremely sensitive to the LIC ionization.  The following 
discussion is based on
equilibrium assumptions for the Local Bubble plasma.  Alternate
models, including models of non-equilibrium cooling, are found in the
literature but not discussed here.

While the nature and origin of the Local Bubble is the subject of
ongoing debate, at least a few facts are agreed upon.  The first is
that the Solar System and the low density clouds around it 
are embedded within a large, $\sim50 - 200$ pc radius, cavity that has
a very low density of both neutral and ionized gas.  Even allowing for
large thermal pressure variations in the ISM, we expect that this
volume is filled with a very low density ionized plasma, which
provides some degree of pressure support for the cavity.  The view
that held sway for many years was that the Local Bubble was filled
with hot, relatively high pressure gas with a temperature of $\sim
10^6$ K and pressure of $P/k_\mathrm{B} \sim 10^4$ cm$^{-3}$~K.  These
conclusions were drawn from observations of the diffuse soft X-ray
background (SXRB), which was mapped over the entire sky at low
spectral and spatial resolution but high sensitivity at energies from
$\sim 70$ eV to several keV by the Wisconsin group using sounding
rockets (e.g. McCammon et al. 1983).

The ROSAT all-sky survey map, which has higher spatial resolution and
sensitivity than the Wisconsin group maps, has made it clear that a
substantial amount of the SXRB at high latitude comes from beyond the
boundaries of the Local Bubble.  X-ray shadows made by clouds outside the
Local Bubble have provided information about the fraction of the emission
coming from within the cavity.

\nocite{Cravens_2000,Cravens_etal_2001} \nocite{Pepino_etal_2004}

Another complication regarding the source of the soft X-ray emission is
that it may be contaminated by X-ray emission from gas much closer to Earth.  
As discussed by
Cox (1998), and further explored by Cravens (2000), Cravens et al. (2001),
\nocite{Cox_1998} 
and Pepino et al. (2004), a potentially substantial contribution to the SXRB
may be coming from the heliosphere and geocorona.  The emission mechanism
involves charge transfer reactions of highly charged ions such O$^{+7}$ in the
solar wind with neutral H or He either in the interstellar wind or geocorona.
After the charge transfer, the ion is in a highly excited state and radiatively
decays to ground by emitting Lyman or K-shell photons.  The heliospheric
emission is expected to peak in emissivity at about $1-10$ AU because of the
combination of the spatial dependence of the ionization of inflowing neutral
atoms and the density of the solar wind.  Current estimates are still quite
uncertain, but put the contribution of the charge transfer photons to the
observed SXRB at about $10-50$\%
(\cite{Wargelin_etal_2004,Cravens_etal_2001}).

It is important to note that even if the heliospheric and geocoronal soft
X-ray emission is at the upper end of current estimates, they are unimportant
for LIC ionization because the flux in the LIC, which is much further away
($\gtsim 100\times$), is insignificant.  Moreover, the emission observed by
\emph{ROSAT} and the Wisconsin sounding rockets ($\gtsim 70$ eV) is
substantially harder than the emission that is directly responsible for LIC
ionization ($\sim 13.6 - 54.4$ eV). Thus the observed SXRB is relevant to the
photoionization of the cloud only insofar as it provides information on the
hot gas that is emitting EUV photons that ionize the LIC.

The available information on the temperature of the hot gas in the
Local Bubble, which we still believe to be responsible for most of the
observed SXRB, particularly near the Galactic plane, is mostly at low
spectral resolution.  The all sky maps of \emph{ROSAT} in the energy
band near 1/4 keV have been used to infer an effective plasma
temperature near $10^6$ K (e.g.\ \cite{Kuntz+Snowden_2000}) under the
assumption of a plasma in collisional ionization equilibrium (CIE).
The two higher spectral resolution data sets that are available (DXS,
\cite{Sanders_etal_2001} and XQC, \cite{McCammon_etal_2002}) are for
very limited portions of the sky and at very low spatial resolution,
and are not consistent with this model. In fact these spectra do not
fit any currently known model for the temperature and ionization state
of the plasma.  One thing that does seem clear at this time is that
iron must be significantly depleted in the plasma because the bright
Fe line complex near 70 eV is observed to be much fainter than
predicted.  Perhaps this should not be surprising, since in the LIC
$\sim$95\% of the Fe is missing from the gas and presumably depleted
onto dust grains, while in cold clouds $\sim$99\% of the Fe is
depleted.

Disregarding the problems with models of the emission spectrum, under
the assumption that there is a CIE hot plasma filling the LB, the
implied density is roughly $5\times 10^{-3}$ \cc\ and pressure of more
than $P/k_B = 10^4$ \cc K.  This is a high thermal pressure for the
ISM, though certainly not unheard of, and may be typical of hot gas in
the ISM.  If the LIC is only supported by thermal pressure, however,
there would seem to be a substantial mismatch between its pressure,
$P/k_B \sim 2300$ \cc K, and that of the LB (\S \ref{sec:pressure}).
This mismatch may be fixed by magnetic pressure, though the size of
the heliopause appears to require a field somewhat below that
necessary to make up the difference (\nolinebreak \cite{Ratkiewicz_etal_1998}).
That limit can also be avoided for a magnetic field that is nearly
aligned with the direction of the interstellar wind
(\cite{Florinski_etal_2004}), since in this case the heliosphere
confinement is much less affected by the field.  If a substantial
portion of the SXRB originally attributed to the hot gas is from
charge transfer emission, then models that properly take that into
account will necessarily imply lower pressures for the LB, which helps
to resolve this discrepancy.  Note, though, that the emissivity goes
as $P^2$, so a 50\% reduction in emission results in only a factor of
$1/\sqrt{2}$ reduction in $P$.

\section[Neighborhood Interstellar Medium]{Neighborhood ISM:  
Cluster of Local Interstellar Clouds}  \label{sec:clic}

The ISM surrounding the Sun is flowing through the LSR with a bulk
velocity of $-19.4\pm4.6$ \kms\ and an upstream direction
\glong$\sim$331\deeg, \glat$\sim -5$\deeg, towards the
Scorpius-Centaurus Association (SCA).  This warm, low density gas
consists of cloudlets defined by velocity, which we denote the
cluster of local interstellar cloudlets, CLIC.  The upwind direction of
the bulk velocity of the CLIC in the LSR suggests an origin related to
stellar activity in the SCA (\nolinebreak \cite{Frisch:1981,FrischYork:1986,Frisch:1995},
also see \S \ref{sec:origin}).  Low column densities are found for ISM
within 30 pc, \NHI$<$10$^{19}$ \cmtwo.  The CLIC temperature, \NHI, and
kinematics resemble global warm neutral medium (WNM) detected by radio
\HI\ 21 cm data (\S \ref{sec:global}).

If we restrict the star sample to objects within $\sim$10.5 pc, the mean
\NHI\ for CLIC cloudlets is $\langle$\NHI$\rangle \sim 6.4 \times
10^{17}$ \cmtwo\ based on \HI\ or \DI\ UV data (\S
\ref{sec:basic}, and Table \ref{tab:clic}).  For \nHI\ 
similar to the LIC density at the solar location,
\nHI=0.2 \cc, the mean cloud length is 1.0 pc, and the mean
time for the Sun to cross them is $\sim$68,000 years (Table
\ref{tab:clic}).  The LSR solar velocity is $13-19$ \kms\ (\S
\ref{sec:lsr}), and the solar galactic environment over timescales of
3 Myrs will be regulated by ISM now within $\sim$60 pc, provided there
are no undiscovered high velocity  ($V>$20 \kms) clouds with suitable
trajectories.
\begin{table}[h]
\caption[Characteristics of ISM within 10 pc.]{The Characteristics of ISM within 10 pc of the Sun}
\begin{tabular}{l c  }
\hline
LSR Upwind direction $^{\rm (a)}$ & \glong$\sim$331.0\deeg, \glat$\sim$--5.1\deeg  \\
LSR Velocity$^{\rm (a)}$ & $V$=--19.4$\pm$4.6 \kms  \\
No. stars sampling ISM within 10 pc$^{\rm (b)}$& 20  \\

Sightline averaged \HI\ space density & \nHIavg = 0.07 \cc  \\
No. velocity components$^{\rm (c)}$ & 30  \\
Component averaged, \NHI & $\langle$\NHI$\rangle$= $6.4\times 10^{17}$ \cmtwo  \\
Component averaged cloud thickness, $L$, &  $\langle L \rangle$ = 1.0 pc  \\
\hspace{0.5in} for \nHI=0.2 \cc  & \\
Component averaged cloud crossing & $\langle  T_{\rm L}  \rangle$ = 68,000 years  \\
\hspace{0.5in} time, $T_{\rm L}$ &   \\
Galactic center hemisphere filling factor, \ff & 0.40 \\
Anti-center hemisphere filling factor, \ff & 0.26 \\
Solar entry into the CLIC &  (44,000--140,000)/\nHIavgtwo\ yra \\
Solar entry into the LIC &  $<$40,000/\nHIavgtwo\ yra \\
Solar exit from LIC &  next $\sim$3700/\nHIavgtwo\ years \\
\hline
\end{tabular}
\newline
Notes: 
The unit ``years ago'' is represented as ``yra''. 
\nHIavgtwo\ is the average \HI\ density in units of 0.2 \cc.
(a)  From Table \ref{tab:velocity}.
(b) The HD numbers of these stars within 10 pc are:
10700, 17925, 20630, 22049, 23249, 26965, 39587, 48915, 61421, 62509, 48915B, 115617, 128620, 128621, 131156, 155886, 165341, 187642, 197481, 201091, 209100. 
Sources of data for these stars are given in \S \ref{sec:basic}.
(c) Based on UV data with resolution $\sim$3 \kms, generally.
\label{tab:clic}
\end{table}

The first evidence of interstellar gas within 15 pc was an anomalously
strong \CaII\ line observed towards Rasalhague ($\alpha$ Oph), formed in what is now
known as the G-cloud that is widespread in the galactic-center
hemisphere (\cite{Adams:1949,MunchUnsold:1962}).  The first spectral
data of Ly$\alpha$ emission from interstellar gas inside of the
heliosphere, obtained by \emph{Copernicus}, showed the similarity of ISM
velocities inside of the heliosphere and towards nearby stars
(\cite{AdamsFrisch:1977,McClintocketal:1978}).

The basic properties of ISM forming the galactic environment of the Sun
were discovered with \emph{Copernicus}, including the widespread
presence of partially ionized gas (\S \ref{sec:wim}), the asymmetrical
distribution of local ISM showing higher column densities towards the
galactic center hemisphere (\cite{BruhweilerKondo:1982,FrischYork:1983}), and the
discrepancy between the velocity of ISM inside of the solar system and
towards the nearest star $\alpha$ Cen
(\cite{Landsmanetal:1984,AdamsFrisch:1977}).  The shocked history of the
nearest ISM was revealed by enhanced abundances of Fe and other
refractory elements, which indicated the destruction of interstellar
dust grains by interstellar shocks
(\cite{SnowMeyers:1979,Frisch:1979,Frisch:1981,Crutcher:1982,York:1983}).
The first data showing the shift between the velocities of \HI\ inside
of the heliosphere and towards nearest star $\alpha$ Cen, now
interpreted as partially due to the hydrogen wall, were obtained by
\emph{Copernicus} and IUE (\nolinebreak \cite{Landsmanetal:1984}), although the
hydrogen wall contribution was not recognized as such until the models
of \HI\ deceleration in the heliosheath were constructed
(\cite{Gayleyetal:1997,LinskyWood:1996}).

Optical or UV data are now available for $\sim$100 stars sampling nearby
ISM (see \cite{FGW:2002,RLIII,Woodetal:2005}, and references in these
papers).  Very high-resolution optical data, $\sim$0.3--0.5 \kms, are
available for $\sim$40 nearby stars, and high resolution UV data,
$\sim$3 \kms\ for an additional $\sim$65 stars.  Component blending
between local and distant ISM usually prevent the use of data from
distant stars.  The \HI\ \Lya\ line is always saturated, even for low
column density sightlines, so we use \DI\ as a proxy for \HI, with a
ratio \DI/\HI=1.5 $\times 10^{-5}$ that is valid for local ISM
(\cite{VidalMadjarFerlet:2002,Linsky:2003,Lehneretal:2003}).  These
data, combined with radiative transfer models of ionization gradients,
give the velocity, composition, temperature, and morphology of the CLIC.

\subsection[Partially Ionized Gas]{Warm Partially Ionized Medium, WPIM }
\label{sec:clicwim}

Charged and neutral particles have different interactions with the
heliosphere, and the ionization gradient of the cloud affects the
heliosphere as it traverses a tenuous cloud.  Over 30 years ago
\emph{Copernicus} discovered that \NII\ is widespread towards nearby
stars, \NII/\NI$\sim$1, indicating that EUV photons capable of
ionizing hydrogen, nitrogen, and oxygen penetrate to the interiors of
tenuous clouds and showing that the nearest ISM is partially ionized,
\chiH$>$0.1 (\cite{RogersonIII:1973}).  This discovery contradicted
the classic view of fully neutral or fully ionized ISM derived from
observations of denser ISM, $\log \NHI\ (\cmtwo) >20$, where electrons
originate from low FIP, $< 13.6$ eV, abundant elements such as C. The
ionizations of H, O, and N, with $\mathrm{FIP} = 13.6$, 13.6, and 14.5
eV respectively, are coupled by charge exchange.  The exception is near
a cloud edge where high EUV flux causes \NI/\HI\ to dip because
photoionization dominates charge exchange ionization for \NI\ (\S
\ref{sec:rt}, Fig.  \ref{fig:nHInHeI}).  Partially ionized gas in the
CLIC is established by N fractional ionizations of
\chiN$\sim$0.27--0.67, derived from FUV and UV data on \NI\ and \NII\
towards stars with $\log \NHI (\cmtwo) < 19$, such as HZ 43, HD 149499B,
WD 0549+158, WD 2211-495, and $\eta$ UMa
(\cite{Lehneretal:2003,Frischetal:2005uma}).

Among these stars the most highly ionized sightlines are HD 149499B
and WD2211-495.  Both of these hot stars have detected \OVI\
absorption, which appears to be interstellar, although a stellar origin
can not be ruled out entirely (\cite{Oegerleetal:2005}).  These stars
are in the upwind hemisphere of the CLIC (\S \ref{sec:bulkflow}),
with HD 149499B at \glong=330\deeg, \glat=--7\deeg, d=37 pc, and
WD2211-495 at \glong=346\deeg, \glat=--53\deeg, d=53 pc.  In contrast,
ISM towards WD 1615-154 is primarily neutral with \chiN$\sim$0.05
(\glong=359\deeg, \glat=+24\deeg, d=55 pc).  Several stars sample
mainly the LIC, such as WD0549+158 near the downwind direction
(\glong=192\deeg, \glat=--5.3\deeg, d=49 pc), where \NHI$\sim 5 \times
10 ^{17}$ \cmtwo\ is comparable to the expected LIC column density.

The H and He ionizations are found from EUV spectra of white dwarf
stars, and comparisons of fluxes at the \HeI\ and \HeII\ ionization
edges at 504 \AA\ and 229 \AA\ with atmosphere models.  Well observed
stars such as HZ 43, GD 153, and WD 0549+158 show \NHI/\NHeI$ \sim
9.8-15.8$, instead of the ratio of 10/1 expected for neutral gas with
a $\mathrm{H/He}=10$ cosmic abundance.  When lower quality data are
included, the variation is larger, \NHI/\NHeI=9--40
(\nolinebreak \cite{Dupuisetal:1995,Kimbleetal:1993a,Frisch:1995,Vallerga:1996,Wolffetal:1999}).
Ratios of $\mathrm{\HI/\HeI}>10$ indicate that He is more highly
ionized than H, and the hardness of the radiation field implied by
this discovery is discussed in \S \ref{sec:rtradiation}.  

The ratios \RMg\ and \RC\ serve as ionization diagnostics and give
\nel\ values that are independent of abundance uncertainties.  The
\MgI\ abundance is enhanced by dielectronic recombination in warm ISM,
$T>6000$ K, and \MgII\ is the dominant ionization state.  The electron
densities are given by \nel=$C_{\rm Mg}$($T , \Lambda$) \NMgI/\MgII\ and
\nel=$C_{\rm C} $($T$) \NCIIstar/\NCII.  The quantity $C_{\rm Mg} $($T$,$\Lambda$)
is the ratio of the temperature dependent photoionization and recombination 
rates, and depends on both cloud temperature ($T$) and
the FUV radiation field ($\Lambda$).  $C_{\rm C} $($T$) is the
temperature sensitive ratio of the \CII\ fine-structure collisional
de-excitation to excitation rates and is independent of the radiation
field.  For more details, see York and Kinahan
(1979). \nocite{YorkKinahan:1979} The observed ratios in the CLIC of
\RC=50--200 and \RMg=200--450 are consistent with the predictions of
equilibrium radiative transfer models of low column density gas such
as the LIC (section \S \ref{sec:rt}). These equilibrium models predict
\nel=0.06--0.12 \cc, with the best value for the CLIC being $\sim$0.1
\cc. Some caution is required when using observed \CII\ column
densities, however, since the \CII\ 1335 \AA\ line is generally
saturated, which may bias \RC\ towards smaller values.

Neutral Ar, FIP=15.8 eV, is a valuable ionization diagnostic since it
is observed inside the solar system, in the form of the anomalous
cosmic ray component seeded by interstellar \ArI, and also in FUV observations 
of nearby stars.  The ratio log \ArI/\HI\ in the ISM, $\sim -5.8\pm 0.3$ dex, 
is below values seen towards B-stars by factors of two or more
(\cite{JenkinsetalAr:2000,Lehneretal:2003}).  The
recombination coefficients of \ArI\ and \HI\ are similar, but the
\ArI\ photoionization cross section exceeds that of \HI\ by $\sim$10.
Argon depletion is expected to be minimal, so Ar/H towards
nearby stars is interpreted as conversion to unobserved \ArII.  The RT
models predict $\ArII/\ArI = 1.8 - 2.9$, $\ArI/\mathrm{Ar} = 0.16 -
0.30$, and $\log \ArI/\HI = -5.98~\mathrm{to}  -6.17$ for the LIC at the solar
location, compared to observed ACR values of $\log \ArI/\HI =
-5.97\pm0.16$ after filtration effects in heliosheath regions are
included (see FS for more discussion and references).

A classic ionization diagnostic uses the strong optical lines
of the trace ionization species \NaI\ and \CaII and the
assumption that ionization and recombination rates balance:
\begin{equation}
n({\rm Na}^\circ) \Gamma({\rm Na}^\circ) = n({\rm Na}^+) \alpha({\rm Na}^+) n({\rm e})
\end{equation}\label{eq:1}
Implicit is the assumption that $N$(\NaI)=$L~ n({\rm Na}^\circ)$ \cmtwo,
or analogously that $n$(\NaI)/$n$(\NaII) is constant, where $L$ is the cloud
length.  $\Gamma({\rm Na^\circ})$ is the total ionization rate for
\NaI$\rightarrow$\NaII, and $\alpha({\rm Na^\circ}) $ is the total
recombination rate for \NaII$\rightarrow$\NaI\ (e.ge.g.
\cite{Spitzer:1978,Pottasch:1972b}).  In principle \CaII\ and \CaIII\
can be substituted for \NaI, and \NaII, respectively. \NaII\ and
\CaIII\ are unobservable.  Since \NaI\ (and \CaII) are trace
ionization states in warm gas, some assumption is required for the
element abundances, $[\frac{Na}{H}]$ (and $[\frac{Ca}{H}]$):
\begin{equation}
n({\rm Na}^\circ) \Gamma({\rm Na}^\circ) = n({\rm H}) \lbrack\frac{Na}{H}\rbrack \alpha({\rm Na^+}) n({\rm e}) 
\end{equation}
The electron densities derived from \NaI\ and \CaII\ data may be
unreliable for low column density warm clouds because of large
ionization corrections and highly variable abundances, of $\sim$40 and
$\sim$1.6, respectively, for Ca and Na (\nolinebreak \cite{Welty23:1999}, \S
\ref{sec:basic}).  Calcium ionization levels are temperature sensitive,
and \CaIII/\CaII$>$1 for $T>4,000$ K and \nel$<$0.13 \cc.
Electron densities of \nel=0.04--0.19 \cc\ are found for
the CLIC within 30 pc from \MgII/\MgI\ and \NaI\
(\cite{Frischetal:1990,LallementFerlet:1997}).

\begin{figure}[ht]
\begin{center}
\includegraphics [width=2.5in]{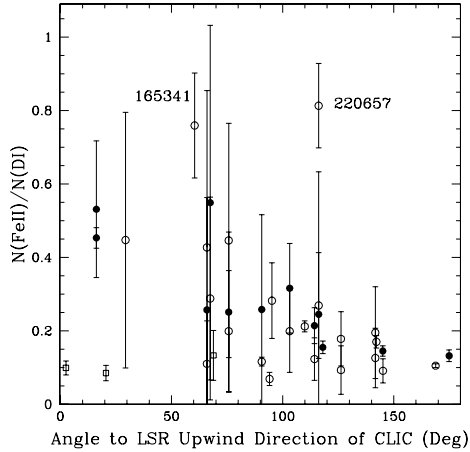}
\end{center}
\caption[The \FeII/\DI\ Ratio as a Function of Angle to LSR Upwind Direction.] {\label{fig:fed} The ratio \NFeII/\NDI\ varies
between the upwind and downwind direction of the CLIC, and over
spatial scales of $\sim 3$ pc.  \NFeII/\NDI\ is plotted against the
angle (in degrees) between the background star and the LSR upwind
direction of the CLIC ($\ell = 331.4^\circ,  b =-4.9^\circ$,
Table \ref{tab:velocity}).  Clouds with observed velocities
within $\pm 1$ \kms\ of the projected LIC velocity are plotted as
filled circles, while open circles represent CLIC components at other
velocities.  The open squares represent lower resolution data towards
$\alpha$ Oph, $\lambda$ Sco and HD 149499B.  The ratio \NFeII/\NDI\
for the LIC differs between the direction of $\alpha$ Cen AB, at a
distance 1.3 pc and angle of 16\deeg, and Sirius, at a distance of 2.7
pc and angle of 103\deeg, although uncertainties overlap.  Velocities
are drawn from the data compilations in the RL and Frisch et
al. (2002) surveys.\nocite{FGW:2002} }
\end{figure}

The ratio \FeII/\DI\ presents clear evidence for variations in the ISM
over spatial scales of 1--3 pc, however, it is unclear whether the
variation arises from an ionization gradient or from a gradient in the
Fe abundance close to the Sun.  Ionization and depletion both affect
\FeII/\DI, since \FeII\ is the dominant ionization state of Fe for
both neutral and warm ionized gas (SF02) and \DI\ traces only neutral
gas.  Fig. \ref{fig:fed} shows \FeII/\DI\ for CLIC velocity
components, plotted as a function of the angle, $\theta$, between the
star and the LSR upwind direction of the CLIC (Table
\ref{tab:velocity}).  The ratio \NFeII/\NDI\ for cloudlets within
$\sim$4 pc differs by $\sim$54\% over spatial scales of 4 pc between
upwind stars such as $\alpha$ Cen AB, $\theta \sim$16\deeg, and
downwind stars such as Sirius, $\theta \sim$ 103\deeg, although
uncertainties overlap.  Similar Fe abundance variations are found in
thin ($\sim 0.1-0.9$ pc) disk clouds towards Orion
(\cite{Welty23:1999}).  The Fe abundance is highly variable in the
ISM, and varies between cold dense and warm tenuous disk clouds by
factors of $\sim$4--6 (\cite{SavageSembach:1996,Welty23:1999}).  
The local \FeII/\DI\ trend may be from a combination of ionization
and abundance variations.  The first explanation for the physical
properties of local ISM attributed the CLIC to a superbubble shell
around the SCA that is now approaching the Sun from the upwind
direction (\cite{Frisch:1981}, \S \ref{sec:origin}).  Dust grain
destruction in the shocked superbubble shell would have restored 
elements in silicates, which can be destroyed by shocks with speeds
$>50$ \kms, back to the gas phase (\cite{Slavinetal:2004,Jones:1994}).  This
effect is echoed in extended distances to the upwind edge of the CLIC
obtained from \CaII\ data (Figure \ref{fig:clic}).

A diffuse \Htwo\ region appears to be close to the Sun in the
LSR upwind direction of the CLIC, based on ISM towards $\lambda$ Sco,
and the high value \chiN$\sim$0.66 towards HD 149499B (WD 1634-593, \cite{Lehneretal:2003}).
About 98\% of the neutral gas towards $\lambda$ Sco belongs
to the CLIC, and shows a HC velocity of $-26.6$ \kms.  
A diffuse \Htwo\ region, with a diameter of $\sim$30\deeg\ and 
\nel$\sim$0.1--0.3 \cc, is also present at $\sim  -17.6$ \kms\
(using optical data to correct \cop\ velocities to the HC scale,
\cite{York:1983}).  This diffuse \Htwo\ region may  
extend in front of both the HD 149499B and $\lambda$ Sco sightlines.
The HD 149499B sightline samples the ISM approaching the solar system 
from the LSR CLIC upwind direction.

Highly ionized gas is observed in the downwind  CLIC.
Interstellar \SiIII\ is seen towards $\epsilon$ CMa but not
towards $\alpha$ CMa (Sirius), $\sim$12\deeg\ away
(\cite{GryJenkins:2001,Hebrardetal:1999}).  The limit on \SiIII\ towards Sirius
is a factor of 15 below the $\epsilon$ CMa LIC component, with
$N($\SiIII$)=2.3\pm0.2\times 10^{12}$ \cmtwo.  
This ion is particularly
interesting because it is predicted to have quite a low column density both in
the warm LIC gas, because of its high charge transfer rate with \HI, and in an
evaporative boundary where it becomes quickly ionized in the interface (see
SF02).  Gry and Jenkins speculate that \SiIII\ is formed
in an outer layer of the LIC behind Sirius (2.7 pc).  
However this scenario
requires that the second local cloud observed towards both stars, which is
blue-shifted by $\sim -7$ \kms\ from the LIC (hence the name Blue Cloud, BC),
is a clump embedded in an extended LIC.  Since \SiIII\ is
also detected for the BC towards $\epsilon$ CMa but not Sirius (but with large
uncertainties), the origin of the \SiIII\ towards $\epsilon$ CMa
may instead require nonuniform interface layers such as turbulent
mixing layers (see below).


\subsection[Dynamical Properties]{Dynamical Characteristics of Nearby ISM}
\label{sec:bulkflow}

The heliosphere radius in the upwind direction is approximately
proportional to the relative velocity between the Sun and ISM, so that
the heliosphere is modified by variations in the ISM velocity.
Absorption line data give the centroid of the radial component of a cloud
velocity, integrated over a cloud length, but UV data do not resolve
all of the velocity structure (\S \ref{sec:absline}).  The coherent
motion of nearby ISM is found from these velocity centroids.  With the
exception of the LIC, which is detected inside of the heliosphere, only
radial velocities can be measured.  Thus the 3D motions of other
cloudlets must be inferred from observations towards several stars.

It has been known for some time that the Sun is immersed in ISM
flowing away from the Scorpius-Centaurus Association and towards the
Sun (\cite{Frisch:1981,Crutcher:1982}).  The CLIC kinematical data can
be interpreted as a coherent flow
(\cite{FrischYork:1986,Vallergaetal:1993}), or individual clouds can
be identified with similar velocities towards adjacent stars (\cite{Lallementetal:1986,FGW:2002}).  Clouds
that have been identified are listed in Table \ref{tab:clouds}.
Among these clouds is the G-cloud, in the galactic center hemisphere,
which was first identified in optical data long ago
(\cite{Adams:1949,MunchUnsold:1962}).  The Sun is located at the
leading edge of the stream of CLIC gas (\cite{Frisch:1995}).

\begin{figure}[ht] 
\begin{center}
\includegraphics[width=4.0in,angle=0.]{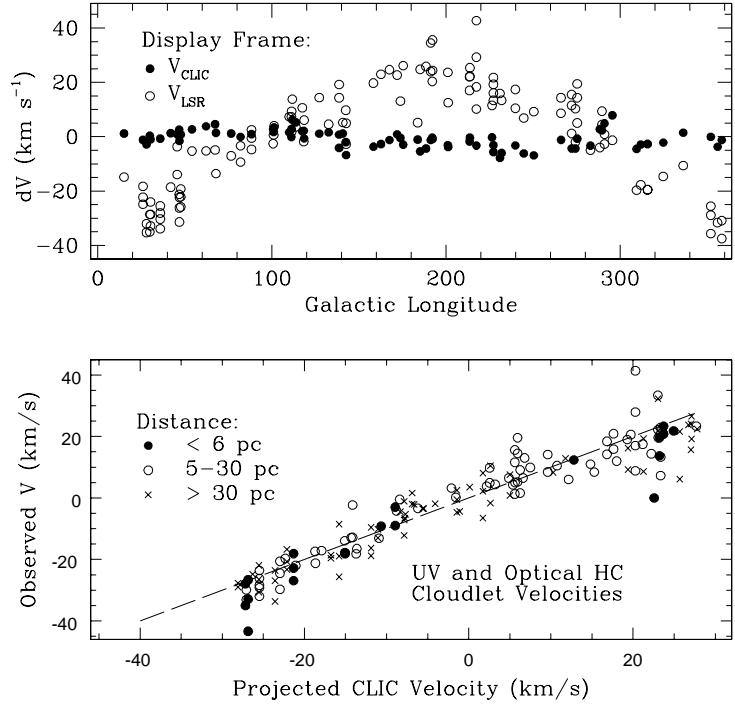}
\end{center}
\caption[Bulk Flow of ISM past Sun.]{ \label{fig:flow} {Top:} The
star longitude is plotted against the velocities, dV, of CLIC absorption
components after transforming the absorption components into the
\Vbf\ frame (filled circles), and into the LSR (open circles, Table
\ref{tab:velocity}).  Clearly \Vbf\ provides a better fit to the CLIC
data than the LSR velocity frame.  \lsrstd\ is used here, and only
components within 50 pc of the Sun are plotted.  {Bottom:} The
observed heliocentric velocities, V, of absorption components tracing the CLIC are
plotted against the expected CLIC velocity projected towards each star
(Table \ref{tab:velocity}).  The
components near the upwind (approaching gas, $V<0$ \kms) and downwind 
(receeding gas, $V>0$ \kms) directions, which are expected to have the maximum
values of projected $|V|$, show indications of a decelerating flow.  }
\end{figure}

Kinematics of nearby ISM show three characteristics.  The first is
that the ISM within $\sim$30 pc flows past the Sun.  The bulk flow
velocity, \Vbf, can be derived from the Doppler-shifted radial
velocities of $\sim$100 interstellar absorption line components
towards $\sim$70 nearby stars (\cite{FGW:2002}).  The resulting \Vbf\
corresponds to a heliocentric velocity \mbox{--28.1}$\pm$4.6 \kms\ and
upwind direction (\glong,\glat)=(12.4\deeg,11.6\deeg)
(\cite{FGW:2002}).  An additional uncertainty of $\sim 1$ \kms\ may be
introduced by a biased sample, since there are more high-resolution
optical data for stars in the upwind than in the downwind directions.  The CLIC motion in
the LSR is then found by subtracting the solar apex motion from the
CLIC heliocentric vector. Table \ref{tab:velocity} gives the LSR CLIC
velocity for two choices of the solar apex motion.
Fig. \ref{fig:flow} (top) displays the velocities of individual
absorption components in the rest frame of the CLIC versus the rest
frame of the LSR.  Obviously the CLIC velocity is more representative of
the nearby gas.

The second property is that distinct clouds in the flow
contribute to the $\pm$4.6 \kms\ dispersion of \Vbf\ (Tables
\ref{tab:clouds} and \ref{tab:velocity}).  Among the clouds are the Apex
Cloud, within 5 pc and centered towards the direction of solar apex
motion, and the G-cloud, which is seen towards many stars in the upwind
hemisphere.  The Apex and G-cloud data indicate that temperature varies
by a factor of 2--6 inside of the clouds (Table \ref{tab:clouds}, \S
\ref{sec:pressure}, RL).  Evidently, for constant pressure the clouds
are either clumpy, or there is a dynamically significant magnetic field.
For densities similar to the LIC, \nHI$\sim$0.2 \cc, the clouds fill only
$\sim33$\% of nearby space and the flow of warm gas is fragmented,
rather than a streaming turbulent homogeneous medium (\S
\ref{sec:clicmorphology}).  The interpretation of CLIC kinematics as a
fragmented flow is supported by the fact that the upwind directions for
the CLIC, LIC, Apex Cloud, and GC are all within 20\deeg\ of each other.

A third characteristic is that the flow appears to be decelerating.
This is shown in Fig. \ref{fig:flow}, bottom.  Stars in the
upwind and downwind directions are those with projected CLIC velocities
of $<0$ \kms and $>0$ \kms, respectively.  The velocities of clouds
closest to the upwind direction are approaching the Sun compared to
\Vbf, while those closest to the downwind direction lag \Vbf, as would
be expected from a decelerating flow.  This deceleration is seen even
for the nearest stars, $d<6$ pc, indicating the pileup of ISM is close
to the Sun, which is consistent with the fact the Sun is in the
leading edge of the flow.

\begin{table}[h]
\begin{center}
\caption[Nearby Interstellar Clouds.]{Nearby Interstellar Clouds }
\begin{tabular}{l c c c c }
\hline
{Cloud}&\multicolumn{2}{c}{Velocity} &   {Location} & {Temp.}\\
{} & {HC} & {LSR } &  {D, \glong\ \glat} & {}\\
&{(\kms)}&{(\kms)}& {(pc,\deeg\ \deeg)} &(10$^3$ K)\\
\hline \hline
LIC   & --26.3$\pm$0.3 &  --20.7  & 0., 50\deeg$\rightarrow$250\deeg\ (all \glat) &6.4$\pm$0.3 \\
G & --29.3$\pm$4.0 & --21.7 & 1.4,  0\deeg $\pm$90\deeg\ 0\deeg $\pm$60\deeg\ & 2.7--8.7   \\
Blue & 10$\pm$1 & \ddot\ & 3,  233\deeg$\pm$7\deeg\ 10\deeg$\pm$2\deeg& 2.0--5.0 \\
Apex  &  --35.1$\pm$0.6  &  --24.5  & 5,  38\deeg$\pm$10\deeg\ 9\deeg$\pm$14\deeg & 1.7--13.0  \\
Peg/Aqr &  --4.5$\pm$0.5 & \ddot\ &  30, 75\deeg$\pm$13\deeg\ --44\deeg$\pm$5\deeg\  & \ddot\ \\
\hline 
\end{tabular}
\label{tab:clouds}
\end{center}
\flushleft
Notes:  The LSR velocities are based on the Standard solar apex motion.
The cloud distances are upper limits.
The Apex cloud is also known as ``Panoramix'' or the ``Aql-Oph cloud''. 
References: Table \ref{tab:velocity}, Witte (2004),\nocite{Witte:2004}
FGW, Frisch (2003),
Lallement et al. (1986), Lallement and Bertin (1992),GJ, RL. 
\nocite{LallementBertin:1992}
\nocite{Lallementetal:1986}\nocite{Frisch:2003apex} 
\end{table}

Note that because the 3D LIC
velocity vector is known, a 3D correction for the solar apex motion is
made, and the ``true'' downwind direction in the LSR differs from the direction
measured with reference to the solar system barycenter.

\subsection[The Distribution of Nearby ISM]{The Distribution of Nearby ISM}
\label{sec:clicmorphology}

The distribution of ISM within 5--40 pc is dominated by the CLIC,
while over larger scales it is dominated by the Local Bubble (\S
\ref{sec:localbubble}).  The $\geq$15\deeg\ diameter cloud observed
towards the Hyades stars, $\sim$40--45 pc away, is not included in the CLIC
discussion here because \NHI\ data are unavailable, and because the nonlocal ISM may
be inside the cluster (\cite{Linsky:2001Hyades}).

\begin{figure}[!t]
\caption[Distribution of Nearby Interstellar Gas.] {\label{fig:clic} 
\emph{See figures 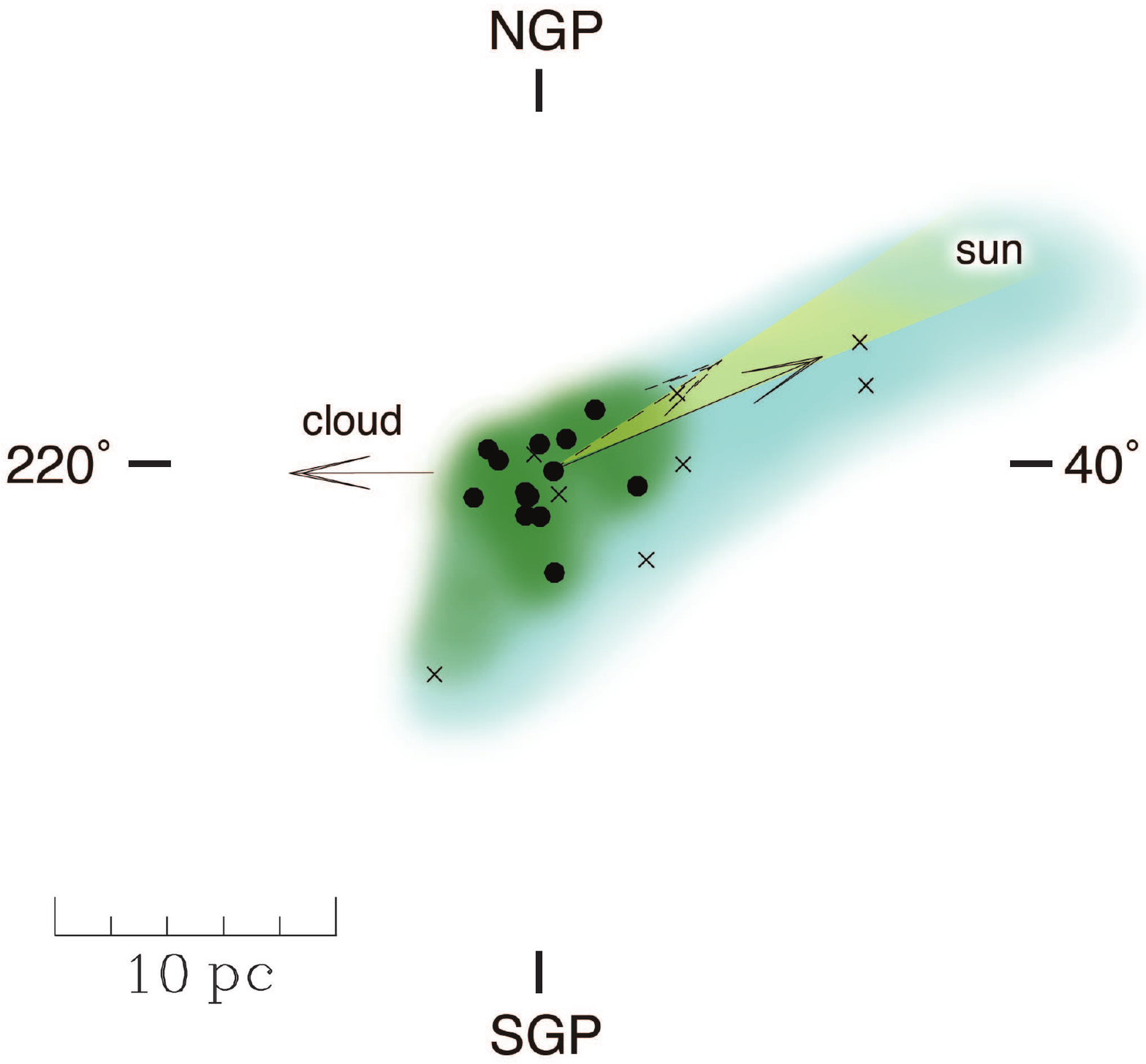 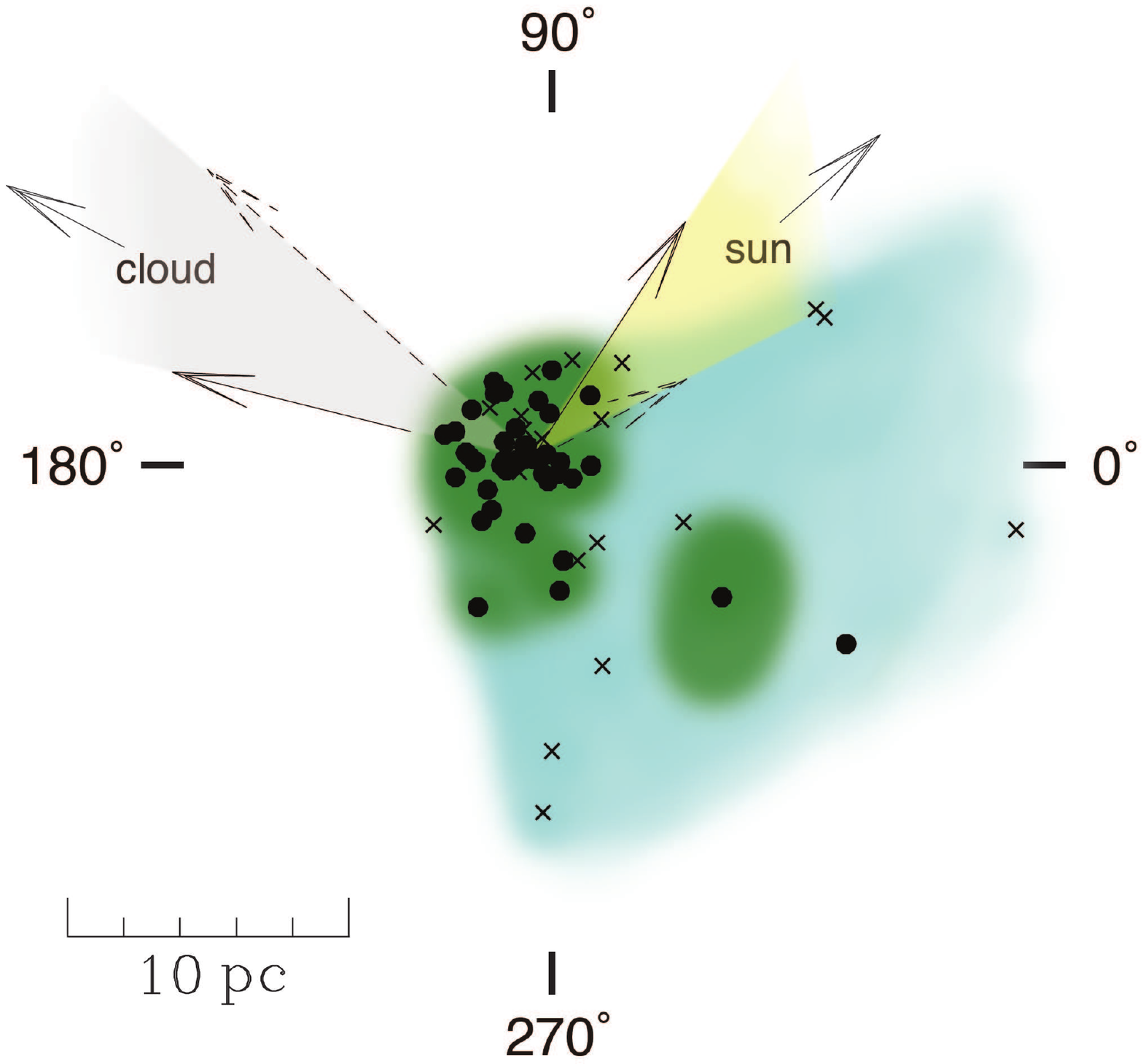.}
The distance to the CLIC edge for \nHI$\sim$0.2 \cc\ and a continuously distributed
ISM.  The dots give distances derived from \NDI\ and \NHI\ data, and
the x's show \CaII\ distances.  The extended \Htwo\ region found 
towards $\lambda$ Sco (\S \ref{sec:clicwim}, \cite{York:1983}) 
is shown as the origin of the excess cloud length towards the 
galactic center indicated by the \CaII\ data.
It is interpreted to indicate fully ionized gas near the Sun because
the diffuse \Htwo\ region seen towards $\lambda$ Sco is at CLIC
velocities.  The arrows give the motions based on the standard (solid)
and Hipparcos (dashed) solar apex motions for both the Sun and
CLIC.  Top: The edges of the CLIC are shown
projected onto the galactic plane.  Bottom: The CLIC distribution is
shown for a meridian cut 50\deeg\ wide in longitude, and extending
between \gl=40\deeg\ and 220\deeg\ (the plane of the solar
apex motion).  The CLIC LSR motion is nearly
perpendicular to the solar apex motion.
}
\end{figure}

The solar apex motion is compared to the distribution of CLIC gas
in Fig. \ref{fig:clic}, where the distance to the cloud edge for 
the CLIC is shown projected onto the galactic plane, and 
in a vertical plane perpendicular to the galactic plane and
aligned along the solar apex motion.  The angular width of the
vertical display, which extends from \glong=40\deeg$\pm$25\deeg\ to
\glong=220\deeg$\pm$25\deeg, includes the directions of both the Hipparcos and Standard
solar apex motions (Table \ref{tab:velocity}).  The downwind direction
of \Vbf\ and direction of solar motion (Table \ref{tab:velocity}) are 
shown by the arrows.  The
distance of the CLIC edge in the direction of a star is given by
\NHI/\nHIavg, where \NHI\ is the total interstellar column density
towards the star, and a uniform density similar to LIC values is
assumed, \nHIavg=0.2 \cc.  This distribution is based on \DI, \HI,
(dots) and \CaII\ (crosses) data.  The high resolution optical \CaII\
data provide excellent velocity resolution, but conversion to \NHI\ is
uncertain because of possible variable abundances.  The value
\NCaII/\NHI=10$^{-8}$ \cc\ is used.  The solar apex motion is shown by
the arrows pointing right, while the arrows pointing left show the two
LSR velocity vectors of the CLIC bulk flow, for the Standard (solid)
and Hipparcos (dotted) solar apex motions, respectively.

The CLIC shape is based on data in Hebrard et al.
(1999), RL, Dunkin and Crawford (1999), Crawford et
al. (1998), FGW, and Frisch and Welty (2005).  The \NHI\ column
densities are either estimated from \NDI\ (\DI/\HI$ =1.5\times
10^{-5}$), or based on the saturated \HI\ \Lya\ line, or
estimated from $N$(\CaII) using \NCaII/\NHI$=10^{-8}$.  
The \NCaII/\NHI\ conversion factor is based on CLIC data 
towards nearby stars such as $\alpha$ Aql, $\eta$ UMa, $\alpha$ CMa, 
and is uncertain because Ca depletion varies strongly.  Most Ca is 
\CaIII\ for warm gas, if $T>4,000$ K and \nel$<$0.13 \cc.
Radiative transfer models of the LIC predict that \CaIII/\CaII$\sim$40--50, so small temperature 
uncertainties may produce large variations in \CaII/\HI.  

With the possible exceptions of $\alpha$ Oph, interstellar column
densities for stars within 30 pc are less than $\log\NHI (\cmtwo) \sim
18.5$ (\cite{FrischYorkFowler:1987,Woodetal:2000}).  Neutral CLIC gas
does not fill the sightline to any nearby star if the ISM density is
\nHI$\sim$0.2 \cc.  The CLIC extends farther in the galactic center
hemisphere than the anti-center hemisphere, as traced by \DI\ and \CaII\
data.  The most puzzling sightline is towards $\alpha$ Oph (14 pc),
where \CaII\ is anomalously strong and the 21-cm \HI\ emission feature
at the same velocity suggests $\log \NHtot (\cmtwo) \sim 19.5$ dex  (see
\ref{sec:next}).  High column densities ($\log \NHI (\cmtwo) \sim
18.75$) are also seen towards HD 149499B, 37 pc away in the LSR upwind
CLIC direction, and towards LQ Hya, where strong ISM attenuation of a
stellar \HI\ \Lya\ emission feature causes the poor definition of the
line core and wings. 

The percentage of a sightline filled with ISM offers insight into
the ISM character, and is given by the filling factor, \ff.
Restricting the discussion of the ISM filling factor to the
nearest 10 pc, we find that $\sim$67\% of space may be devoid of \HI.
If \nHI$\sim$0.2 \cc, then \ff$\sim$0.33 for ISM within 10.5 pc.
For galactic center and anti-center  hemispheres, respectively,
\ff$\sim$0.40 and \ff$\sim$0.26.  Mean cloud lengths are similar
for both hemispheres.  The highest values are \ff$\sim$0.57, towards
$\alpha$ Aql (5 pc) and 61 CygA (3.5 pc), and the lowest values are
\ff$\sim$0.11 towards $\chi^1$ Ori (8.7 pc), which is $\sim$14\deeg\ from
the downwind direction.  The sightline towards Sirius (2.7 pc,
43\deeg\ from the downwind direction) has \ff$\sim$0.26.  
These filling factors indicate that the neutral ISM does not fill
the sightline towards any of the nearest stars,
including towards $\alpha$ Cen where \ff$\sim$0.5, unless instead the true
value for 
$\langle $\nHI$\rangle$ is much smaller than 0.2 \cc.

\subsection[Cloud Pressure and Magnetic Field]{Cloud Temperature, Turbulence, and Implications for Magnetic Pressure}  \label{sec:pressure}

The basic thermal properties of the CLIC are presented in the Redfield
and Linsky (2004) survey of $\sim50$ cloudlets towards 29 stars with
distances $1-95$ pc.  Cloudlet temperatures (\T) and turbulence ($\xi$)
values are found to be in the range $ T = 1,000-13,000$ K and
$\xi$=0--5.5 \kms\ for clouds within $\sim$100 pc of the Sun.  The
mean temperature is 6680$\pm$1490 K, and the mean turbulent velocity
is 2.24$\pm$1.03 \kms.  From these values, RL estimate the mean
thermal (\Pth/$k=n T$) and mean turbulent (\Ptu/k=0.5$\rho \xi^2$/k)
pressures of 2,280$\pm$520 K \cc\ and 89$\pm$82 K \cc, respectively,
by assuming \nHI$ = 0.1$ \cc\ and \nel=0.11 \cc.  
The thermal pressure calculation includes
contributions by \HI, \HeI, electrons, protons, and assumes that He is
entirely neutral.  The pressure will be underestimated by $\sim$15\%
if He is $\sim$50\% ionized as indicated by radiative transfer models
(\S \ref{sec:rt}).  For comparison, if these clouds have 
ionization similar to the LIC, then \Pth$\sim$2300 \cc\ K.

These cloud temperatures are determined from the mass dependence of
line broadening using the Doppler parameter, \bdop, so spectral data
on atoms or ions with a large spread in atomic masses are needed.  In
practice, observations of the \DI\ \Lya\ line are required for an
effective temperature determination that distinguishes between thermal
and nonthermal broadening.

\begin{figure}[ht] 
\begin{center}
\includegraphics [height=2.3in,width=2.3in,angle=0]{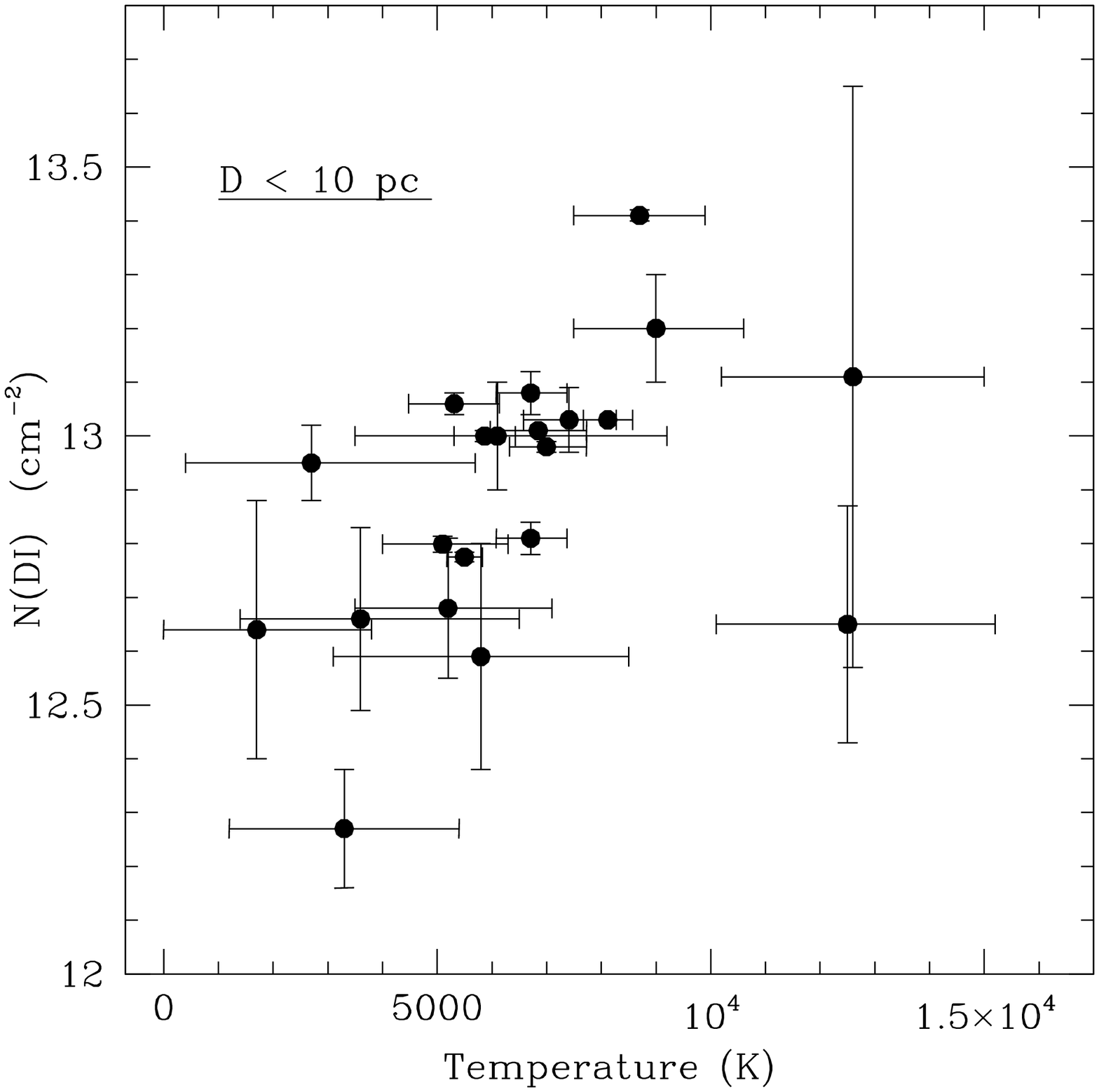}
\includegraphics [width=2.3in,height=2.3in,angle=0]{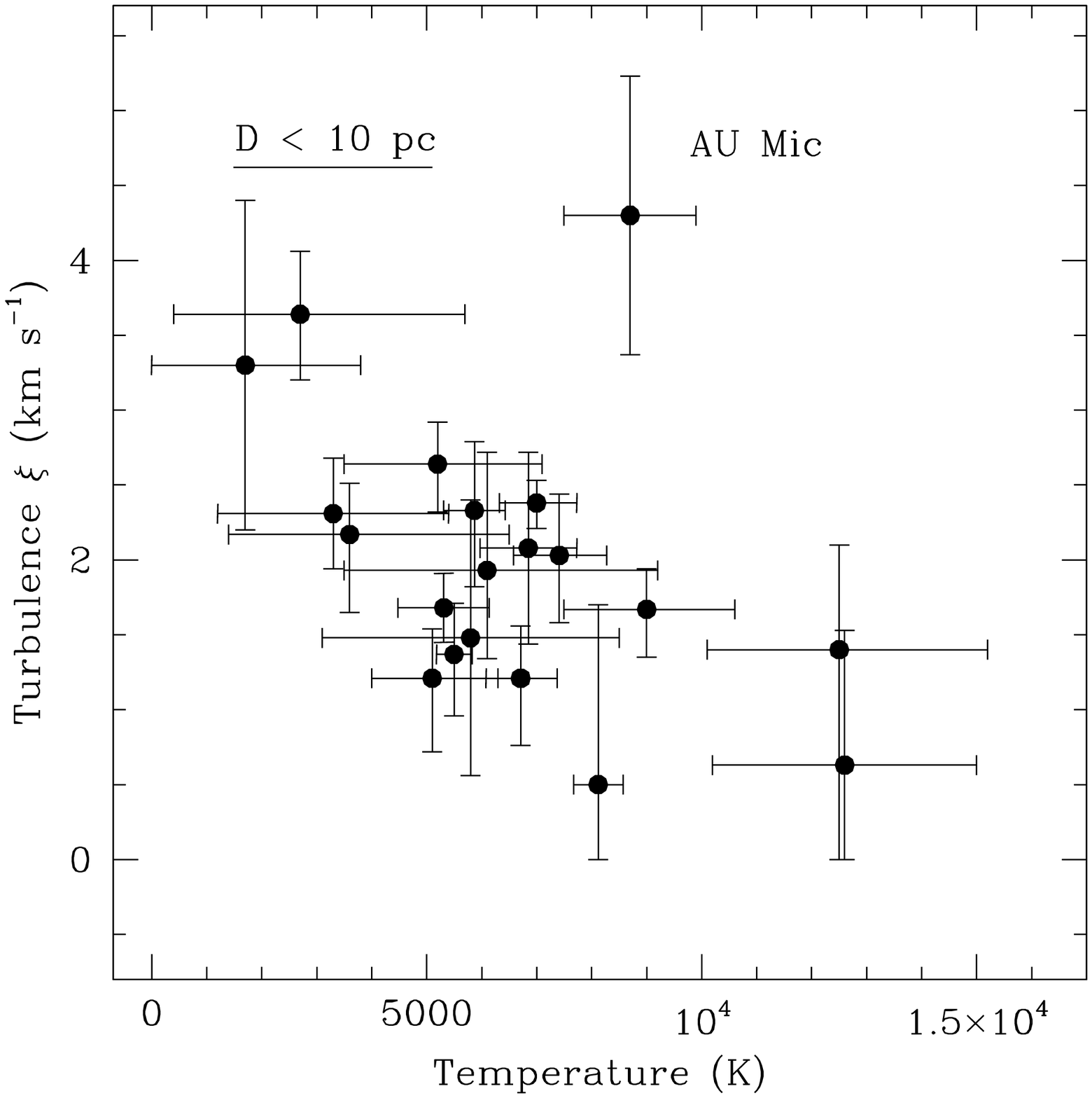}
\end{center}
\caption[Cloud Temperature and Turbulence.]{\label{fig:rl} Temperature
versus turbulence $\xi$ (right) and temperature versus column density
\NDI\ (left ) for interstellar absorption components seen towards
stars within 10 pc of the Sun (based on data in RL).  }
\end{figure}

When the star sample is restricted to objects within 10.5 pc,
the cloud temperature is found to be anticorrelated with turbulence,
and to be correlated with \NDI\ (Fig. \ref{fig:rl}).  From a larger sample
of components, RL have
concluded that the \T--$\xi$ anti-correlation is significant.  However
the likelihood that unresolved velocity structure is present in these
UV data allows for the \T--$\xi$ anti-correlation to contain some
contribution from systematic errors.  High resolution optical data
show that velocity crowding for interstellar Maxwellian components
persists down to component separations below 1 \kms\ (\S
\ref{sec:absline}), so that the weak positive correlation between \T\
and \NDI, and negative correlation between \T\ and $\xi$ may result
from unresolved component structure.

There are no direct measures of the magnetic field strength in the LIC,
but the field strength is presumed to be non-zero based on observations of polarized
starlight for nearby stars, which may originate from magnetically
aligned grains trapped in interstellar magnetic field lines draped over the
heliosphere (\nolinebreak \cite{Frisch:2005L}).  
The thermal properties of the CLIC have implications for pressure 
equilibrium and magnetic field strength.  The
magnetic field strength and density fluctuations can be constrained
using equipartition of energy arguments.  If clouds in the CLIC are in
thermal pressure equilibrium with each other, $P_{\rm Th}/k = n T$,
where $n$ is the total number of neutral and charged particles in the
gas, 
and if magnetic field $B=0$, then the temperature range of $T
\sim 10 ^3 - 10 ^4 $ K found by RL indicates that densities must vary
by an order of magnitude.  Since particle number densities vary by a
factor of $\sim$2 as the cloud becomes completely ionized, most of the
temperature variation must be balanced either by variations in the
mass-density or in the magnetic field strength if the cloud is in
equilibrium.  If the CLIC has a uniform total density, and if thermal
pressure variations are balanced by magnetic pressure, \Pb/$k=B^2 /8 k
\pi$, then magnetic field strengths must vary by factors of $\sim$3 in
the CLIC.  
If the star set is restricted to objects within $\sim$10 pc, a temperature and 
turbulence range of \T=1,700--12,600 K and $\xi=0-5.5$ \kms\ are 
found, with mean values of $6,740 \pm 2800$ K and $\xi =1.9 \pm 1.0$ \kms.  

A rough estimate is obtained for the 
magnetic field strength in the LIC by assuming equipartition between
thermal and magnetic energies, and using the results
of the RT models that predict neutral and ion 
densities (see \S\ref{sec:rt}).  The first
generation of models gives a LIC thermal energy density of \Eth/$k \sim 3600$
\cc\ K for $T=6340$ K, and including \HI, \HII, \el, \HeI, and \HeII.
Lower total densities and ionization in the second generation of 
models reduce the thermal energy density somewhat.
Equipartition between thermal and magnetic energy density gives \Eth=\Eb\ 
and \Eb/$k  = B^2 / 8 k \pi $.  These assumptions then give
\B $\sim 3.1-3.8$ $\mu$G for the LIC.  

The interstellar magnetic field strength in the more extended CLIC can be guessed using
the RL value for the mean thermal
pressure of 2280 \cc\ K, and assuming that the mean magnetic and thermal pressures 
are equal.  For this case \B$\sim$2.8 $\mu$G.  If
these clouds are, instead, in pressure equilibrium with the Local Bubble
plasma, then ($P_{\rm Th} +$\Pb/$)k = P_{\rm LB}/k \sim 5 \times 10 ^3$ K \cc\
(\S \ref{sec:lbradiation}), and magnetic field strengths are $\sim$3.1 $\mu$G.
In contrast, for \Pb$\sim$\Ptu\ then \B$\sim$0.6 $\mu$G.  Based on
equipartition of energy arguments, typical field strengths of $B \sim$3 $\mu$G
seem appropriate for the CLIC, with possible variations of a factor of 3.

The LIC turbulence appears to be subsonic.  Treating the LIC as
a perfect gas, the isothermal sound speed is \Vsound$\sim 0.09
\sqrt{T} \sim$ 7.1 \kms, and turbulent velocities are 0.5--2.7
\kms\ (\cite{Hebrardetal:1999,GryJenkins:2001}, RL).  The Alfven velocity
is given by \Valf$\sim 2.2 B_\mu /\sqrt{n{\rm (p)}}$, where $B_\mu$ is
the interstellar magnetic field in $10^{-6}$ G ($\mu$G), \Valf\ is in \kms, and the
proton density \np\ is in \cc.  For gas at the LIC temperature (6,300
K), the Alfven velocity exceeds the sound speed for 
$\B_\mu >1.3 ~\mu$G.  The velocity of the Sun
with respect to the LIC (26.3 \kms) is both supersonic and
super-Alfvenic for interstellar field strengths $B < 3.7$ $\mu$G.


\section[Radiative Transfer Models of Partially Ionized Gas]{Radiative Transfer Models of Local Partially Ionized Gas} \label{sec:rt}

Radiative transfer (RT) effects dominate the ionization level
of the tenuous ISM at the Sun.
The solar environment is dominated by low opacity ISM, 
\NHI$ \ltsim 10^{18.5} $ \cmtwo.  In contrast to dense clouds where only
photons with $\lambda >$912 \AA\ penetrate to the cloud interior, the
low column density ISM near the Sun is partially opaque to H-ionizing
photons and nearly transparent to He-ionizing photons.  At 912 \AA\
the cloud optical depth $\tau \sim$1 for \logNHI$ \sim 17.2$ \cmtwo,
and at the \HeI\ ionization edge wavelength of 504 \AA, $\tau \sim$1
for \logNHI $\sim 17.7 $ \cmtwo.  The average \HI\ column and mean
space densities for stars within 10 pc of the Sun are
$\langle$\NHI$\rangle \sim 10^{18}$ \cmtwo\ and \nHIavg$\sim 0.07$
\cc, so the heliosphere boundary conditions and the ratio \nHI/\nHII\
vary from radiative transfer effects alone as the Sun traverses the
CLIC (Fig. \ref{fig:nHInHeI}).  Warm, $T>5000$ K, partially ionized gas
is widespread near the Sun and is denoted WPIM (\S \ref{sec:clicwim}).
Charged interstellar particles couple to the interstellar magnetic
field and are diverted around the heliopause, while coupling between
interstellar neutrals and the solar wind becomes significant inside of
the heliosphere itself.  The density of charged particles
in the ISM surrounding the Sun supplies an important constraint
on the heliosphere, and this density varies with the radiation
field at the solar location, which is now described.

\subsection[Local Radiation Field]{The Local Interstellar Radiation Field}\label{sec:rtradiation}

\begin{figure}[hb] 
\includegraphics [width=2.4in]{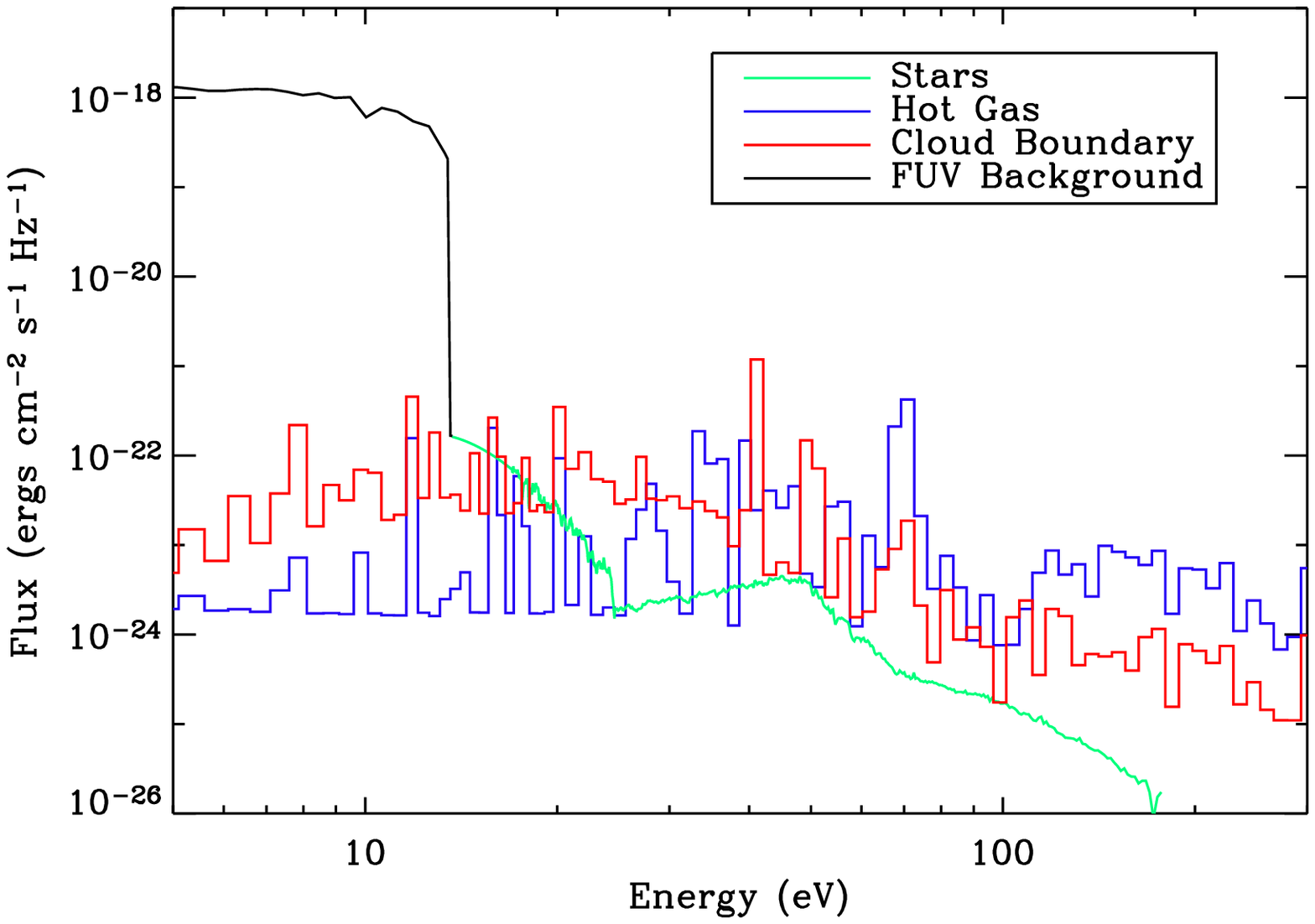}
\includegraphics [width=2.4in]{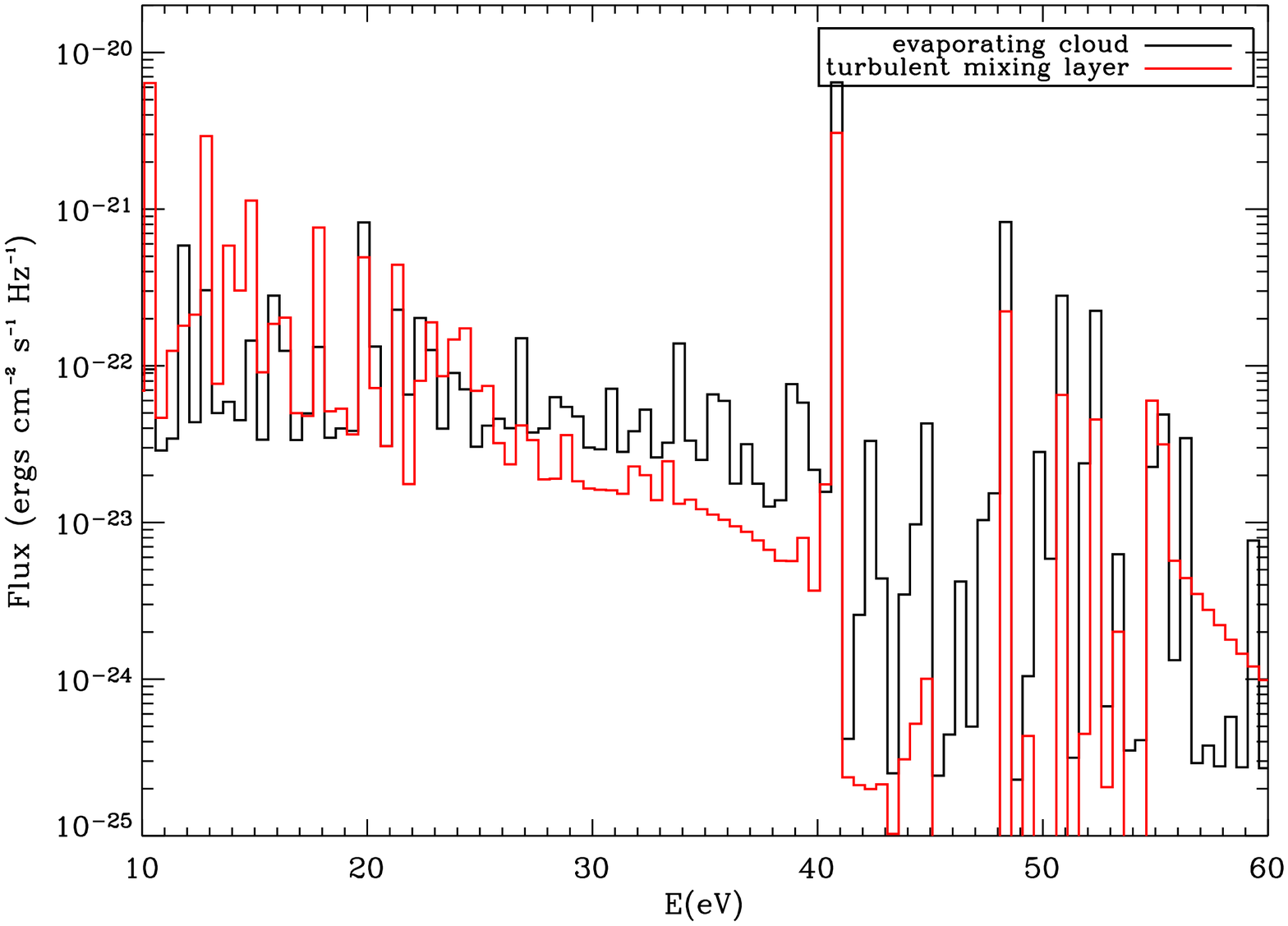}
\caption[The Local Interstellar Radiation Field.]{
\label{fig:interface}  
{Left:}
Components of the interstellar radiation incident on the local interstellar
cloud complex.  Contributions from stars show the EUV flux from nearby white
dwarf and B stars, after deabsorbing the corresponding \NHI\ to the cloud
surface.  The cloud boundary flux produced by emission from an evaporative
interface between the local gas and the hot gas of the Local Bubble is shown
(\cite{SlavinFrisch:2002}).  { Right:} Radiation field from a turbulent mixing
layer interface compared to that of a conductive interface.  
Instabilities at the boundaries of tenuous clouds may lead to quite
different radiation fields inside of the clouds, particularly near 584 \AA\
where \HeI\ is ionized.
}\end{figure}

The interstellar radiation field is a key ingredient of cloud
equilibrium and ionization at the solar position.  This radiation field
has four primary components: A. The FUV background, mainly from distant
B stars, B.  Stellar EUV emission from sources including nearby white
dwarfs and B stars ($\epsilon$ CMa and $\beta$ CMa); C. Diffuse EUV and
soft X-ray emission from the Local Bubble hot plasma, as we discussed in
\S \ref{sec:lbradiation}; and D. Additional diffuse EUV emission thought
to originate in an interface between the warm LIC/CLIC gas and the Local
Bubble hot plasma (\S \ref{sec:lbradiation}).  This last component is
required because, although radiative transfer models show that the
stellar EUV and Local Bubble emission account for most LIC ionization,
it is not sufficient to account for the high He ionization inferred
throughout the cloud from the \emph{EUVE} white dwarf data
(\cite{ChengBruhweiler:1990,Vallerga:1996,Slavin:1989}).  The spectra of
these radiation sources are shown in Fig. \ref{fig:interface}.

The interstellar radiation flux at the cloud surface must be inferred
from data acquired at the solar location, together with models of
radiative transfer effects.  LISM column densities are so small that
dust attenuation is minimal, e.g. for the LIC $A_{\rm V} < 10^{-4}$
mag, and fluxes longwards of $\lambda \sim 912$ \AA\ are similar at
the solar location and cloud surface (with the exception of Ly $\alpha$
absorption at 1215.7\AA).  For wavelengths $\lambda <912$
\AA\ the situation is different, however, and the spectrum hardens as
it traverses the cloud because of the high \HI-ionizing efficiency of
$800-912$ \AA\ photons.  Thus a self-consistent analysis is required
to unravel cloud opacity effects, and extrapolate the EUV radiation
field observed at the Sun to the cloud surface.  The observational
constraints on the $200-912$ \AA\ radiation field are weak, partly due to
uncertainties in \NHI\ towards $\epsilon$ CMa and partly due to the
difficulty in observing diffuse EUV emission, which
allows some flexibility in introducing physical models of the cloud
interface.

A thin layer of intermediate temperature ISM is expected to exist in
the boundary between the CLIC and the LB hot plasma.  This layer will
emit radiation with a spectrum and flux dependent on the underlying
physical mechanisms.  Models for this interface emission indicate it
radiates strongly in the energy band $E=20-35$ eV, which is important
for both \HeI\ and \NeI\ ionization.  The exact physical processes at
work in interface regions are unclear, but possibilities include
thermal conduction, radiative cooling, and shear flow, which lead to
evaporative interface boundaries (\cite{Cowie+McKee_1977}), cooling
flows (\cite{Shapiro+Benjamin_1991}) or turbulent mixing layers
(\cite{Slavin_etal_1993}).  All of these boundary types produce
intermediate temperature gas ($T\sim 10^{4.5} - 10^{5.5}$ K) that
radiates in the EUV, although ionization levels, and thus the spectrum
and intensity of the emission, depend on the detailed physics.
Parameters that constrain interface properties include the strength
and topology of the interstellar magnetic field, \Bis, the hot gas
temperature, and the relative dynamics of the hot and warm gas.  The
magnetic field affects the RT models because \Bis\ reduces the
evaporative flow by inhibiting thermal conduction in directions
perpendicular to field lines and at the same time supports the cloud by
magnetic pressure.  Since the total (magnetic $+$ thermal) pressure is
roughly constant in an evaporative outflow, the magnitude of the
magnetic field is an important factor in determining the pressure in any
evaporative flow that might be present.  Fig.  \ref{fig:interface}
shows examples of the EUV spectrum produced by an evaporative
interface and a turbulent mixing layer.


\subsection[Radiative Transfer Models of the Local Cloud]{Radiative Transfer
Models of the Local Cloud and other Tenuous ISM \label{sec:rtmodels}}

Our radiative transfer models are constrained by observations
of local ISM towards nearby stars, and \emph{in situ} data 
from LIC neutrals that have penetrated the heliosphere.   The \emph{in situ}
data includes direct detection of \HeI, observations of solar \Lya\
florescence from \HI\ in the heliosphere, and observations of
the pickup ion and
anomalous cosmic ray populations that are seeded by interstellar neutrals
(see Moebius et al., Chapter 8).
Generally elements with FIP$<$13.6
eV are fully ionized in tenuous clouds, while elements such as H, O,
N, Ar, He, and Ne with 13.6$\ltsim$FIP$\ltsim$25 eV are partially
ionized.  Neutrals from these partially ionized species enter the
heliosphere, where they seed the pickup ion and anomalous cosmic ray
populations measured by instruments on various spacecraft.  Column
densities towards nearby stars constrain sight-line integrated values,
and permit the recovery of \nHII/\nHI\ as a function of distance to
the cloud surface (Fig. \ref{fig:rt}).  While the LIC temperature is
determined directly from Ulysses observations of \HeI, the densities
\nHI\ and \nHII\ at the solar location vary as the Sun moves through
the LIC, and must be determined from radiative models.

The detailed attention paid here to LIC radiative transfer models is
motivated by the facts that \nHI\ and \nHeI\ at the solar location are
important boundary conditions of the heliosphere, and that ionization
gradients in the CLIC are factors in reconstructing the 3D cloud
morphology from data.  An extensive study of the LIC ionization has
resulted in a series of $\sim$50 radiative transfer models appropriate
for tenuous ISM such as the LIC and CLIC
(\cite{SlavinFrisch:2002,FrischSlavin:2003,FrischSlavin:2005cospar}).
These models use the CLOUDY code (\cite{Ferland:1998}), an
interstellar radiation field based on known sources, and models for
the interface expected between the LIC and Local Bubble plasma (also
see \S \ref{sec:lbradiation}).  Boundary conditions for outer scales
are provided by ISM data integrated over the LIC, which can be subject to
ionization gradients.  On inner scales the boundary conditions are
given by \emph{in situ} observations of ISM at the heliosphere.
Both sets of constraints are important for evaluating the
RT properties of the surrounding ISM where large \HI/\HII\ gradients
are found.  The data sets used here include \nHeI, pickup ions, and
anomalous cosmic rays inside of the heliosphere, and absorption line
data of the LIC towards the downwind star $\epsilon$ CMa (GJ, FS).

These RT models will be updated in the future, as the density,
composition, LIC magnetic field, radiation field, and interface
regions between the LIC and Local Bubble plasma become better
understood.  The LIC magnetic field enters indirectly through the
contribution of magnetic pressure to the cloud interface. The magnetic
pressure in the cloud helps to determine the thermal pressure and thus
cooling in the interface, so that further studies of \Bis\ near the
Sun will provide insight into the characteristics of the interface
radiation.  One of the significant results of our study is that the
spread in predicted neutral and ion densities demonstrates that low
column density ISM can be in equilibrium at a range of ionization
levels, and that the ionization itself is highly sensitive to the
radiation field, interface characteristics, and other cloud
properties.

\begin{figure}[ht] 
\centerline{\includegraphics*[width=4.2in]{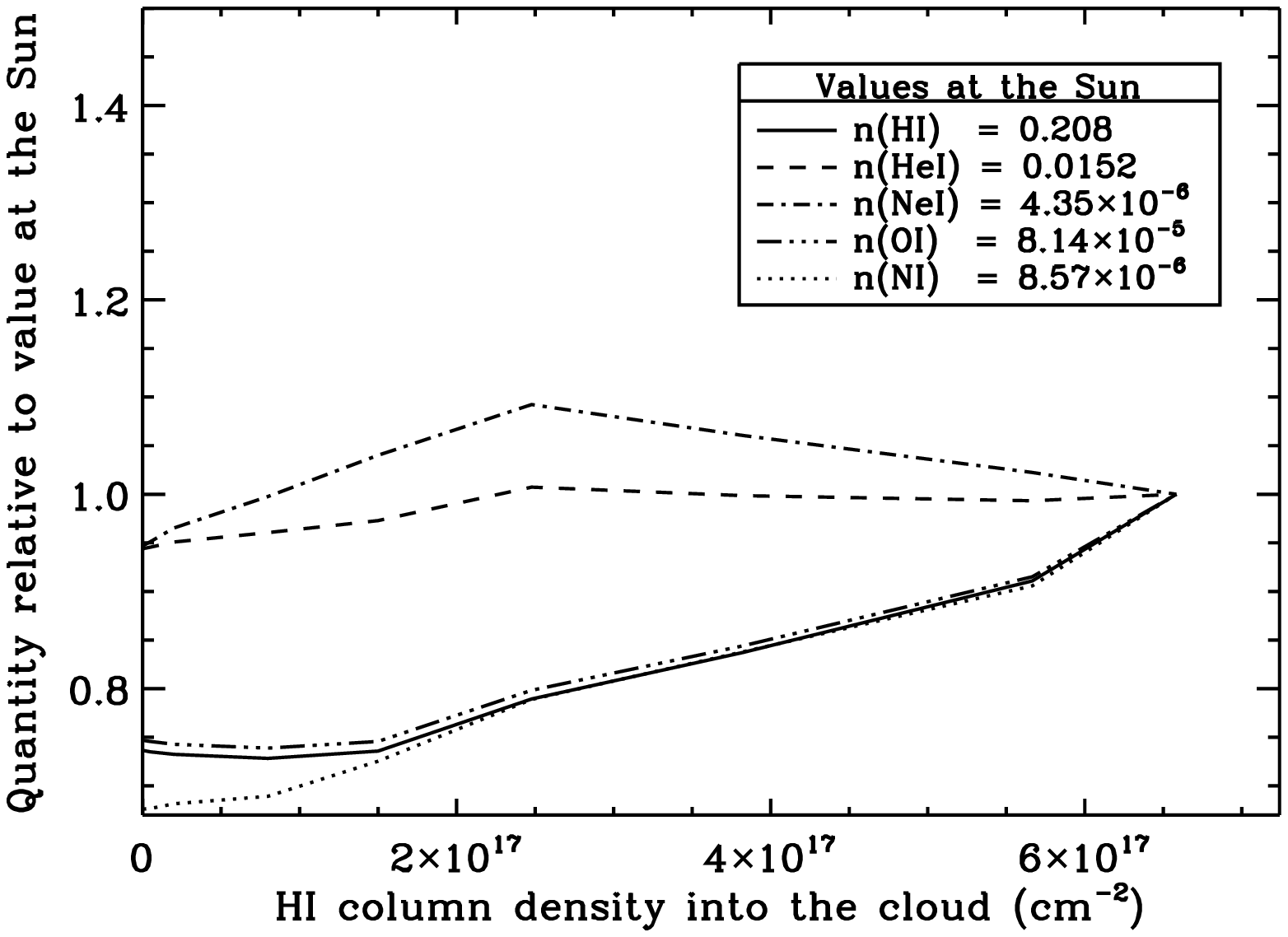}}
\caption[Radiative Transfer Effects in an Interstellar Cloud.]{\label{fig:nHInHeI} Variation of
neutral densities between the Sun and cloud surface for Model 2 from
Slavin and Frisch (2002).\nocite{SlavinFrisch:2002} Shown are
variations in neutral column densities between the Sun (\NHI
=6.5$\times$ 10$^{17}$ \cmtwo) and cloud surface (\NHI=0) for \HI,
\HeI, \NeI, \OI, and \NI.  At the heliopause, \nHeI$\sim$0.015 \cc,
\nHI$\sim$0.22 \cc, \nel$\sim$0.1 \cc.  }
\end{figure}


\subsection[Model Results]{Results of Radiative Transfer Models}
\label{sec:results}

Two generations of radiative transfer (RT) models have been developed,
with the focus on matching data for the LIC inside of the heliosphere,
and matching data on nearby ISM in the downwind direction where the 
brightest point sources of EUV radiation ($\epsilon$ CMa and $\beta$ CMa) 
and low column densities are found.  
Both sets of models are constrained by LIC data obtained inside
of the heliosphere, such as for pickup ions, \HeI, and anomalous
cosmic rays.  The PUI and ACR populations are seeded by neutral ISM flowing into
the heliosphere, and are subject to filtration losses in the
heliosheath regions (Moebius et al., Zank et al., this volume).

\begin{figure}[t] 
\centering{\includegraphics [width=3.2in]{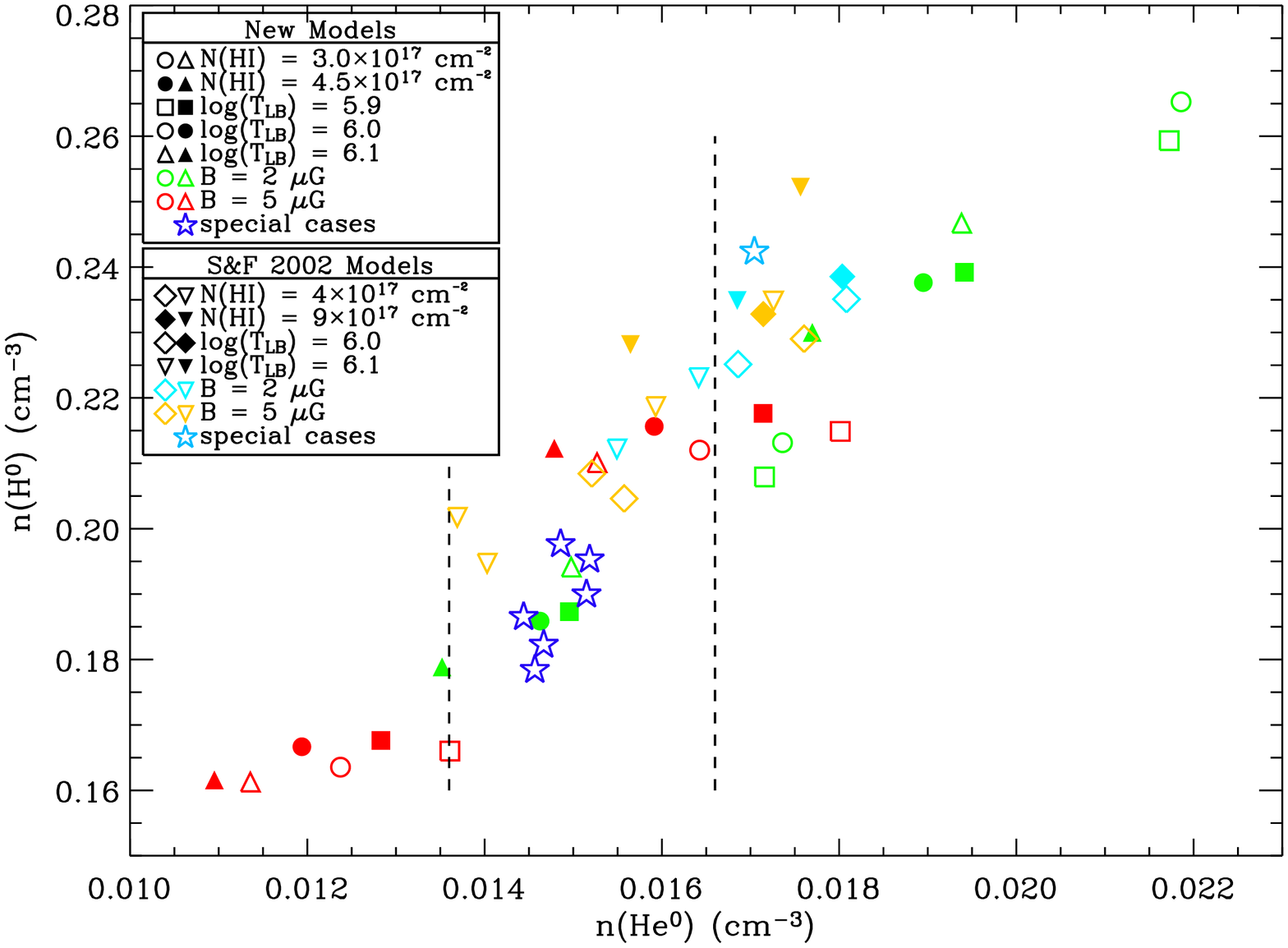} 
\includegraphics [width=3.2in]{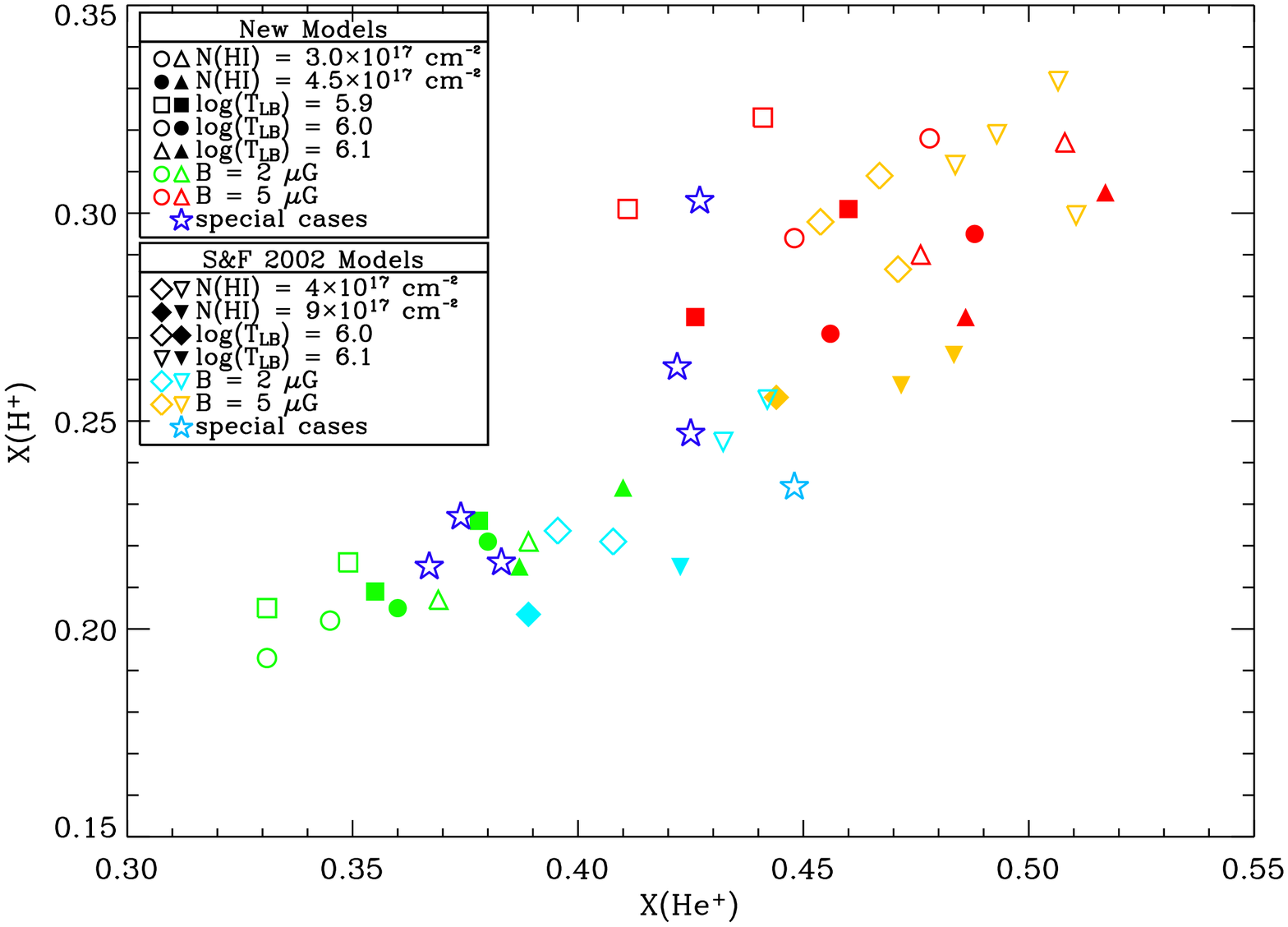} }
\caption[Results of Ionization Models for Tenuous ISM.]{\label{fig:rt} {Top:} 
H neutral density vs.\ He neutral density, both at the Solar location for a large
set of model calculations.  The symbol colors indicate the assumed
magnetic field strength in the cloud, the fill indicates the assumed HI
column density, and the shape indicates the assumed temperature of the
hot gas of the Local Bubble.  Stars represent special parameter sets 
which do not fall on the grid of model parameters, but rather are chosen to
better match the data.  The density and uncertainties of \HeI\ are
plotted as vertical lines \cite{Moebiusetal:2004}.  Note that a range of
n(HI) values are consistent with the n(\HeI) measurements.  The
radiative transfer models that provide the best agreement with all
available data on the ISM also 
yield a consistent estimate of \nHI$=0.19-0.21$ \cc. {Bottom:} Same as left plot
but for the \HII\ and \HeII\ ion fractions.  Note: Some of the
differences not explained for by $N($\HI$)$, $T_\mathrm{LB}$, or $B$ are
due to differences in the assumed total H densities.} 
\end{figure}

The ionization state of the pristine ISM outside of the heliosphere is
obtained from the RT models, using slightly different constraints for
the first and second generations of models.  The first generation of 25
models are constrained by the local ISM towards $\epsilon$ CMa, i.e. the
sum of the LIC and blue shifted (BC) clouds, and by ISM byproducts
inside the heliosphere.  The ISM within $\sim$1--2 pc towards $\epsilon$
CMa is divided between two clouds, with projected HC velocities of 10
\kms\ (the BC) and 18 \kms\ (the LIC), and with a total \NHI$<10^{18}$
\cmtwo.  The LIC and BC are also observed towards Sirius ($\alpha$ CMa,
2.7 pc).  Two models, Model 2 and Model 8, provide a good match to all
data except for the cloud temperature inside of the heliosphere.  The
predicted temperature at the heliosphere is $\sim$3,000 K higher than
found from \emph{in situ} \HeI\ data, $T$= 6,400$\pm$340 K.  A possible
explanation for this difference is that the abundances of coolants such
as \CII\ are incorrect in the models, particularly since $N$(\CII)
generally has large measurement uncertainties.  A second mismatch occurs
because predicted $N$(\SiIII) values are lower than observed values,
which suggests that interface models require additional processes such
as a turbulent mixing layer.  Models 2 and 8 predict that \nHI$\sim$0.20
\cc, \nel$\sim$0.1 \cc, and ionizations of $\chi$(H)$\sim$0.30, and
$\chi$(He)$\sim$0.49 are appropriate for the solar location
(\cite{SlavinFrisch:2002,FrischSlavin:2003}).  The mean cloud density to
the downwind surface is \nHIavgLIC$\sim$0.17 \cc.

The second set of models is constrained by data on the LIC towards
$\epsilon$ CMa (excluding the BC) and \emph{in situ} ISM.  The best of these
models are in good agreement with the \emph{in situ} data, including
cloud temperature data.  However, these models require a high C
abundance, possibly in conflict with solar abundances, and fail to
predict \RMg\ in the LIC.  The second set of models require that the
LIC and BC have different ionization levels, which is similar to the
findings of Gry and Jenkins (2001) but remains unexplained.  LIC
properties predicted by these models, which are still under study, are
\nHI=0.19 \cc, $\chi$(H)=0.22, \nel=0.06 \cc, \nHeI=0.015 \cc, and
$\chi$(He)=0.37
 
The results of the RT models that are significant for heliosphere
studies are summarized in Figures \ref{fig:nHInHeI} and \ref{fig:rt}.  
The ionizations of H, He, O, N, and Ne  throughout the LIC are shown 
in Fig. \ref{fig:nHInHeI}, with variations of $\sim$25\% between the cloud 
surface and the Sun.  Fig. \ref{fig:rt} shows the extent of ionization 
states possible under equilibrium conditions in tenuous ISM.  Minor 
variations in the physical assumptions input to the RT models result 
in a continuum of ionization levels that are in equilibrium for low 
density clouds like the LIC.  Depending on the RT model constraints, 
H ionization levels are 0.19--0.23, while He ionization levels are 
0.32--0.53.  \emph{These models show that ionization levels might vary 
between clouds in the CLIC, so that variations in the heliosphere
boundary conditions are expected as the CLIC flows past the Sun (see \S
\ref{sec:clic}).}

\section[Passages through Nearby Clouds]{Passages through Nearby Clouds}\label{sec:lic}

The transition of the Sun from the near void of the Local Bubble, and
into the stream of tenuous ISM flowing away from the Scorpius-Ophiuchus
Association, can be probed with the kinematics and column densities of
nearby clouds, combined with models of the volume density \nHI.
Specifically, cloud dimensions are assumed to be $\propto$\NHI/\nHIavg.
Radiative transfer models of tenuous ISM show that cloud ionization,
\chiH, varies with column density.  Since the densities of individual
cloudlets are not currently available, \nHI\ determined by LIC radiative
transfer models is extrapolated to other nearby clouds (\S \ref{sec:rt}).
With this simple approach, we see that the Sun appears to have entered
the CLIC within the past $\sim$140,000 years. The Sun is located close
to the downwind edge of the CLIC (\cite{Frisch:1995}).  CLIC physical
properties are discussed in \S \ref{sec:clic}.

The most predictable passage of the Sun into an interstellar cloud is
the epoch at which the Sun entered the LIC, the cloud now surrounding
the Sun.  The LIC is the only cloud with an accurate 3D velocity vector,
which is found from Ulysses measurements of interstellar \HeI\ inside of
the solar system.  The LIC HC velocity is 26.3$\pm$0.4 \kms, and the
downwind direction is $\lambda=74.7\pm0.5$\deeg,
$\beta$=--5.2$\pm$0.2\deeg\ (ecliptic coordinates, \cite{Witte:2004}).
Several estimates are given for the epoch of the Sun's entry into the
LIC.  Both the data and radiative transfer models, upon which the
estimate relies, need further refinement before answers are conclusive.
We test several models of the poorly known LIC shape.  The observed
cloud velocity and the projected LIC velocity towards the $\chi ^1$ Ori,
8.7 pc away and $\sim$15\deeg\ from the downwind direction, differ by
$\sim$2.4 \kms, suggesting non-LIC gas is contributing (Table
\ref{tab:lic}).

The distance to the cloud edge is given by \NHI/\nHIavg, where \nHIavg\
is the average space density of \HI\ in the cloud.  The density of the
LIC at the solar location is \nHI$\sim$0.2 \cc, and a mean density
\nHIavgLIC$\sim0.17$ \cc\ is found from radiative transfer
models of the LIC in the downwind direction (\S \ref{sec:rt}).  When the
cloud structure is not fully resolved, or several velocity components
are present, there may be gaps in the cloud that are not incorporated
into the adopted value for \nHIavg.  We present several estimates below
for the entry of the Sun into the LIC, with results that differ by
factors of three.  However, the answers are consistent with the Sun's
entry into the LIC sometime within the past $\sim$47,000 years for \nHIavg=0.17 \cc, and possibly
quite recently within the past thousand years.

The difference between the LIC and upwind gas velocities, of $\ltsim 2$
\kms, including towards the nearest star $\alpha$ Cen, suggest that the
properties of the cloud surrounding the Sun will change rather soon, in
$\sim$3,800 years.  

\subsection[First Encounter of Sun with Local ISM ]{First Encounter of Sun with the Local ISM } \label{sec:enterclic}

The projected LIC velocity and \NHI\ towards nearby stars in the
downwind direction (Table \ref{tab:lic}), indicates that the Sun
entered the CLIC sometime within the past 44,000--140,000/\nHIavgtwo\
years, where \nHIavgtwo\ is the average cloud density in units of 0.2
\cc\ (\S \ref{sec:rt}, Fig. \ref{fig:rt}).  The value \nHIavgtwo$\sim$1 
is reasonable for nearby tenuous ISM, and is consistent with radiative
transfer models of the LIC.  This transition of the Sun
out of the deep vacuum of the Local Bubble, and into the higher density 
CLIC ISM, would have been accompanied by the appearance of neutral 
interstellar gas, pickup ions, anomalous cosmic rays, and dust in the heliosphere. 
The geological record of cosmic ray radioisotopes should have sampled
this transition (see the discussions of cosmic rays and the spallation product
radioisotopes in Chapters 9, 10, and 12). 

The simplest estimate for the date of entry of the Sun into the CLIC
is to ignore non-radial motions for all clouds and look only at the
nearby stars in the anti-apex direction.  ISM towards the three stars
$\alpha$ CMaA, $\alpha$ CMaB, and $\epsilon$ CMa are useful for this
estimate, giving \NHI=5--7$\times 10^{17}$ \cmtwo\ (based on data and
models in \cite{Hebrardetal:1999,FrischSlavin:2003}).  The distance to
the cloud edge in the anti-apex direction is $\propto
$\nHIavgtwo$^{-1}$ pc.  The most distant of the two cloudlets seen in
these sightlines is a blue-shifted cloud, with a heliocentric radial
velocity of 9.2--13.7 \kms\ (Table \ref{tab:velocity}).  We then find
that the Sun entered the CLIC within the past
(59,000--120,000)/\nHIavgtwo\ years.

Using instead the ISM towards the stars $\chi ^1 $ Ori and $\alpha$ Aur, 
which are near the heliocentric downwind direction, and for a
distance to the cloud edge given by \NHI/\nHIavgtwo, the observed
radial velocities (Table \ref{tab:lic}) indicate that the Sun entered the CLIC
(44,000--140,000)/\nHIavgtwo\ years ago.

For these estimates, possible non-radial cloud motions and gaps between
clouds are ignored.  The cloud gas filling factor in the downwind
direction is \ff$\sim$0.26 in \S \ref{sec:clic}.  If the downwind CLIC
consists of a series of wispy clouds with similar velocities, then
\nHIavgtwo$\ll$0.2 \cc, and the first solar encounter with the CLIC could
have occurred up to $4 \times 10^5$ years ago based on the $\chi^1$ Ori
data.

\subsection[Entering the Local Cloud]{Entering the Local Cloud}\label{sec:enterlic}

In principle the entry date of the Sun into the surrounding cloud, LIC, can be determined
precisely since the full 3D velocity of the LIC is known from
observations of \HeI\ inside of the heliosphere (Table
\ref{tab:velocity}).  The LIC shape is uncertain because components at
the LIC velocity are not resolved in many sightlines. Although there
may be high density contrasts from low speed waves in the LIC, for
instance acoustic waves will have velocities $\sim$7 \kms, we take the
approach that local clouds are defined by the component structure and
look at several geometric LIC models to estimate the epoch the Sun entered the
LIC.  ISM data for downwind stars are summarized in Table
\ref{tab:lic}.

The simplest estimate (A) uses \NHI\ and the LIC velocity towards
$\chi ^1$ Ori, $\sim$15\deeg\ from the downwind LIC direction.
Estimate (B) is based on the closest observed star in the downwind direction,
$\alpha$ CMa, which is $\sim$45\deeg\ from the downwind direction, 
combined with the assumption that 
the normal to the downwind cloud surface is parallel to the LSR LIC velocity.  
This assumption results in an entry epoch that
varies with the assumed LSR (\S \ref{sec:lsr}), and ignores data for
more distant downwind stars.  Estimate (C) approximates the downwind
LIC surface as a flat plane, defined by the distance of the LIC edge
towards any three stars from Table \ref{tab:lic}, moving through space
at the LIC velocity.  The final estimate (D) relies on the Colorado
LIC model (RL0), which however does not incorporate the most recent
data for the LIC, but can be expected to substantially improve in the
future.

A.  The values for the LIC \NHI\ and the projected LIC velocity
towards $\chi ^1$ Ori (Table \ref{tab:lic}) indicate that the
Sun entered the LIC $\sim$40,000/\nHIavgtwo\ years ago, where \nHIavgtwo\ is the
mean \nHI\ to cloud surface in units of 0.2 \cc.  Radiative transfer
models indicate \nHIavgtwo$\sim$0.85 (\S \ref{sec:rt}).  The observed cloud
velocity towards $\chi ^1$ Ori differs from the projected LIC velocity
by 2.4$\pm$1.0 \kms, so that additional cloud gas is present in this
sightline that is not at the LIC velocity.  
In addition, the turbulent contribution to the
line broadening is $\xi=2.38^{+0.15}_{-0.17}$ \kms\ towards $\chi ^ 1$
Ori (\S \ref{sec:absline}, \cite{RLIII}), indicating unresolved
cloudlets may be present.

\begin{figure}[ht] 
\caption[A Model for the Closest ISM.]{
\label{fig:amsci} 
SEE 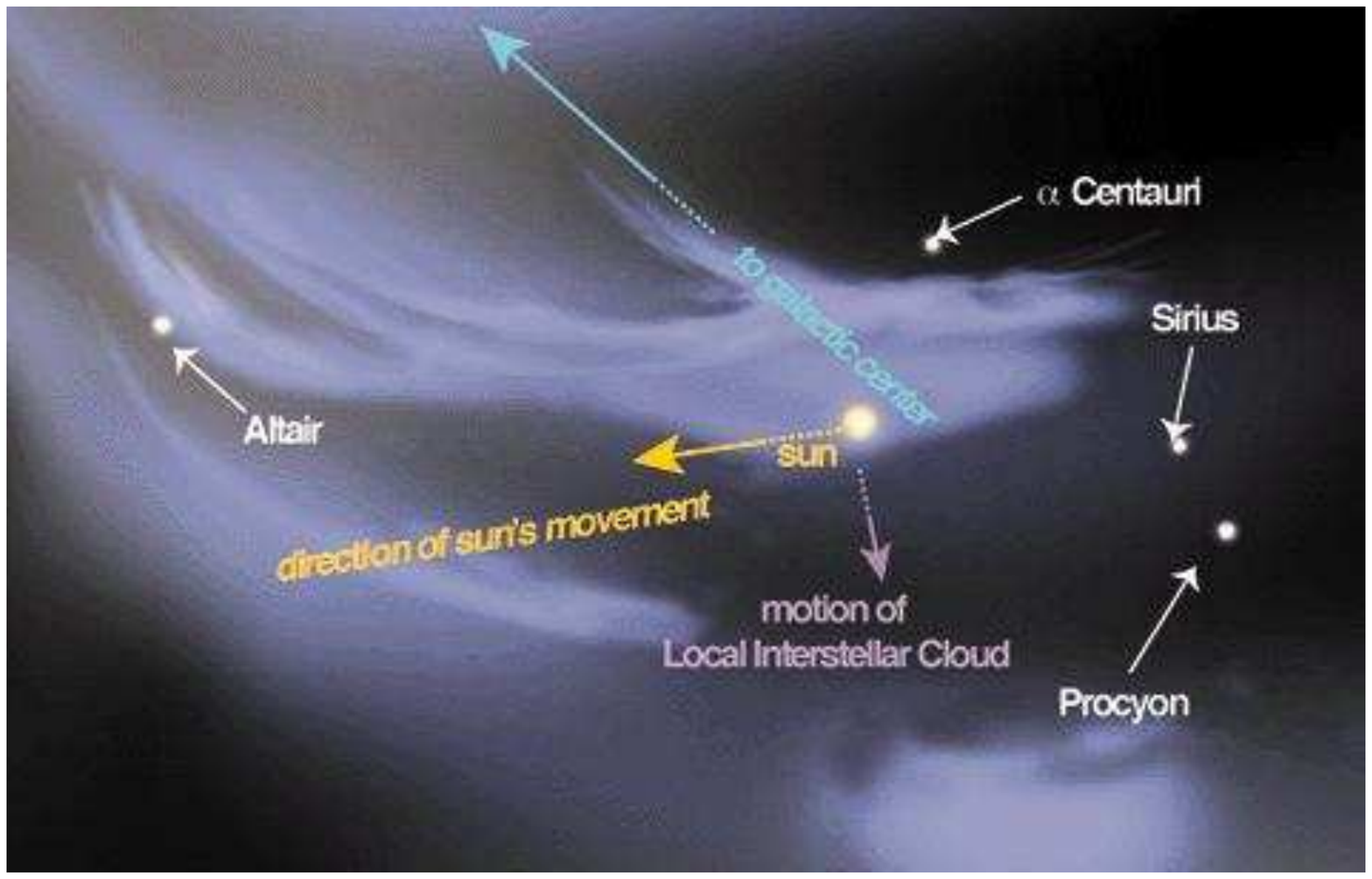.
The distribution of nearby ISM is shown based on the assumption
that the LSR motion of the LIC is parallel to the normal of the
cloud surface (model B, \S \ref{sec:enterlic}).  
The Sun and LIC motions are in the LSR.
Figure from Frisch (2000), courtesy of \emph{American Scientist}.
\nocite{Frisch:2000amsci}
 }
\end{figure}

B.  For this estimate we use LIC data towards the nearest star in the
downwind direction for which ISM data are available, $\alpha$ CMaAB
(Sirius).  The LIC downwind surface is assumed to be oriented such
that the surface normal is parallel to the LIC LSR velocity, based on
the model of Frisch (1994, Figure
\ref{fig:amsci}).\nocite{Frisch:1994} Note that the ``downwind''
nomenclature is referenced with respect to the Sun, and the ``downwind
surface'' is the leading edge of the LIC as it moves through the LSR.
Since the LIC surface is referenced to the LSR, the solar apex motion
is a variable (Table \ref{tab:velocity}, \S \ref{sec:lsr}).  The Sun
then encountered the LIC within the past $\sim$13,500/\nHIavgtwo\
years or $\sim 10^3$/\nHIavgtwo\ years, for the Hipparcos or Standard
solar apex motions, respectively.  The present distance to the LIC
downwind surface is $\sim 0.01-0.3$ pc for this model. The velocity of
most of the interstellar gas observed towards $\chi$ Ori and $\alpha$
Aur differs by a small amount, 0.7--1.4 \kms, from the projected LIC
velocity (Table \ref{tab:lic}), and therefore is located in a separate
cloud.  Because the LIC also extends less than $\sim 10 ^4$ AU in the
upwind direction, this model implies that the LIC is filamentary or
sheet-like, similar to global low density ISM (\S \ref{sec:global}).
This model results in an interstellar magnetic field direction, \Bis,
which is approximately parallel to the cloud surface (perpendicular to
the surface normal).

\begin{table}[h]
\caption[Stars that Sample the Downwind LIC.]{Stars Sampling the Downwind LIC}
\begin{tabular}{l c c c  c c }
Object & \glong, \glat & Dist. & V$_{\rm observed}$  & V$^{proj}_{\rm LIC}$ & log\NHI \\ 
       & (\deeg, \deeg) & (pc)   &  (\kms)  & (\kms) &(\cc)   \\
\hline
$\alpha$ Aur & 162.6, 4.6 & 12.9  & { \it 21.5$\pm$0.5}  & 23.0$\pm$0.3
&18.26  \\
LIC \HeI\  &  183.3, --15.9 &  0  &  {\it 26.3$\pm$0.3} & {\it 26.3$\pm$0.3} & --
\\   
$\chi ^1$ Ori & 188.5, --2.7 & 8.7 & { \it 23.1$\pm$0.7}  &  25.5$\pm$0.3 &
17.80   \\
$\alpha$ CMi &  213.7, 13.0 &  3.5  &  { \it 20.5$\pm$1.0}, 24.0 &
19.6$\pm$0.2  & 17.90   \\
$\alpha$ CMaA & 227.2, --8.9 & 2.7 & 13.7, { \it 19.5$\pm$0.3} & 19.1$\pm$0.2
& 17.25  \\  
$\alpha$ CMaB & 227.2, --8.9 & 2.7 & 11.7, { \it 17.6$\pm$1.5} & 19.1$\pm$0.2
& 17.63  \\  
$\epsilon$ CMa & 239.8, --11.3  & 132  & 9.2, { \it 16.2$\pm$1.5} &
15.1$\pm$0.2 & $\sim$17.54  \\
\hline
\end{tabular}
\newline
Notes:  
The LIC velocity component is in $italics$.  V$^{proj}_{\rm LIC}$ is the
LIC velocity projected towards the star.  All velocities are heliocentric.
The column densities and data are from
Witte (2004), Hebrard et al. (1999), Frisch \& Welty (2005), RL, GJ, SF02, and FS.
\nocite{Witte:2004}
\nocite{Hebrardetal:1999}
\nocite{FrischWelty:2005} 
\nocite{RLIII}
\nocite{GryJenkins:2001} \nocite{SlavinFrisch:2002,FrischSlavin:2003}
\label{tab:lic}
\end{table}

C.  The third estimate models the downwind cloud surface as a locally
flat surface moving with the LIC velocity, based on the model in
Mueller et al. (2005).\nocite{Muelleretal:2005} We then use any three
stars in Table \ref{tab:lic} to define this plane, and the time at
which the Sun crosses this plane marks the entry of the Sun into the
LIC.  For this model, the Sun entered the LIC between 33,000/\nHIavgtwo\ to
36,000/\nHIavgtwo\ years ago.

D.  The final estimate relies on the Colorado LIC model, based on
density \nHIavg=0.1 \cc\ and a LIC velocity determined from UV
absorption lines towards nearby downwind stars (RL).  This UV LIC
vector differs by $\sim$0.6 \kms\ in velocity, and $\sim$0.8\deeg\ in
direction, from the LIC velocity given by data on interstellar \HeI\
inside of the solar system (\cite{Witte:2004}).  The Colorado Model
Web calculator gives a distance to the downwind surface of 4.5 pc,
indicating the Sun entered the LIC $\sim$170,000 years ago for a
relative Sun-LIC velocity of 26.3 \kms, or alternatively
$\sim$85,500/\nHIavgtwo\ years ago.  The column density towards the downwind
direction is predicted to be \NHI=$10^{18.14}$ \cmtwo, however, which
differs substantially from recent values for $\chi ^1$ Ori (Table
\ref{tab:lic}).  We therefore disregard this estimate, but expect
further improvements in the Colorado model to significantly increase
our understanding of the LIC morphology.

\subsection[Future Cloud Encounters]{Future Cloud Encounters} \label{sec:next}

The best limits on the distance to the upwind edge of the LIC are found
from observations of 36 Oph, 6 pc away and $\sim 16$\deeg\ from the
heliosphere nose.  The ISM velocity towards 36 Oph is --28.5$\pm$0.6
\kms, which corresponds to the G-cloud velocity (Table
\ref{tab:clouds}).  The limit on \NHI\ at the projected LIC velocity of
25.7 \kms\ is \NHI$ < 6\times 10^{16}$ \cmtwo\ (\cite{Wood36Oph:2000}).
These numbers indicate that the Sun will exit the LIC sometime within
the next $\sim$3700/\nHIavgtwo\ years, and that the distance to the upwind LIC
surface is $<$0.10 pc.  Support for this
result is provided by the ISM velocity towards the nearest star $\alpha$
Cen, which is $\sim$50\deeg\ from the HC nose direction.  Towards
$\alpha$ Cen, the observed cloud velocity and the projected LIC velocity
differ by $\sim$1 \kms\ (\cite{Landsmanetal:1984,LinskyWood:1996}).
Column densities of \NHI=17.6--18.0 \cmtwo\ were originally found
towards $\alpha$ CenAB (\cite{LinskyWood:1996}), which indicate that
\nHIavg\ may be similar to the LIC density of $\sim$0.2 \cc.  Later
interpretations of the data favor the low end of this column density
range (\nolinebreak \cite{Woodetal:2005}), indicating that the sightline is only
partially filled with warm gas.


Possibilities for the next cloud to be encountered by the Sun include
the G and Apex clouds (Table \ref{tab:clouds},
\cite{Frisch:2003apex}).  Both clouds have LSR upwind directions,
(\glong, \glat)=(348$\pm$3\deeg, 5$\pm$5\deeg), that are approximately
perpendicular to the LSR solar apex motion (\S \ref{sec:lsr}).
Observations of $\alpha$ Cen and $\alpha$ Aql show that the G-cloud is
within 1.3 pc, and the Apex cloud is within 5 pc of the Sun, giving
upper limits for an encounter date of $\sim$45,000 and $\sim$175,000
years, respectively.  The column densities of the G-cloud are $\sim$3
times larger than Apex cloud column densities, which are 
\NHI$\sim 10^{17.47}$ \cmtwo\ towards 70 Oph and $\alpha$ Aql.  Typical
G-cloud column densities are \NHI$\sim 10^{17.92}$ \cmtwo.

Both clouds are inhomogeneous.  Only the Apex cloud is seen towards AU Mic
(d=9.9 pc, \glong=12.7\deeg, \glat=--36.8\deeg).  Only the G-cloud is seen
towards $o$ Ser (d=13 pc, \glong=13.3\deeg, \glat=9.2\deeg) and $\nu$ Ser
(d=59 pc, \glong=10.6\deeg, \glat=13.5\deeg).  Both clouds are seen
towards $\alpha$ Aql (d=5 pc, \glong=47.7\deeg, \glat=--8.9\deeg) and $\alpha$
Oph (d=14 pc, \glong=35.9\deeg, \glat=22.6\deeg).

The G-cloud \CaII\ component (\NCaII=1.58$\times 10^{11}$ \cmtwo)
towards $\alpha$ Oph is extraordinarily strong for a nearby star, and
it implies a density of \nHI$>$5 \cc\ in the G-cloud, which is
$\sim$15 times the density of the LIC (\cite{Frisch:2003apex}).  This
estimate is found by using the G-cloud limit of
$N$(\CaII)/$N$(\HI)$<7.1 \times 10^{-9}$ towards $\alpha$ Cen in order
to estimate \NHI\ for the G-cloud in the $\alpha$ Oph sightline, and
by assuming the G-cloud is entirely foreground to $\alpha$ Cen, giving
a distance limit of 1.3 pc.  However, should the G-cloud temperature
be colder towards $\alpha$ Oph than towards $\alpha$ Cen, or should a
significant amount of the G-cloud be beyond the distances of 36 Oph
and $\alpha$ Cen, then the requirement for a high density becomes
diminished.


\section[The Solar Environment and Global ISM]{The Solar Environment and
Global ISM}
\label{sec:global}

The global properties of ISM in the solar neighborhood offer hints
about potential variations in the ISM that might impact the
heliosphere.

Over the past $\sim$50 years, radio and UV data have slowly disproved
simple ISM models consisting of gravitationally clumped clouds
embedded in a substrate of warm or hot low density gas.  Evidence
supporting a filamentary nature for low column density ISM, with \NH$<
10^{21}$ \cmtwo, is provided by \HI\ 21 cm observations of high
latitude \HI\ filaments, and also by UV observations that are
interpreted as high space density (\nHI$>$10 \cc), low column density
( \NHI$< 10 ^{19}$ \cmtwo\ and \NHII$< 10^{18}$ \cmtwo) nearby gas
(\cite{Hartmann:1997,Welty23:1999,Welty:2002zeta}).  Towards 23 Ori,
warm low velocity diffuse gas has densities in the range \nHI=15--20
\cc, and cloud thicknesses, $L$, are 0.7--0.9 pc.  The densities of
cold clouds, with temperatures $\leq$100 K, are slightly lower,
\nHI=10--15 \cc.  Diffuse ionized gas is present at intermediate
velocities, $|V_{\rm LSR}|\geq 20$ \kms, \nel$\le$5 \cc, and $L$=
0.0006--0.04 pc.  Thin diffuse ionized clouds, with \nel$\sim$0.17
\cc, $L$= 0.2--2.7 pc, and similar velocities, are observed towards
Orion stars spread over $>$15\deeg, which indicates that these clouds
must be thin sheets or filaments.

Winds of massive stars and supernova in young stellar associations
eject energy into the surrounding ISM, and evacuate cavities in the
ISM that are surrounded by irregular shells of WNM.  These shells,
which may overlap or be filled with X-ray plasma, are traced by \HI\
21-cm emission, and explain some of the ISM filamentary structure
(\cite{MacLowMcCray:1988,CoxSmith:1974,Dickey:2004,HTII}).  The CLIC
appears to be a fragment of a superbubble shell from one epoch of star
formation in the SCA, and $\alpha$ Oph may be in a direction that is
tangential to this shell (\cite{Frisch:1995}).

\subsection[Warm Ionized and Partially Ionized Gas]{Warm Ionized and
Partially Ionized Gas}\label{sec:wim}

Interstellar plasma and neutrals interact with the heliosphere in
fundamentally different ways, and small variations in the interstellar
radiation field convert neutral gas to plasma. Nearby WPIM was discussed
in \S \ref{sec:clicwim}, and we now compare warm interstellar plasma
over 100 pc scales and show that the CLIC and global observed diffuse
ionized gas are similar.  We conclude that dense ionized ISM has not
been, and will not be, part of the immediate solar environment over
short timescales.  However, diffuse WPIM similar to the LIC and CLIC is
much more widespread.

Dense fully ionized ISM is not found close to the Sun, although
diffuse ionized gas surrounds many hot stars at distances of $>$150
pc, such as $\beta$ CMa bordering the giant ionized Gum Nebula,
$\beta$ Cen, which energizes the interior of Loop I, and several stars
in Upper Scorpius.  Generally \Htwo\ regions surround hot O, early B,
and white dwarf stars that emit strongly in the EUV, $\lambda<$912
\AA.  The solar path is unlikely to traverse an \Htwo\ region
surrounding an O--B1 star over timescales of $ \pm $4 Myrs because
there are no hot stars or dense \Htwo\ regions nearby. The closest
\Htwo\ region surrounds the high latitude star $\alpha$ Vir, 80 pc
away.  White dwarf stars are relatively frequent near
the Sun.  Over 25 white dwarf stars with surface temperatures $T> 10
^4$ K are found within 20 pc, and $>40$\% of these stars are hotter
than 15,000 K.  Small Stromgren spheres will surround the hottest
white dwarf stars, but the densities will be very low for nearby stars
(\cite{TatTerzian:1999}).

Diffuse ionized gas is widespread, in contrast to dense ionized gas.
It is traced by weak optical recombination lines such as the H$\alpha$
line, N\,\textsc{ii}, He\,\textsc{i}, He\,\textsc{ii}, O\,\textsc{i},
and Si\,\textsc{ii} (\cite{WHAM:2003},\cite{Reynolds:2004}), UV
absorption lines
(\cite{Welty23:1999,Holbergetal:1999,Lehneretal:2003}, GJ), and EUV
observations of \HeI\ and \HeII\ towards white dwarf stars (\S
\ref{sec:clic}).  Pulsar dispersion measures trace the interactions
between pulsar wave packets and electrons
(e.g. \cite{TaylorCordes:1993,CordesLazio:2002,Armstrongetal:1995}).
These data reveal a low density, ionized component, \nel$\sim$0.1 \cc,
which fills $\sim$20\% of the volume of a 2-kpc-thick layer around the
Galactic disk.  The diffuse \HII\ has been mapped in the northern
hemisphere by WHAM, however the LIC emission measure is an order of
magnitude below the WHAM sensitivity (e.g. \cite{Reynolds:2004}).
About 30\% of diffuse \HI\ and \HII\ gas are found in spatially and
kinematically associated clouds with densities $\sim$0.2--0.3 \cc\
(\cite{ReynoldsTufteHeiles:1995}).  If the \HI\ and \HII\ were
spatially coincident in these clouds then \chiH\ would be $\sim 0.4$,
but Reynolds et al.\ argue that the \HI\ and \HII\ are spatially
separated.  Despite the evidence from spatial maps and line widths,
however, we believe that the existence of some true WPIM (perhaps at
lower ionization levels) cannot be ruled out in these regions. These
clouds are associated with large filamentary or sheet-like structures,
and exhibit cloud densities and hydrogen ionization levels that are
similar to LIC values. In other directions, \Halpha\ line emission
indicates that H is highly ionized in the WIM, \chiH$\sim$1.  Emission
at 588 \AA\ from \HeI\ shows that for the WIM \chiHe$\sim$0.3--0.6.
Ratios of \chiHe/\chiH\ are $\sim$0.3--0.6 are found, versus $\sim$1.6
for the LIC, implying that the LIC radiation field is harder than for
most of the WIM, perhaps because of lower column densities.  The WIM
has a continuum of temperatures, 5000--10,000 K, and ionization levels,
as demonstrated by emission in [N\,\textsc{ii}], [O\,\textsc{i}], and
[S\,\textsc{ii}] lines that trace temperature and ionization.

The interstellar plasma is structured over all spatial scales, and
these plasma clumps appear to be similar to the LIC.  The plasma is
partially opaque to radio emission in the energy range 0.1 to 10 MHz,
and as a result synchrotron emission probes the WIM clumpiness
(\cite{KulkarniHeiles:1988,PetersonWebber:2002}).  Models for the
propagation of low frequency synchrotron emission through a clumpy WIM
fit the radio data better than do propagation models, that assume a
uniformly distributed parallel slab model for the WIM.  These clumps
have density \nel$\sim$0.2 \cc, fill 8\%--15\% of the disk, and have a
free-free opacity at 10 MHz that is consistent with WIM temperatures
as low as 4500 K.  The dispersion, refraction and scintillation of
pulsar wave packets indicates that the WIM is turbulent over scale
sizes $10^{-2} - 10^{2}$ AU, or even larger, $\sim 10 ^2$ pc, if
Faraday rotation and plasma density gradients are included
(\cite{Armstrongetal:1995}).

The WIM properties are consistent with predictions of the radiative
transfer models, which find $\chi$(\HII)=0.15--0.35, $T \sim 5,000-9,000$
K, and \nHI=0.2--0.3 \cc\ for low column density clouds (Fig.
\ref{fig:rt}, \cite{SlavinFrisch:2002}). For very low column densities,
$<$10$^{18}$ \cmtwo, a continuum of ionization and temperature levels
are expected from variations in the cloud column density and the EUV
radiation field.  The WNM and WIM in the CLIC may show similar
variations, if the \NHI$\sim$10$^{19}$ \cmtwo\ clouds represent clusters
of lower column density objects.  The uncertain role of turbulence in
cloud evolution allow this possibility.

\subsection[Neutral Interstellar Gas and Turbulence]{Neutral Interstellar 
Gas and Turbulence}\label{sec:cnmwnm}

Radio and UV observations of low \NHI\ cloudlets in the solar vicinity
indicate that dense, low column density, cloudlets may be hidden in
the CLIC flow.  Radio 21-cm data indicate ISM of this type is
widespread, and has column densities and kinematics similar to CLIC
ISM.  

Two populations of interstellar clouds, WNM and CNM, are found based
on observations of the collisionally populated H\,\textsc{i} hyperfine
21-cm line towards radio-continuum sources.  The recent Arecibo
Millennium Survey used an ON-OFF observing strategy to survey the
properties of \HI\ clouds towards $\sim$80 radio continuum sources,
including emission and and opacity profiles, which were then fit with
$\sim$375 Gaussian components.  The detailed attention paid to sources
of noise and line contamination yielded a data set that provides a new
perspective on the statistics of low \NHI\ CNM and WNM clouds
(\nolinebreak \cite{HTI,HTII}).

The CNM and WNM components observed in the Arecibo survey 
show that $\sim$25\% of the ISM \HI\ mass is contained in 
clouds traveling with velocities $\ge$10
\kms\ through the local standard of rest.
When the solar apex motion of 13--20 \kms\ is included,
this means that Sun-cloud encounters with relative velocities 
larger than 25 \kms\ are quite likely, as we might have guessed
because of the LIC heliocentric velocity of --26.3 \kms.  

For 21-cm absorption components with spin
temperature \Ts, the opacity is $\tau \propto$\NHI/\Ts\ and the line
width is FWHM$\propto$\sigV, where \sigV\ is the dispersion in cloud
velocity.  Therefore the turbulent ($\Delta V_{\rm turb}$) and thermal
contributions to \sigV\ can be distinguished for the CNM.  Spin
temperatures, \Ts, for the CNM are typically $\sim$20--100 K.

The Arecibo survey discovered that 60\% of the diffuse ISM is WNM, and
$\sim$40\% is CNM.  Median column densities for the WNM and CNM
towards sources with $|$\glat$|>$10\deeg\ are \NHI=1.3 $\times$
10$^{19}$ \cc\ and 5.2 $\times$ 10$^{19}$ \cc, respectively.  The
finding that CNM column densities are typically lower than WNM
values was unexpected.  The column density weighted median spin
temperature for the CNM is 70 K, although minimum temperatures are 20 K or
lower.  Upper limits on WNM kinetic temperatures are 15,000 K or
higher, with a typical value of 4,000 K.  The turbulent velocities
derived for CNM components indicate that the turbulence must be highly
supersonic, with a turbulent Mach number $M_{\rm turb} =3 \Delta
V_{\rm turb}^2 / C^2_{\rm s} \sim$3, where $C^2_{\rm s}$ is the sound
speed and the factor of 3 converts to three-dimensional turbulence
(\cite{Heiles:2004HI}).  The densities of the CNM observed towards 23 Ori
are $>$10 \cc\ (\cite{Welty23:1999}).

\begin{figure}[ht] 
\caption[Velocities of Interstellar Warm and Cold H$^\circ$ Clouds.]{
SEE figures 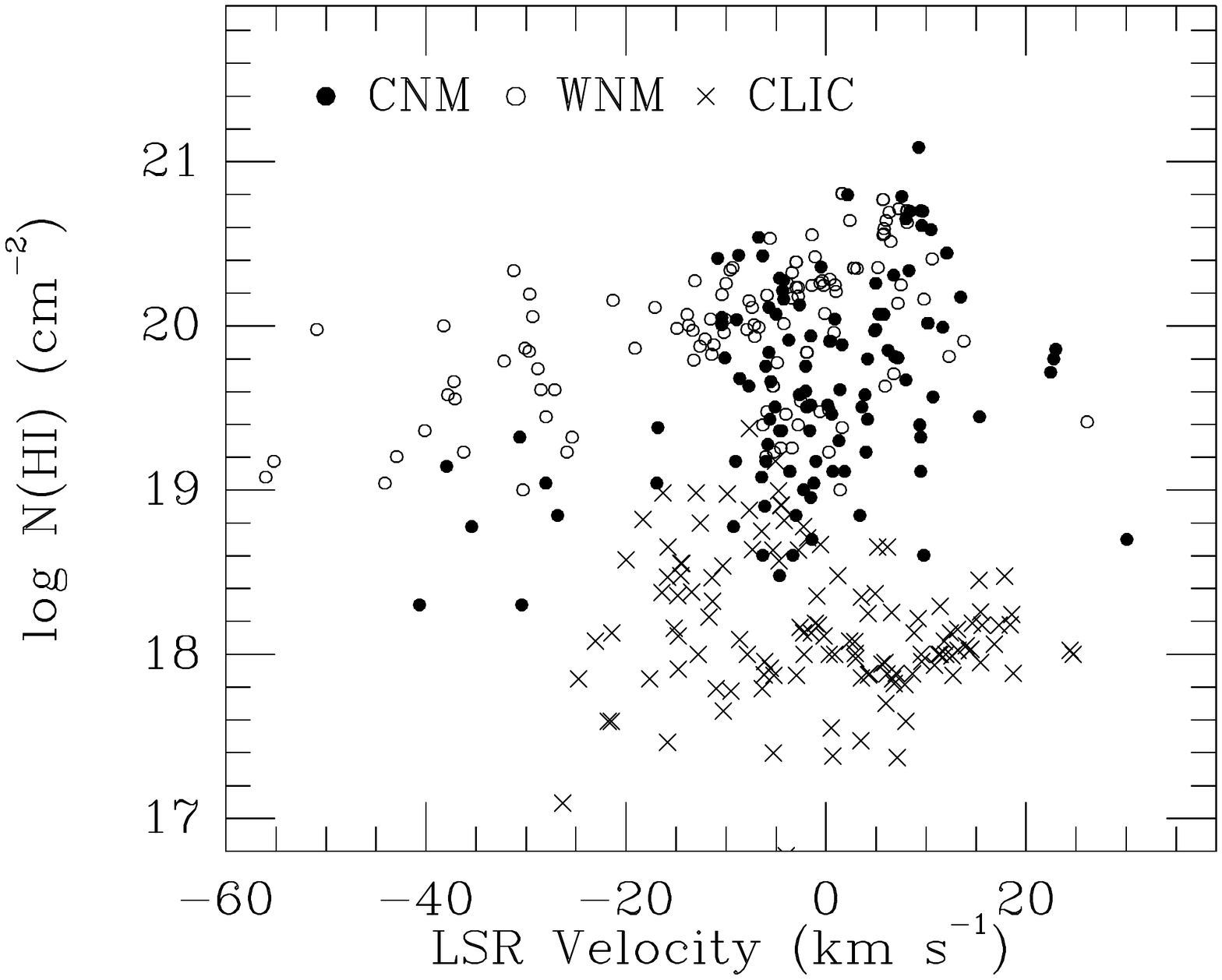 and 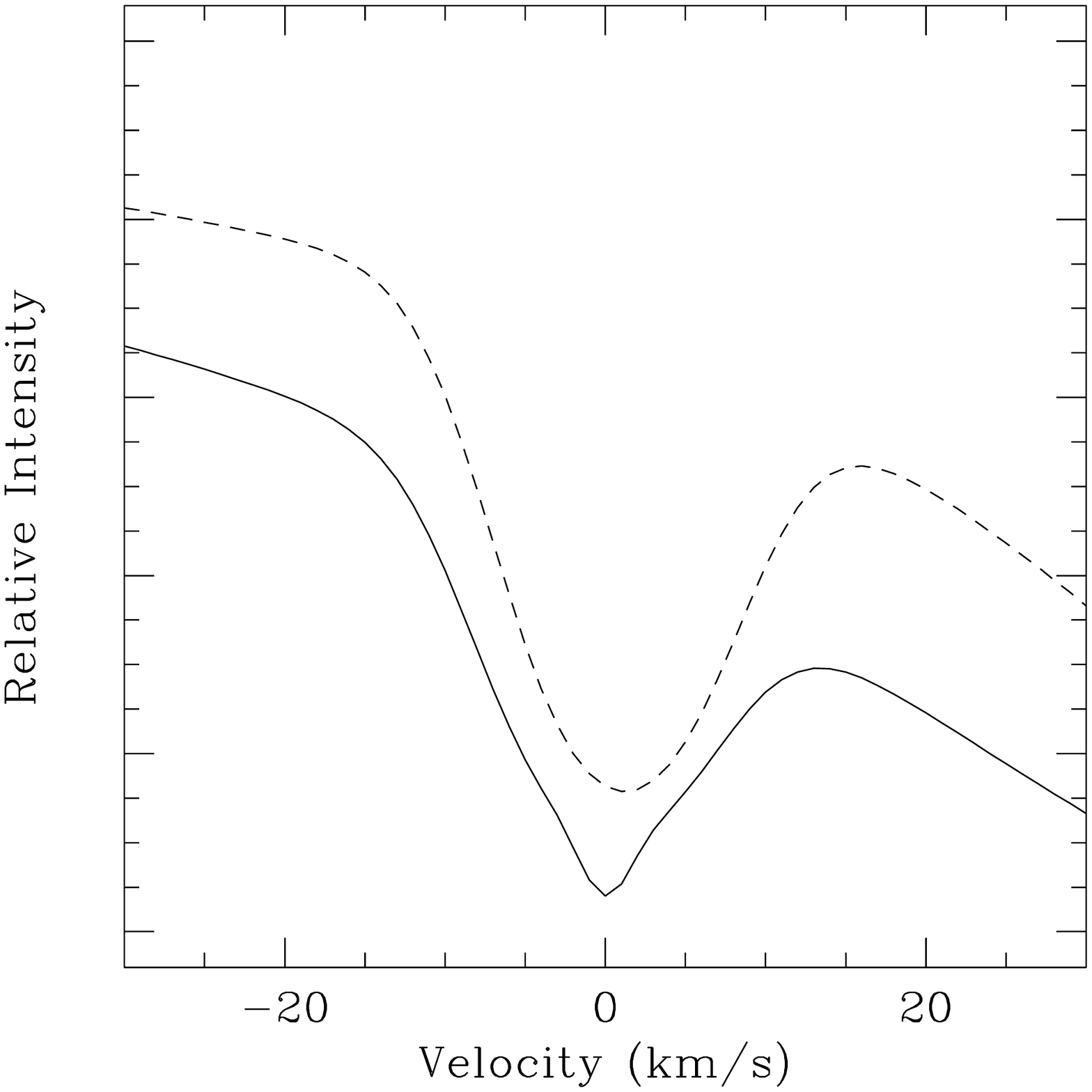.
\label{fig:arecibo} Left:  CLIC velocities, for d$<$50 pc, are plotted
against CNM and WNM velocities for sightlines with
$|$\glat$|>$25\deeg\ in the Arecibo Millennium Survey \cite{HTII}.  The
Arecibo clouds are within $\sim$425 pc if the \HI\ layer thickness is
185 pc.  The CLIC \NHI\ is based on \DI/\HI$=1.5 \times 10 ^{-5}$,
\CaII/\HI$=8.3 \times 10 ^{-9}$, or \MgII/\HI$= 6.7 \times 10 ^{-6}$, and \lsrstd\ is
used.  Right: This composite profile shows that some CNM components can
be lost in the WNM at UV resolutions.  The solid line shows a \DI\
1215 \AA\ absorption component formed by two clouds at the same
velocity (0 \kms) and column density (\NDI$=10^{13.17}$ \cmtwo), but
different temperatures (50 K and 7,000 K).  The dashed line shows WNM
only, with \NDI$=10^{13.17}$ \cmtwo\ and $T=7000$ K.  This theoretical
profile incorporates a Gaussian instrument resolution of FWHM 3 \kms\
appropriate for UV data.  Turbulent broadening is not included.
 }
\end{figure}

The velocities of CLIC, WNM, and CNM
components are compared in Fig. \ref{fig:arecibo}.  Only
high latitude WNM and CNM data with $|$\glat$|>$25\deeg\ are included,
in order to avoid distant gas near the galactic plane.  Some of
the CNM and WHM components are found at velocities lower 
than --25 \kms, which may be partly due to infalling \HI\ gas at high
latitudes.  Otherwise, the CLIC kinematics and column densities
overlap generic WNM and CNM values.

Of special interest are very low column density cold CNM,
or ``tiny'' clouds, found for a broad range of velocities.
An example is a component with \Ts=16.6 K, \NHI=2
$\times$10$^{\rm 18}$ \cmtwo\ and velocity $-40.6$ \kms\ observed towards 3C
225.  Another example is a component with \Ts=43 K, \NHI$ = 3 \times 10^{18}$
\cmtwo, and velocity --4.6 \kms\ observed towards 4C 32.44.  In addition,
CNM components with \NHI$\sim$10$^{18}$ \cc, \Ts$\sim$30--90 K, and 
velocities --30 to 0 \kms, are confirmed towards several other sources
(\cite{StanimirovicHeiles:2005,BraunKanekar:2005}).

CNM ISM is detected in the UV towards 23 Ori, with low column
densities, weak ionization, and low temperatures:  \NH$\gtsim 2$ x $10 ^{19}$
\cmtwo, \Ts$\sim$60--150 K, and $\sim$1\% ionization.  Cold clouds with
lower column densities may be difficult to resolve at UV resolutions. The
densities of these clouds are found from pressure considerations using
C\,\textsc{i} and C\,\textsc{ii} fine structure lines
(\cite{Jenkins:2002}), and yield \nH$\sim$10-15 \cc\ and a cloud
thickness of $\sim$0.5 pc (\cite{Welty23:1999}).  These components are
formed in the Orion-Eridanus Bubble shell.  Similar features are
observed in other Orion stars.  This CNM is either filamentary or
sheet-like, and it is a plausible template for the CNM in
Fig. \ref{fig:arecibo}.

The evident crowding of absorption components in velocity space may
introduce problems with the interpretation of UV data (\S
\ref{sec:absline}).  CNM and WNM clouds with low \NHI\ and similar
velocities would be unresolved at a nominal 3 \kms\ UV resolution, as
shown by the theoretical composite \DI\ profile plotted in Fig.
\ref{fig:arecibo}.  This profile is synthesized from the expected
profiles of two clouds at 0 \kms, temperatures 7,000 K and 50 K clouds, 
and each with \NHI=10$^{18}$ \cmtwo.  An instrumental resolution of 3 \kms\
is included, but turbulence is not.  Resolution problems are
worse for heavy elements, since the line FWHM$\propto mass^{\rm -1/2}$.
Therefore local low \NHI\ cold cloudlets can not be ruled out, and if
present would be traversed by the heliosphere in $\sim$1--20,000
years.

\subsection[Magnetic Fields]{Magnetic Fields}  \label{sec:magnetic}

Interstellar magnetic fields interact strongly with the heliosphere,
and modify the propagation of galactic cosmic rays in space.
Unfortunately the uniformity and strength of the interstellar magnetic
field (\Bis) at the Sun are unknown.  We showed that magnetic field
strengths of \Bis$\sim$0.6--3.7 $\mu$G would satisfy several
assumptions about the equilibrium of local ISM (\S
\ref{sec:pressure}).  However, the only direct data on the nearby
magnetic field are the starlight polarization data attributed to
weakly aligned interstellar grains in the nearest $\sim$35 pc
({\cite{Tinbergen:1982,Frisch:1990}).  More recently, it appears as if
these grains are instead trapped in the magnetic wall of the
heliosheath (\cite{Frisch:2005L}).  The global nearby \Bis\ thus
must be used to obtain insight on \Bis\ near the Sun.

The properties of the interstellar magnetic field, including the weak
spatially uniform (\Bu) and stronger random (\Br) components, are
known for both CNM and WIM (e.g.
\cite{Beck:2001,CrutcherHT:2003,Heileszeeman:2004,HeilesCrutcher:2005}).
The CNM median field strength of 6.0$\pm$1.8 $\mu$G is determined from
data on the Zeeman splitting of the H\,\textsc{i} 21 cm line
(\cite{HTIV}), but may not apply to the CLIC where CNM is not yet
detected.  Although \Bis$\sim n$ for densities larger than 10$^3$ \cc,
at lower densities \Bis\ strengths are uncorrelated with cloud density,
in violation of simple flux freezing assumptions.

Pulsar data, synchrotron emission, and starlight polarization data
indicate that the ratios of the random (\Br) and uniform (\Bu)
components of the local field are \Bu/\Br$\sim$0.6--1.0.  Synchrotron
emission of external galaxies indicate that the ratio \Bu/\Br\
increases in interarm regions such as surrounding the Sun, but
decreases in spiral arm regions as the magnetic fields become more
random.  Comparisons of magnetic field strengths determined from
polarized synchrotron emission and pulsar dispersion measure
observations, which interact with \Bis\ and the WIM, give the total
local field strength and the relative strengths of \Bu\ and \Br.
Synchrotron emission gives a total field strength locally of \Bis$\sim
6 \pm 2 ~\mu$G, and the uniform component strength \Bu $\sim 4 ~\mu$G.
Faraday rotation from the propagation of the oppositely polarized
circular components produces a polarization phase shift, given by the
rotation measure RM =$\int_0^L B_{\rm ||}$(x)$n_{\rm e} (x) dx$, which
responds to the parallel component of the magnetic field, $B_{\rm
||}$, weighted by the electron density, \n$_{\rm e}(x)$, and
integrated over pathlength $L$.  \Bis\ is directed away from the
observer for RM$<$0.  Wave packet dispersion in the WIM gives the
dispersion measure DM = $\int_0^L \n_{\rm e} (x) dx$.  The mean local
field, weighted by \nel, is found from the ratio RM/DM and is
\Bpar$\sim 6.5 ~ \mu$G, where \Bu$\sim$\Bpar.

Starlight polarization data show that the local \Bis\ is relatively
uniform, and not strongly strongly distorted or tangled by turbulence.
Locally the total field, \Bis, follows the pitch angle, 9.4\deeg, of
the local spiral arm, and has a curvature radius of $\sim$7.8 kpc
centered on \glong=344.6\deeg.  The starlight polarization vectors
converge near \glong$\sim$80\deeg\ and $\sim$270\deeg, where the
random component of the field, \Br, results in a polarization angle
that is rotated over all angles.

A large excursion from \Bu\ is seen towards within 150 pc, towards
Loop I, where both polarized starlight and synchrotron emission
indicate a magnetic field conforms to the superbubble shell around the
Scorpius-Centaurus Association.  Surveys of \Bpar\ in morphologically
distinct objects such as the Loop I, Eridanus, and Orion supershells,
show that the magnetic field is strong, \Bpar$>5 ~\mu$G, and ordered
throughout these shell features.

\subsection[Origin of Local ISM]{Origin of Local ISM} \label{sec:origin} 

Several origins have been suggested for the CLIC.  CLIC kinematics and
abundance patterns led to the suggestion that the CLIC is part of a
superbubble shell surrounding the Scorpius-Centaurus Association
(\cite{Frisch:1981,FrischYork:1986,deGeus:1992,Frisch:1995}).  The SCA
is surrounded by a well-known supershell formed by epochs of star
formation, winds from massive stars and supernovae
(\cite{Weaver:1979}).  Using superbubble expansion models and the star
formation rates in the SCA, Frisch modeled the CLIC as a fragment of
the shell from the formation of the Upper Scorpius subgroup ($\sim 4$
Myrs old) of the SCA, which has expanded away from the high-density
ISM close to the association to the lower pressure regions around the
Sun (\cite{Frisch:1995}).  A related suggestion attributes the origin
of the CLIC to Rayleigh-Taylor instabilities in the shell of the
younger Loop I supernova remnant ($\sim$0.5 Myrs old), which has
propelled ISM towards the Sun (\cite{Breitschwerdtetal:2000}).  A
third origin suggests that the CLIC originates as a magnetic flux tube
that is originally captured in the inner wall of the Local Bubble
surrounding the Sun, and is driven back into the Local Bubble interior
by magnetic tension (\cite{CoxHelenius:2003}).  This scenario requires
that the CLIC velocity and magnetic fields are parallel, which appears
unlikely because of the asymmetric capture of small interstellar dust
grains in the heliosheath (\cite{Frisch:2005L}).  Each of these
suggestions is consistent with the CLIC kinematics, but should have
different consequences for short-term variations in the solar
environment.

\section[Summary]{Summary} \label{sec:conclusions}

In this chapter we have tried to present the essential information
required to understand short-term variations in the galactic
environment of the Sun, and at the same time make an educated guess
about the times that the Sun has transitioned between different types
of ISM for the past $10 ^3 - 10^5$ years.  It is our hope some day
that the geological radio isotope records can be searched for the
signatures of these transitions, although earlier searches have been
inconclusive (e.g.,
\cite{Frisch:1981,SonettJokipii:1987,Frisch:1997}).

Both the Sun and ISM move through our local neighborhood of the Milky
Way Galaxy, which is described by the LSR rest velocity frame (with
its own uncertainties).  The Sun has been in the nearly empty space of
the Local Bubble interior for several million years.  Sometime in the
past $\sim$140,000/\nHIavgtwo\ years, the Sun exited the nearly
empty Local Bubble and entered an outflow of ISM with an upwind
direction towards the Scorpius-Centaurus Association.  Even more
recently, $\sim$40,000/\nHIavgtwo\ years ago, the Sun entered the
cloud now surrounding us, the LIC.  Within $\sim$4,000 years the Sun
will exit the LIC.  What's next?  The flow of local ISM past the Sun
is bearing down on the Sun at a relative velocity of $\sim$28 \kms,
the cloudlets in this flow are expected to pass over the Sun more
often than roughly once per $\sim$70,000 years.  The next cloud the
Sun is most likely to traverse is either the G-cloud or the Apex
Cloud.  This cloud must be either either inhomogeneous or denser by a
factor of $\sim$30 than the LIC.

Most of this nearby ISM is warm partially ionized gas, not dissimilar
to warm ISM observed elsewhere, except that column densities tend to be
lower.  The properties of this gas are mainly consistent with
predictions of radiative transfer models, which then suggests that the
local gas is close to equilibrium.  If viewed from elsewhere in the
galaxy, the ensemble of ISM close to the Sun could easily pass as warm
neutral material such as observed by H\,\textsc{i} 21 cm data.  The
general characteristics of nearby ISM are similar to the global warm
neutral material, except that column densities are very low so that
\HI\ ionizing photons penetrate to the cloud cores.  The physical
properties of the CLIC vary both between and within clouds.  The low
column densities of this ISM lead to ionization variations inside of a
cloud.  As the Sun passes between clouds, the interstellar ionization,
dynamic pressure or the thermal pressure may vary, possibly all at once.  
As a consequence, the boundary conditions of the heliosphere may experience dramatic
variations as the Sun moves through the CLIC.  The efforts to
understand these variations have just begun.

One of the more intriguing possibilities is that tiny cool ISM
cloudlets are undetected in the nearby warm partially ionized gas
(WPIM).  Such structures, with \NHI$\sim 10 ^{18}$ \cmtwo, are seen
elsewhere, although infrequently.  A new class of cloud models, perhaps
based on macro-turbulence, will be required if such structures exist
locally.  

Most of our conclusions have substantial inherent uncertainties.  The
most prominent is that all distances, such as those to cloud edges,
are based on \NHI, which is moderately well known for most sightlines
discussed here, and an assumption about the mean \HI\ space density,
\nHIavg.  We have a very good idea of the value for \nHIavg\ in the
LIC because of extensive development of radiative transfer models for
the LIC.  Our confidence in the models follows from their ability to
predict both ISM densities inside of the heliosphere, including the
Ulysses satellite values for \nHeI, and ISM column densities towards
nearby stars.  Since column densities are too small locally for
measurements of the fine-structure lines that yield \nHI, we have
assumed that the LIC value for \nHIavg\ applies elsewhere in the CLIC.
In addition, the transition epochs for the Sun are calculated by
assuming that there are no gaps between individual cloudlets, as
defined by velocity.  The LIC \nHIavg\ value is probably
representative of the downwind CLIC, because the Sun is so close to
the downwind edge.  However, we can not rule out gaps between the
cloudlets, and such gaps would alter estimates of the epoch the Sun
entered the CLIC.

One of the most significant improvements will be to make good maps of
nearby ISM.  Such maps would require a new generation of space
instrumentation, capable of high resolution, R$\sim$500,000, and high
signal-to-noise observations of both bright and faint objects over the
spectral range 912--3,000 \AA.  We also need more precise measurements
of the diffuse interstellar radiation field, throughout the full
spectral interval of 2000 \AA\ to 0.1 keV.  Such data would provide a
strong constraint on the interface emission and reduce uncertainties
in the \MgI\ photoionization rate.

In this chapter we have reviewed the
short-term variations in the galactic environment of the Sun.  Our
galactic environment affects the heliosphere, and by analogy the astrospheres 
of nearby stars, so that the space trajectory of a star is a
filter for other planetary systems with conditions conducive
to stable climates for exoplanets (\cite{Frisch:1993a}).
As other papers in this volume demonstrate, the ISM-modulated heliosphere
has a pronounced effect on the cosmic ray flux in the inner heliosphere,
which in turn appears to affect the climate.  Similar processes will
affect the climates of planets outside of the solar system.  Such possibilities make
the study of short-term variations in the galactic environment of the Sun
highly topical.

\vspace{0.2in}
\emph{Acknowledgments:}
The authors thank NASA for 
supporting this research through grants NAG5-11005, NAG5-13107
and NAG5-13558.  This article will appear in the book ''Solar Journey: 
The Significance of Our Galactic Environment 
for the Heliosphere and Earth'',
Springer, in press (2006), editor P. C. Frisch.


\emph{Submitted 19 July 2005; accepted 19 October 2005.}

\end{document}